\documentclass{elsart}
\usepackage{graphicx,amssymb}

\def\bea{\begin{eqnarray}}
\def\eea{\end{eqnarray}}
\def\be{\begin{equation}}
\def\ee{\end{equation}}
\newcommand{\ub}[1]{\underline{#1}}
\def\del{\partial}
\def\g{\gamma}
\def\psibar{\overline{\psi}}

\begin{document}

\begin{flushright}
SLAC-PUB-11400 \\
UMN-D-05-2 \\
SMUHEP/03-14
\end{flushright}

\begin{frontmatter}

\title{Two-boson truncation of Pauli--Villars-regulated Yukawa 
theory\thanksref{thanks1}}
\thanks[thanks1]{Work supported in part by the Department of Energy
under contract numbers DE-AC02-76SF00515, DE-FG02-98ER41087,
and DE-FG03-95ER40908.}

\author[SLAC]{Stanley J. Brodsky,}
\author[UMD]{John R. Hiller,} and
\author[SMU]{Gary McCartor}

\address[SLAC]{Stanford Linear Accelerator Center, Stanford University,
Stanford, California 94309}
\address[UMD]{Department of Physics, University of Minnesota-Duluth,
Duluth, Minnesota 55812}
\address[SMU]{Department of Physics, Southern Methodist University,
Dallas, TX 75275}

\begin{abstract}
We apply light-front quantization, Pauli--Villars regularization,
and numerical techniques to the nonperturbative solution of
the dressed-fermion problem in Yukawa theory in $3+1$ dimensions. 
The solution is developed as a Fock-state expansion truncated to 
include at most one fermion and two bosons.  The basis includes a 
negative-metric heavy boson and a negative-metric heavy fermion 
in order to provide the necessary cancellations
of ultraviolet divergences.  The integral equations for
the Fock-state wave functions are solved by reducing them to
effective one-boson--one-fermion equations for eigenstates
with $J_z=1/2$.  The equations are converted to a matrix equation
with a specially tuned quadrature scheme, and the lowest mass state 
is obtained by diagonalization.  Various properties of the 
dressed-fermion state are then computed from the nonperturbative 
light-front wave functions.
This work is a major step in our development of Pauli--Villars
regularization for the nonperturbative solution of four-dimensional
field theories and represents a significant advance in the
numerical accuracy of such solutions.
\end{abstract}


%
\begin{keyword}
light-cone quantization \sep Pauli--Villars regularization \sep
Yukawa theory
\PACS 12.38.Lg \sep 11.15.Tk \sep 11.10.Gh \sep 11.10.Ef
\end{keyword}

\end{frontmatter}

\section{Introduction}
\label{sec:Introduction}

One of the important unsolved problems in obtaining bound-state 
solutions in quantum field theories such as quantum chromodynamcs
is how to implement ultraviolet renormalization nonperturbatively.  
The central difficulty is that any truncation or approximation of 
the theory which breaks Lorentz symmetries, such as the truncation 
of the Fock space, introduces spurious divergences. 
Quadratic divergences can occur in the approximated theory,
even when the renormalizable perturbative Feynman theory has only 
logarithmic divergences.  Such problems arise even when one uses 
light-front quantization methods which retain the maximal set 
of kinematical Lorentz symmetries.  A systematic light-front 
approach to renormalization, developed by Glazek, Harindranath, Perry,  
and Wilson~\cite{Perry:1990mz,Perry:1993gp,Glazek:1994qc}, is to introduce 
new effective interactions which act as counterterms to control the
ultraviolet behavior of the approximated theory.   However, this 
method is challenging to implement in practice, because the 
large set of new effective interactions, some of which have a 
nonlocal structure, can require so much input data that the
ability to make predictions is seriously compromised. 

Another important nonperturbative approach is to cast the 
bound-state problem in the form of an effective Bethe--Salpeter 
equation and use Schwinger--Dyson methods to construct the effective 
renormalized quark and gluon propagators of the theory, including 
running quark masses consistent with chiral symmetry 
breaking~\cite{Burden:1996nh,Tandy:2004rk}. In principle, this 
method could be used to predict the light-front wave functions 
needed for QCD phenomenology, if one evaluates the 
Minkowski-space Bethe--Salpeter wave functions at fixed light-front 
time.  However, it has been difficult to carry out this program 
in practice since the analyses have only been done in ladder 
gluon-exchange approximation and in Euclidean space.  The full 
structure of nonperturbative renormalization will require
consideration of higher-order kernels.

Lattice gauge theory~\cite{Wilson:2004de} is currently the most 
successful method for solving gauge theories; it has a systematic 
gauge-invariant regularization procedure despite the fact that
nonlinear gauge interactions are introduced and the lattice 
structure itself violates Lorentz symmetries at finite lattice 
size.  However, unlike light-front methods, wave function 
information for bound states and other physical features of hadron 
dynamics have to be obtained indirectly through moments, because the
lattice theory is formulated in Euclidean space.

In our work~\cite{bhm1,bhm2,bhm3,bhm4,TwoParticles,qed} on 
light-front quantization, we have shown that the technique of
Pauli--Villars (PV) regularization~\cite{PauliVillars} of ultraviolet
divergences can be implemented in nonperturbative light-front calculations 
of field-theoretic bound states.  Until now this has been limited
to light-front Fock-space truncations where the resulting wave functions 
could be computed analytically~\cite{TwoParticles,qed} or to numerical 
calculations using discretized light-cone quantization 
(DLCQ) \cite{PauliBrodsky,DLCQreview} with limited precision~\cite{bhm3}.  
In this paper we will test the PV method in some detail and present accurate
results for the dressed-fermion state in Yukawa theory with a Fock
basis which includes three-particle states.

In order to perform our calculations using PV regularization, we first
introduce a sufficient number of PV fields in the underlying 
Lagrangian in order to render perturbation 
theory finite.  However, we must also make sure that our nonperturbative result, 
if expanded in a power series in the coupling constant, would give agreement 
with the usual Feynman series for processes that can be calculated 
perturbatively.  Obtaining such assurance may require the introduction 
of additional PV fields or counterterms, or both.  Paston and 
Franke~\cite{Paston:1997hs} have given a general set of 
rules which determine exactly what combination of PV fields and 
counterterms are needed to assure perturbative equivalence with 
Feynman methods.  In that paper they apply their methods to the case 
of Yukawa theory; in \cite{Paston:2000fq}, Paston, Franke and Prokhvatilov 
apply the methods to give the same information for QCD.  In the case of the 
Yukawa theory that we study here, one PV fermion and one PV boson are sufficient 
to assure perturbative equivalence with Feynman methods without the need 
for counterterms.  The PV regulators always preserve Lorentz invariance 
and may preserve gauge invariance, or at least the Ward identity.  In the
theories studied to date, the PV fields alone have been sufficient to give 
a finite theory, perturbatively equivalent to Feynman theory, without 
breaking Lorentz or gauge invariance; in future work we expect to include 
counterterms, along with the PV fields, in order to restore symmetries 
broken by the PV fields.  

We must truncate the Fock space in order to be able to 
perform a numerical calculation of the spectrum and the wave functions.  
So far, all of our calculations have been for the ground state.  The 
truncation breaks the symmetries of the theory, but since this introduces 
only finite breaking,  we argue that if our answer is close, numerically, 
to the answer without truncation, then the approximated result should 
yield a useful result.  This was shown to be the case in a nonperturbative 
calculation of the electron's magnetic moment in QED~\cite{qed}.

We will work with the theory in light-cone coordinates~\cite{Dirac}, in order
to have a well defined Fock expansion and a simple 
vacuum~\cite{DLCQreview}.  The field-theoretic bound-state problem
for the dressed fermion is then reduced to integral equations for 
the wave functions which appear in the Fock expansion.  The expansion 
is truncated to include states in three sectors: the bare fermion,
the one-boson--one-fermion states, and the two-boson--one-fermion
states.  We neglect fermion-pair contributions.

In Yukawa theory and in QED, both without fermion loops, it is adequate 
to include one PV fermion and one PV boson to regulate the ultraviolet
divergences. At the end of the calculation, 
we wish to take the PV masses large, and there is an ambiguity as to how to 
do that.  We can take the fermion PV mass much larger than the PV boson 
mass, take the PV masses to be about equal, or take the PV boson mass to 
be much larger than the PV fermion mass.  The answer that one obtains depends 
on the choice of this mass ratio~\cite{TwoParticles}.  In the case of QED, 
restoration of gauge 
invariance requires that the PV fermion mass be taken to infinity while the 
PV boson mass remains finite~\cite{qed}. (Although both types of PV 
particles must be 
included to render the theory finite, once the calculations are complete the 
limit of the PV fermion mass going to infinity exists.)  Of course, since 
the limit of the PV fermion mass going to infinity is finite, taking the 
fermion mass sufficiently large but finite will be adequate.  We believe 
that taking the fermion mass to infinity first is the physical limit in 
Yukawa theory as well, but we have no argument as strong as the one in 
the case of QED; in the present paper we shall show results for taking 
the PV fermion mass large first and also for keeping the PV masses equal.  
Due to the truncation of the representation space, there always remain 
uncancelled divergences, so at least one PV mass must remain finite.  

Unlike QED, where the wave function and vertex renormalizations cancel,
Yukawa theory has a logarithmically divergent charge renormalization
even without considering fermion loops and vacuum polarization.  Thus in our 
truncation we must consider renormalization of both charge and fermion mass.  
To handle these renormalizations nonperturbatively, we regulate the theory
and then fix the bare coupling and bare fermion mass by imposing conditions
on the mass and Dirac radius of the dressed-fermion eigenstate.

There are three objectives for the present work:  One is associated with 
the need to take the limit of the PV fermion mass to infinity first, 
leaving the theory regulated by just the PV boson mass.  While the answer 
for infinite PV fermion mass exists, an explicit solution is not known,
and for now we must settle for taking the fermion 
mass very large compared with the PV boson mass.  That choice of masses 
raises a strong possibility of numerical trouble.  Numerical methods work 
best in problems with a single scale, but here we have 
three scales: the physical mass scale, the PV boson mass scale, and the 
PV fermion mass scale.  In this paper we present techniques
which allow us to make accurate 
calculations even when the three mass scales are very different.

A further objective of the present work is associated with the need to 
keep one PV mass, usually the PV boson mass, finite, and therefore the 
need to choose a value for it.  In~\cite{qed} we proposed a method for 
choosing the PV boson mass based on the idea that there are two types 
of error associated with a finite PV mass, and that we should choose a 
value for which neither type of error is too large. (If no such value 
exists then our method will not work.)  Two elements of the proposed 
method of choosing a final PV mass are examined in the present paper.  
One proposal is that one can obtain a reasonable estimate of the weight of 
the true wave function on the lowest excluded Fock sector by performing 
a perturbation calculation using the calculated wave function as the 
unperturbed state.  The other proposal is that the weight of the wave 
function on the first excluded sector provides a useful estimate of 
the percent by which physical parameters would shift if the first 
excluded sector were to be included.  We use the results of a calculation 
including only up to two particles to perform a perturbative estimate of 
the weight of the true wave function on the three-particle states and then 
compare that result with the weight of the wave function in the 
three-particle sector when that sector is included in the calculation.  We 
also calculate a number of physical quantities for both the two-particle 
truncation and the three-particle truncation and compare the percent 
by which they change with the percent of the wave function that is in 
the three-particle sector.

The final objective of the present work is just to study the effects of 
adding an additional sector to the calculation.  This extends earlier
work~\cite{TwoParticles} where a truncation to two particles allowed 
analytic solutions.  We can see whether properties of the solutions
persist with inclusion of the three-particle sectors and whether new 
physics appears.

The PV fermion and PV boson are introduced to the theory in
such a way that the interaction term in the modified Lagrangian
is a coupling between zero-norm combinations of the physical and PV fields.  
This guarantees that instantaneous-fermion terms do not appear in the
light-cone Hamiltonian and only ordinary three-particle vertices remain.
The negative-metric particles also cancel infinities from integrals
over transverse momenta.  The numerical approximation is then
applied to a finite theory at fixed PV masses, and the behavior
of the solution is studied as the PV masses are increased.
The structure of the problem is simpler than in the case of 
quantum electrodynamics~\cite{qed}, not only because the
bosons are massive scalars but also because the kernels of
the integral operators are not plagued by singularities
caused by spurious thresholds.
Here the bare fermion mass is driven to large values rather
than the small values seen in QED; this prevents the singularities.

The solution to the eigenvalue problem for the dressed-fermion
state is obtained by first rearranging the coupled set of
integral equations for the mass eigenvalue problem into a set of
effective equations for the one-boson--one-fermion wave functions.
The kernels of these equations have contributions from intermediate
states containing a single bare fermion and states containing two
bosons and a fermion, including self-energy terms.  
The effective equations can be considered an eigenvalue problem for the 
bare coupling at fixed bare and dressed fermion masses.
The numerical solution provides the wave functions as well as
the bare coupling.  We can use these to compute various properties 
of the dressed fermion, including the magnetic moment
and axial coupling, as well as structure functions.

The traditional DLCQ method~\cite{PauliBrodsky,DLCQreview} 
has limited accuracy in the present case, because the solution
is sensitive to regions of small longitudinal light-cone
momentum fractions, on the order of the reciprocal of the
PV mass squared in units of the physical mass.  
The equal spacings in momentum used in
DLCQ must then be so large in number as to be impractical.
Here we use coordinate transformations and Gauss--Legendre
quadrature to capture these important regions.

The calculation of the solution is simplified 
by working in transverse polar coordinates.  The Fock
expansion is constructed to be an explicit eigenstate of $J_z$.
The dependence of the wave functions on the azimuthal angle
can then be determined exactly and removed from the numerical
calculation.  This reduces the effective dimension of the
numerical problem from three to two.

The notation that we use for light-cone coordinates is
\begin{equation}
x^\pm = x^0+x^3 ,\;\;
\vec{x}_\perp=(x^1,x^2) .
\end{equation}
The time coordinate is $x^+$, and the dot
product of two four-vectors is
\begin{equation}
p\cdot x=\frac{1}{2}(p^+x^- + p^-x^+)
                -\vec{p}_\perp\cdot\vec{x}_\perp .
\end{equation}
The light-cone momentum component conjugate to $x^-$ is $p^+$,
and the light-cone energy is $p^-$.
Light-cone three-vectors are identified by
underscores, such as
\begin{equation}
\ub{p}=(p^+,\vec{p}_\perp) .
\end{equation}
For additional details, see Appendix A of Ref.~\cite{bhm1}
or the review~\cite{DLCQreview}.

We begin in Sec.~\ref{sec:YukawaTheory} by introducing the light-cone
Hamiltonian and dressed-fermion Fock-state expansion for Yukawa theory
and by giving expressions for quantities to be computed from the
Fock-state wave functions.  Previous results for the one-boson
truncation~\cite{TwoParticles} are summarized in Sec.~\ref{sec:OneBoson}\@.
Section~\ref{sec:TwoBoson} contains the analysis of the two-boson
truncation, including a summary of the numerical techniques and
a presentation of the results.  Some concluding remarks are
given in Sec.~\ref{sec:Discussion}\@.  Details of the kernels and the
numerical approximation are left to two appendices.

Somewhat related calculations have been done by
Bylev, G{\l}azek, and Przeszowski~\cite{Bylev},
except that they did not use a covariant regulation
procedure.  Similar work in a purely scalar theory
has been done by Bernard {\em et al}.~\cite{Bernard},
and more recently the two-fermion problem has 
been considered by Mangin-Brinet {\em et al}.~\cite{ManginBrinet}.
For a more formal treatment of dressed
constituents, see~\cite{EffectiveFermions}.

Other light-cone methods that show promise include supersymmetric
DLCQ (SDLCQ)~\cite{SDLCQ} and the transverse lattice
approximation~\cite{TransLattice}.  Both are used specifically
for gauge theories.

\section{Yukawa theory}
\label{sec:YukawaTheory}

We consider Yukawa theory with a PV scalar and a PV fermion.
The action is 
\bea
\lefteqn{S=\int d^4x
\left[\frac{1}{2}(\del_\mu\phi_0)^2-\frac{1}{2}\mu_0^2\phi_0^2
-\frac{1}{2}(\del_\mu\phi_1)^2+\frac{1}{2}\mu_1^2\phi_1^2\right.} \\
 &&+\frac{i}{2}\left(\psibar_0\g^\mu\del_\mu-(\del_\mu\psibar_0)\g^\mu\right)
     \psi_0
  -m_0\psibar_0\psi_0  \nonumber \\
&&\left.
  -\frac{i}{2}\left(\psibar_1\g^\mu\del_\mu-(\del_\mu\psibar_1)\g^\mu\right)
      \psi_1+m_1\psibar_1\psi_1 
      -g(\phi_0 + \phi_1)(\psibar_0 + \psibar_1)(\psi_0 + \psi_1)\right].
\nonumber
\eea
The subscript 0 indicates physical fields and 1, PV fields.  The fermion
masses are denoted by $m_i$, and the boson masses by $\mu_j$.
When antifermions are excluded, the resulting light-cone
Hamiltonian is
\bea \label{eq:YukawaP-}
\lefteqn{P^-=
   \sum_{i,s}\int d\ub{p}
      \frac{m_i^2+\vec{p}_\perp^2}{p^+}(-1)^i
          b_{i,s}^\dagger(\ub{p}) b_{i,s}(\ub{p})} \\
   && +\sum_{j}\int d\ub{q}
          \frac{\mu_j^2+\vec{q}_\perp^2}{q^+}(-1)^j
              a_j^\dagger(\ub{q}) a_j(\ub{q})  \nonumber \\
   && +\sum_{i,j,k,s}\int d\ub{p} d\ub{q}\left\{
      \left[ V_{-2s}^*(\ub{p},\ub{q})
             +V_{2s}(\ub{p}+\ub{q},\ub{q})\right]
                 b_{j,s}^\dagger(\ub{p})
                  a_k^\dagger(\ub{q})
                   b_{i,-s}(\ub{p}+\ub{q})\right. \nonumber \\
      &&\left.\rule{0.5in}{0in}
           +\left[U_j(\ub{p},\ub{q})
                    +U_i(\ub{p}+\ub{q},\ub{q})\right]
               b_{j,s}^\dagger(\ub{p})
                a_k^\dagger(\ub{q})b_{i,s}(\ub{p}+\ub{q})
                    + h.c.\right\},  \nonumber
\eea
where $a_j^\dagger$ creates a boson of type $j$,
$b_{i,s}^\dagger$ creates a fermion of type $i$ and spin $s$,
\be
U_j(\ub{p},\ub{q})
   \equiv \frac{g}{\sqrt{16\pi^3}}\frac{m_j}{p^+\sqrt{q^+}},\;\;
V_{2s}(\ub{p},\ub{q})
   \equiv \frac{g}{\sqrt{8\pi^3}}
   \frac{\vec{\epsilon}_{2s}^{\,*}\cdot\vec{p}_\perp}{p^+\sqrt{q^+}},
\ee
and
\be
\vec{\epsilon}_{2s}\equiv-\frac{1}{\sqrt{2}}(2s,i). 
\ee
The nonzero (anti)commutators are
\bea
\left[a_i(\ub{q}),a_j^\dagger(\ub{q}')\right]
          &=&(-1)^i\delta_{ij}
            \delta(\ub{q}-\ub{q}'), \\
\left\{b_{i,s}(\ub{p}),b_{j,s'}^\dagger(\ub{p}')\right\}
     &=&(-1)^i\delta_{ij}   \delta_{s,s'}
            \delta(\ub{p}-\ub{p}').  \nonumber
\eea
The opposite signature of the PV fields is the reason that
no instantaneous-fermion terms appear in $P^-$; these terms are
individually independent of the fermion mass and cancel
between instantaneous physical and PV fermions.

\begin{figure}[hbpt]
\begin{center}
\includegraphics[width=10cm]{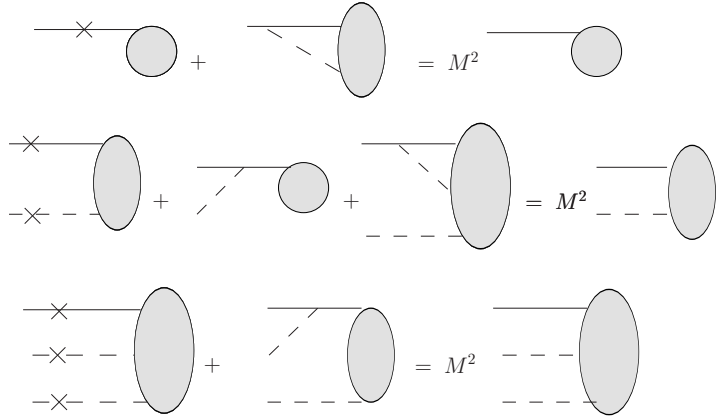} 
\caption{Diagrammatic representation of the first
three coupled equations for the wave functions of
the dressed-fermion state, Eqs.~(\ref{eq:noboson})-(\ref{eq:twoboson})
in the text.  The large blobs represent wave functions, the
crosses represent light-cone energies, the solid lines indicate the 
fermion constituent, and the dashed lines correspond to bosons.
}
\label{fig:coupledeqns}
\end{center}
\end{figure}

We expand the eigenfunction for the dressed-fermion state
in a Fock basis as
\bea
\lefteqn{\Phi_+(\ub{P})=\sum_i z_i b_{i+}^\dagger(\ub{P})|0\rangle
  +\sum_{ijs}\int d\ub{q} f_{ijs}(\ub{q})b_{is}^\dagger(\ub{P}-\ub{q})
                                       a_j^\dagger(\ub{q})|0\rangle}&& \\
 && +\sum_{ijks}\int d\ub{q_1} d\ub{q_2} f_{ijks}(\ub{q_1},\ub{q_2})
       \frac{1}{\sqrt{1+\delta_{jk}}}   b_{is}^\dagger(\ub{P}-\ub{q_1}-\ub{q_2})
                 a_j^\dagger(\ub{q_1})a_k^\dagger(\ub{q_2})|0\rangle 
 +\ldots       \nonumber
\eea
and normalize it according to
$\Phi_\sigma^{\prime\dagger}\cdot\Phi_\sigma
=\delta(\ub{P}'-\ub{P})$.
The wave functions $f$ that define this state must satisfy the coupled 
system of equations that results from the field-theoretic
mass-squared eigenvalue problem $P^+P^-\Phi_+=M^2\Phi_+$, since we work in the
frame where $\vec{P}_\perp$ is zero.  The state has $J_z=+1/2$.
The first three coupled equations are
\bea \label{eq:noboson}
m_i^2z_i&+& \sum_{i',j}(-1)^{i'+j} P^+ \int^{P^+} d\ub{q}
  \left\{ f_{i'j-}(\ub{q})[V_+(\ub{P}-\ub{q},\ub{q})+V_-^*(\ub{P},\ub{q})]
  \right. \nonumber \\
  &&\left.
   + f_{i'j+}(\ub{q})[U_{i'}(\ub{P}-\ub{q},\ub{q})+U_i(\ub{P},\ub{q})]\right\}
   = M^2z_i,
\eea
\bea \label{eq:oneboson}
\lefteqn{
\left[\frac{m_i^2+q_\perp^2}{1-y}+\frac{\mu_j^2+q_\perp^2}{y}\right]
  f_{ijs}(\ub{q}) 
  +\sum_{i'}(-1)^{i'}\left\{
    z_{i'}\delta_{s,-}[V_+^*(\ub{P}-\ub{q},\ub{q})+V_-(\ub{P},\ub{q})] 
                               \right.}\nonumber \\
  && \rule{2in}{0in}\left.
    +z_{i'}\delta_{s,+}[U_i(\ub{P}-\ub{q},\ub{q})+U_{i'}(\ub{P},\ub{q})]\right\}
     \\
    &&+2\sum_{i',k}\frac{(-1)^{i'+k}}{\sqrt{1+\delta_{jk}}}P^+ \int^{P^+-q^+}
    d\ub{q}' \left\{f_{i'jk,-s}(\ub{q},\ub{q}')
       [V_{2s}(\ub{P}-\ub{q}-\ub{q}',\ub{q}')\right. \nonumber \\
       && \rule{2in}{0in} +V_{-2s}^*(\ub{P}-\ub{q},\ub{q}')]
       \nonumber \\
           &&\left.+f_{i'jks}(\ub{q},\ub{q}')
             [U_{i'}(\ub{P}-\ub{q}-\ub{q}',\ub{q}')+U_i(\ub{P}-\ub{q},\ub{q}')]
             \right\} = M^2f_{ijs}(\ub{q}),\nonumber
\eea
and
\bea \label{eq:twoboson}
\lefteqn{\left[\frac{m_i^2+(\vec{q}_{1\perp}+\vec{q}_{2\perp})^2}{1-y_1-y_2}
    +\frac{\mu_j^2+q_{1\perp}^2}{y_1}+\frac{\mu_k^2+q_{2\perp}^2}{y_2}\right]
        f_{ijks}(\ub{q_1},\ub{q_2})} \\
    &&+\sum_{i'}(-1)^{i'}\frac{\sqrt{1+\delta_{jk}}}{2}P^+ \nonumber \\
  && \times\left\{f_{i'j,-s}(\ub{q_1})
[V_{-2s}^*(\ub{P}-\ub{q_1}-\ub{q_2},\ub{q_2})
           +V_{2s}(\ub{P}-\ub{q_1},\ub{q_2})] \right. \nonumber \\
 &&+ f_{i'js}(\ub{q_1})
[U_i(\ub{P}-\ub{q_1}-\ub{q_2},\ub{q_2})+U_{i'}(\ub{P}-\ub{q_1},\ub{q_2})]
 \nonumber \\
 &&  + f_{i'k,-s}(\ub{q_2})
[V_{-2s}^*(\ub{P}-\ub{q_1}-\ub{q_2},\ub{q_1})+V_{2s}(\ub{P}-\ub{q_2},\ub{q_1})]
  \nonumber \\
 &&  \left. + f_{i'ks}(\ub{q_2})
[U_i(\ub{P}-\ub{q_1}-\ub{q_2},\ub{q_1})+U_{i'}(\ub{P}-\ub{q_2},\ub{q_1})]
\right\}+\ldots \nonumber \\  
   && = M^2f_{ijks}(\ub{q_1},\ub{q_2}). \nonumber
\eea
The equations are invariant under Lorenta boosts: $P^+\rightarrow \gamma P^+$.
We represent these diagrammatically in Fig.~\ref{fig:coupledeqns}.

Each wave function has a total $L_z$ eigenvalue of 0 for $s=+1/2$ and
1 for $s=-1/2$.  For the one-boson wave functions, this greatly restricts
the dependence on the azimuthal angle; however, for the two-boson wave
functions, the total $L_z$ eigenvalue can be obtained in an infinite
number of ways by combining different individual orbitals.

We define the physical wave functions as the coefficients of
Fock states containing only particles with positive-definite norm.
This reduction can be achieved by requiring all Fock states
to be expressed in terms of the positive-norm creation operators
$b_{0s}^\dagger$ and $a_0^\dagger$ and the zero-norm
combinations $b_s^\dagger\equiv b_{0s}^\dagger+b_{1s}^\dagger$
and $a^\dagger\equiv a_0^\dagger+a_1^\dagger$.
We then discard any term containing a $b_s^\dagger$ or an $a^\dagger$,
which would be annihilated by the PV-generalized electromagnetic
current, leaving the physical state
\bea
\lefteqn{\Phi_{+ {\rm phys}}=(z_0-z_1)b_{1+}^\dagger(\ub{P})|0\rangle} \\
  &&         + \sum_s \int d\ub{q}
\left(\sum_{ij}(-1)^{i+j}f_{ijs}(\ub{q})\right)
     b_{0s}^\dagger(\ub{P}-\ub{q})a_0^\dagger(\ub{q})
              |0\rangle  \nonumber \\
 && +\sum_s\int d\ub{q_1} d\ub{q_2} 
 \left(\sum_{ijk}\frac{(-1)^{i+j+k}}{\sqrt{1+\delta_{jk}}}
      f_{ijks}(\ub{q_1},\ub{q_2})  \right) \nonumber \\
 &&\rule{2in}{0in}  \times   b_{0s}^\dagger(\ub{P}-\ub{q_1}-\ub{q_2})
                 a_0^\dagger(\ub{q_1})a_0^\dagger(\ub{q_2})|0\rangle +\ldots
                 \nonumber
\eea
It is normalized by
\bea \label{eq:normint2}
\lefteqn{1=(z_0-z_1)^2+\sum_s\int d\ub{q}
     \left|\sum_{ij}(-1)^{i+j}f_{ijs}(\ub{q})\right|^2} && \\
 && +\sum_s\int d\ub{q_1} d\ub{q_2} \left|\sum_{ijk}(-1)^{i+j+k}
     \frac{\sqrt{2}}{\sqrt{1+\delta_{jk}}}f_{ijks}(\ub{q_1},\ub{q_2})\right|^2
     +\ldots \nonumber
\eea
{}From this state we can compute a boson structure function
\bea \label{eq:fB}
\lefteqn{f_{Bs}(y)=\int d\ub{q} \delta(y-q^+/P^+)
   \left|\sum_{ij}(-1)^{i+j}f_{ijs}(\ub{q})\right|^2}&& \\
   &&+\int d\ub{q_1} d\ub{q_2} \sum_{n=1}^2\delta(y-q_n^+/P^+)
   \left|\sum_{ijk}(-1)^{i+j+k}\frac{\sqrt{2}}{\sqrt{1+\delta_{jk}}}
       f_{ijks}(\ub{q_1},\ub{q_2})\right|^2+\ldots ,
   \nonumber
\eea
which is defined as the probability density for finding a constituent boson of 
longitudinal momentum fraction $y$ when the constituent fermion has helicity $s$.
The form factor slope $F'_1(0)$ is given by~\cite{bhm1}
\bea \label{eq:Fprime}
\lefteqn{F'_1(0)=-\sum_s\int d\ub{q}
     \left|\frac{y}{2}\vec{\nabla}_\perp
                   \sum_{ij}(-1)^{i+j}f_{ijs}(\ub{q})\right|^2} && \\
 && -\sum_s\int d\ub{q_1} d\ub{q_2} \sum_l
     \left|\frac{y_l}{2}\vec{\nabla}_{l\perp} \sum_{ijk}(-1)^{i+j+k}
     \frac{\sqrt{2}}{\sqrt{1+\delta_{jk}}}f_{ijks}(\ub{q_1},\ub{q_2})\right|^2
     -\ldots ,
    \nonumber
\eea
with $\vec{\nabla}_{l\perp}=\hat{x}\frac{\partial}{\partial q_{lx}}
+\hat{y}\frac{\partial}{\partial q_{ly}}$.
Since no transverse cutoff is used, this expression is no longer the 
approximation that it was in earlier work~\cite{bhm1,bhm2,bhm3}.
The Dirac radius of the state is given by $R=\sqrt{-6F'_1(0)}$.
{}From the two-body wave function, we compute a distribution function
\be \label{eq:phi}
\phi(x,Q_\perp)\equiv\int^{Q_\perp}\frac{d^2k_\perp}{\sqrt{16\pi^3}}
    \sum_{ij}(-1)^{i+j}f_{ij+}(1-x,-\vec k_\perp)
\ee
and its moments
\bea \label{eq:moments}
f_M&\equiv&\int_0^1 dx\phi(x,Q_\perp),\;\;
\bar{x}\equiv\int_0^1 dx\,x\phi(x,Q_\perp)/f_M, \\
\bar{x^2}&\equiv&\int_0^1 dx\,x^2\phi(x,Q_\perp)/f_M. \nonumber
\eea
We can also calculate the axial coupling constant
\bea \label{eq:gA}
\lefteqn{g_A=(z_0-z_1)^2+\sum_s (-1)^{s-1/2}\int d\ub{q}
     \left|\sum_{ij}(-1)^{i+j}f_{ijs}(\ub{q})\right|^2} && \\
 && +\sum_s (-1)^{s-1/2}\int d\ub{q_1} d\ub{q_2} \left|\sum_{ijk}(-1)^{i+j+k}
     \frac{\sqrt{2}}{\sqrt{1+\delta_{jk}}}f_{ijks}(\ub{q_1},\ub{q_2})\right|^2
     +\ldots  \nonumber
\eea
and the anomalous magnetic moment~\cite{BrodskyDrell}
\bea \label{eq:kappa}
\lefteqn{\kappa=-M\sum_s\int d\ub{q} 
  \left(\sum_{ij}(-1)^{i+j}f_{ijs}^{\uparrow *}(\ub{q})\right)
   y\left(\frac{\partial}{\partial q_x}+i\frac{\partial}{\partial q_y}\right)
     \left(\sum_{ij}(-1)^{i+j}f_{ijs}^\downarrow(\ub{q})\right)} && 
     \nonumber \\
 && -M\sum_s\int d\ub{q_1} d\ub{q_2} 
    \left(\sum_{ijk}(-1)^{i+j+k}\frac{\sqrt{2}}{\sqrt{1+\delta_{jk}}}
                   f_{ijks}^{\uparrow *}(\ub{q_1},\ub{q_2})\right) \\
 && \rule{0.5in}{0mm} \times
    \sum_l \left[y_l\left(\frac{\partial}{\partial q_{lx}}
                        +i\frac{\partial}{\partial q_{ly}}\right)\right]
     \left(\sum_{ijk}(-1)^{i+j+k}
     \frac{\sqrt{2}}{\sqrt{1+\delta_{jk}}}
         f_{ijks}^\downarrow(\ub{q_1},\ub{q_2}) \right) \nonumber \\
 &&  -\ldots
    \nonumber
\eea
Here an up or down arrow indicates wave functions associated with
a dressed fermion having a $J_z$ value that is
positive or negative, respectively.
The wave functions with an up arrow are the same as those
without an arrow; the coupled system of equations is
constructed and solved for this case.  The wave functions
for the opposite spin are related as 
$f_{\ldots\pm}^\downarrow=\mp f_{\ldots \mp}^{\uparrow *}$,
as can be established by comparing the equations that they
each satisfy.

Once the coupled equations are
truncated to a finite system, we have a well-defined problem
with cancellations of infinities between any infinite integrals.  
The PV particles are kept in the basis to provide these cancellations.
A one-boson truncation produces an analytically solvable problem,
which we explored in \cite{TwoParticles} and discuss briefly here in 
Sec.~\ref{sec:OneBoson}\@.  Less severe truncations produce larger
coupled systems that require numerical techniques for their solution.
Given an accurate discretization and the consequent finite matrix
eigenvalue problem, one can compute mass eigenvalues and 
associated wave functions.  The bare parameters, i.e. the bare coupling
$g$ and the bare mass $m_0$ of the positive-norm fermion, 
can be fixed by fitting ``physical'' constraints.  Here we specify
the dressed fermion's mass $M$ and radius $R$; the fit to a 
chosen value of $R$ is determined by an iterative root-finding
scheme.

The discretization must be chosen carefully.
When the PV masses are large, the integrals become sensitive
to small momentum fractions, of order $\mu_0^2/\mu_1^2$ and
$\mu_0^2/m_1^2$.  This makes traditional DLCQ~\cite{PauliBrodsky}
impractical, because it divides the longitudinal momentum 
into segments of equal size; the number of segments then
grows as $\mu_1^2/\mu_0^2$ and $m_1^2/\mu_0^2$, making the
matrix diagonalization problem impossibly large.  We instead
employ a discretization based on Gauss--Legendre quadrature
and certain variable transformations, as discussed
more fully in Sec.~\ref{sec:TwoBoson} and Appendix~B\@.  
The derivatives needed for computing the radius and the anomalous
moment are estimated from cubic-spline fits to 
$\left[(m_i^2+q_\perp^2)/(1-y)+(\mu_j^2+q_\perp^2)/y\right]
  f_{ijs}(y,q_\perp)$ as a function of $q_\perp$ at fixed $y$.
The use of the multiplier reduces the variation in $f$ and
makes possible a better fit.  Standard finite-difference 
approximations are not useful, because the quadrature scheme
introduces very unequal spacings.

\section{One-boson truncation}
\label{sec:OneBoson}

\begin{figure}[hpbt]
\begin{center}
\begin{tabular}{cc}
\includegraphics[width=6.5cm]{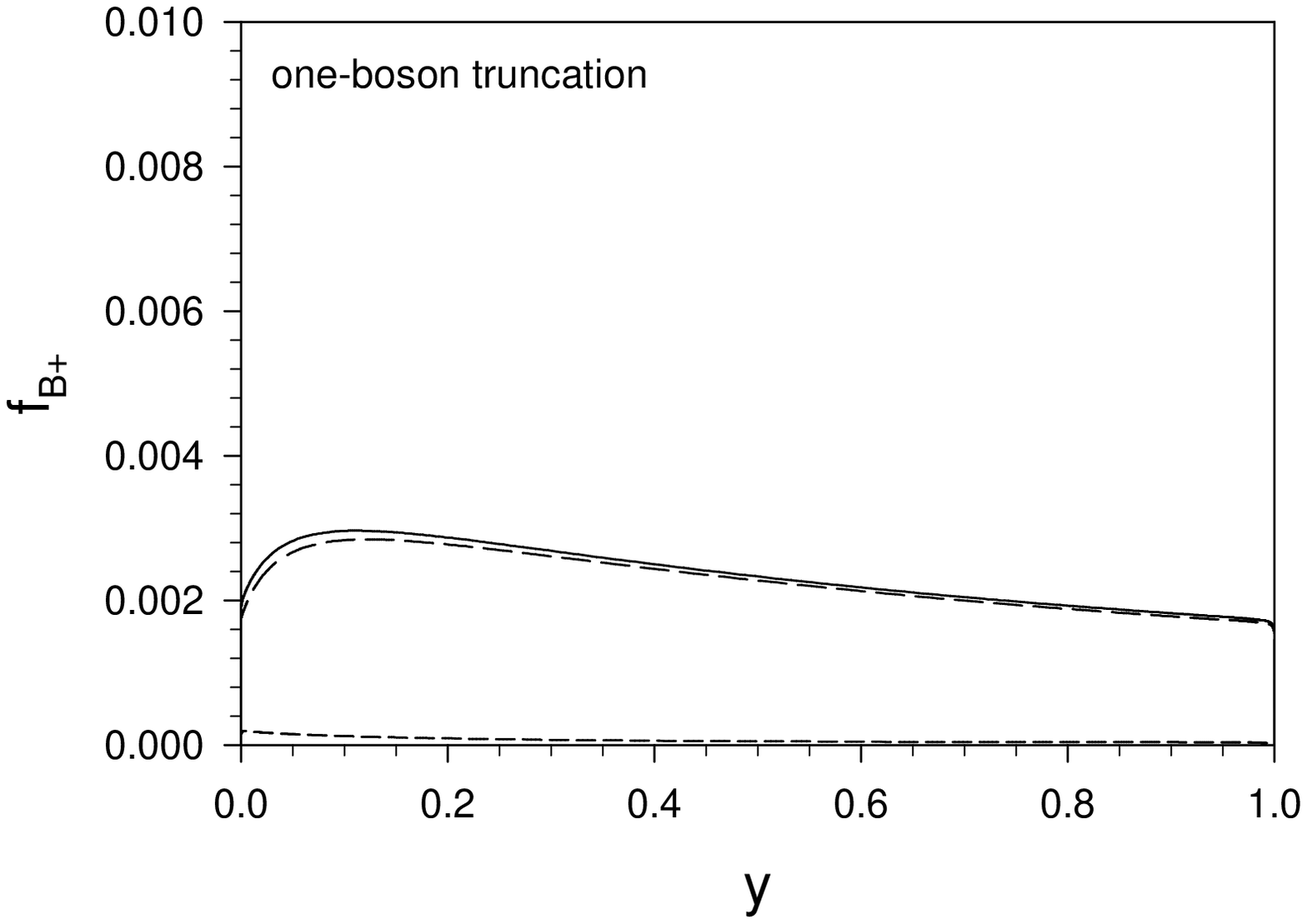} &
\includegraphics[width=6.5cm]{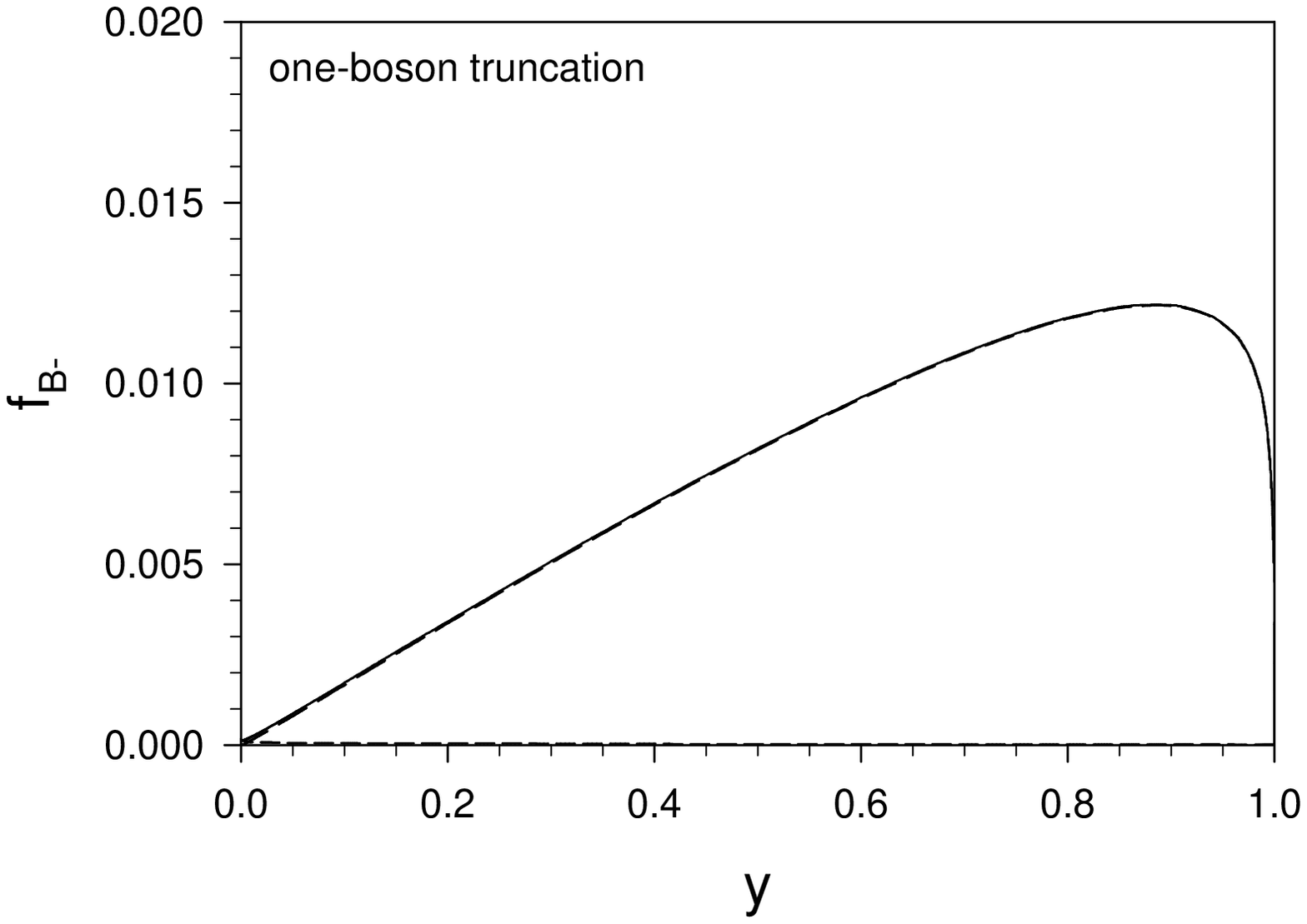} \\
(a) & (b) \\ \vspace{0.05in} \\
\includegraphics[width=6.5cm]{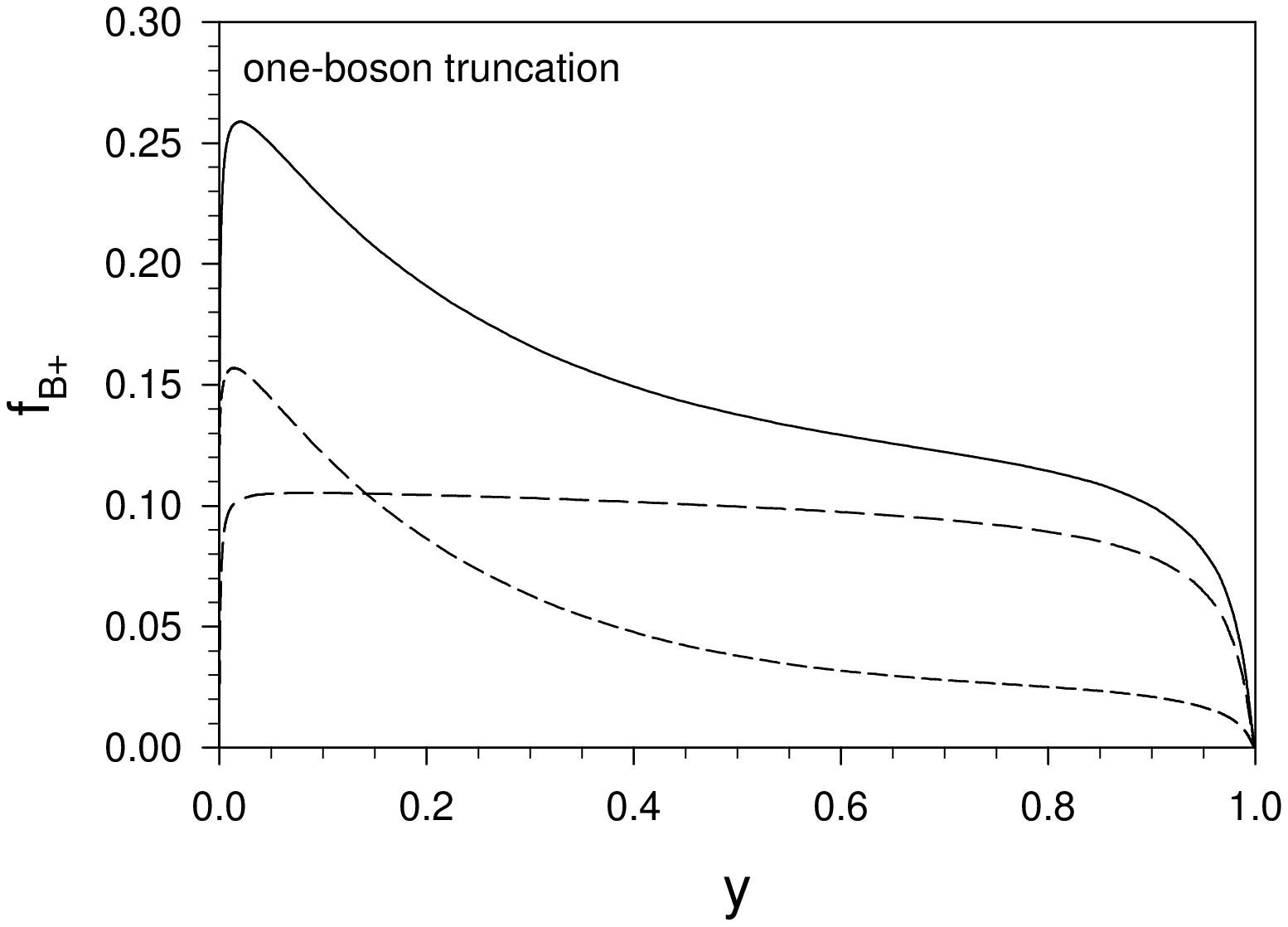} &
\includegraphics[width=6.5cm]{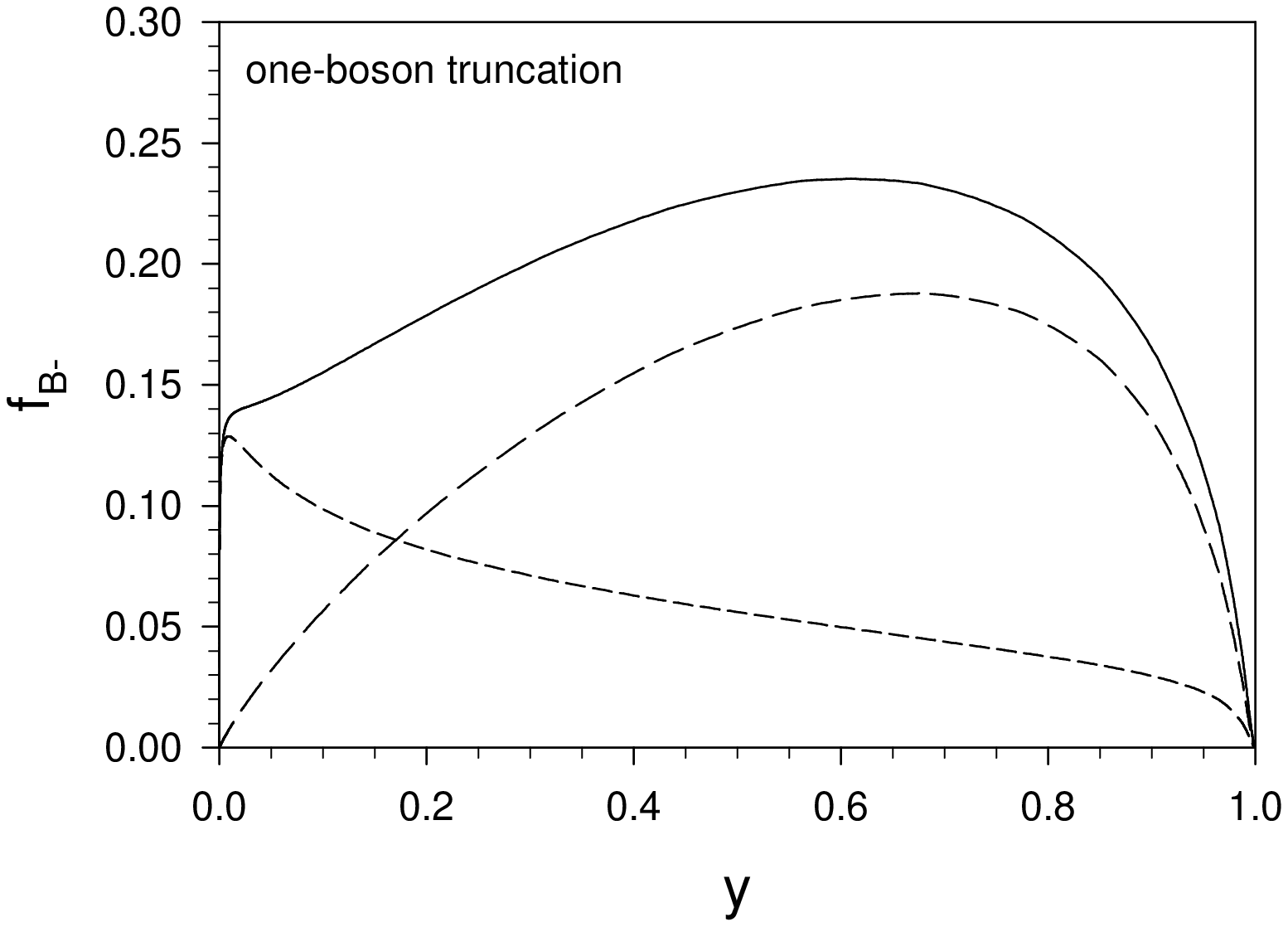} \\
(c) & (d)
\end{tabular}
\caption{Bosonic structure functions $f_{B\pm}(y)$ as defined in
Eq.~(\ref{eq:fB}) of the text, where $y = x_B = q^+/P^+$,
for the dressed-fermion state
with mass $M=\mu_0$ and radius $R=0.01/\mu_0$.  The wave functions 
are computed from
a truncation to one constituent boson, with the two-boson
contribution then computed perturbatively.  The solid line includes
both contributions.  The long dashes show the one-boson contribution,
and the short dashes show the two-boson contribution.
For (a) and (b), the PV masses are $m_1=\mu_1=2000\mu_0$,
and for (c) and (d), they are $m_1=50000\mu_0$ and $\mu_1=500\mu_0$.
}
\label{fig:1bfb}
\end{center}
\end{figure}

\begin{figure}[hpbt]
\begin{center}
\begin{tabular}{cc}
\includegraphics[width=6.5cm]{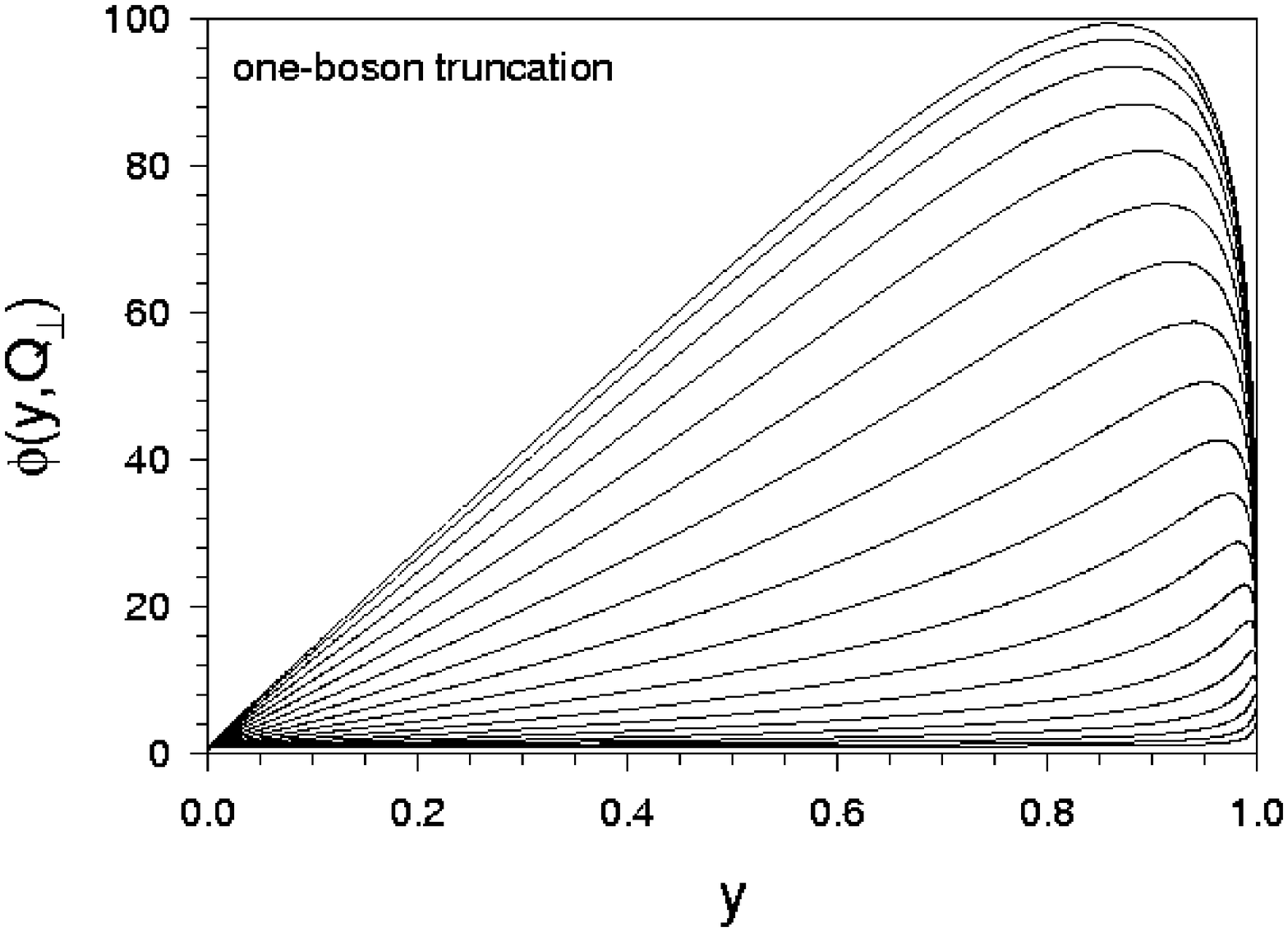} &
\includegraphics[width=6.5cm]{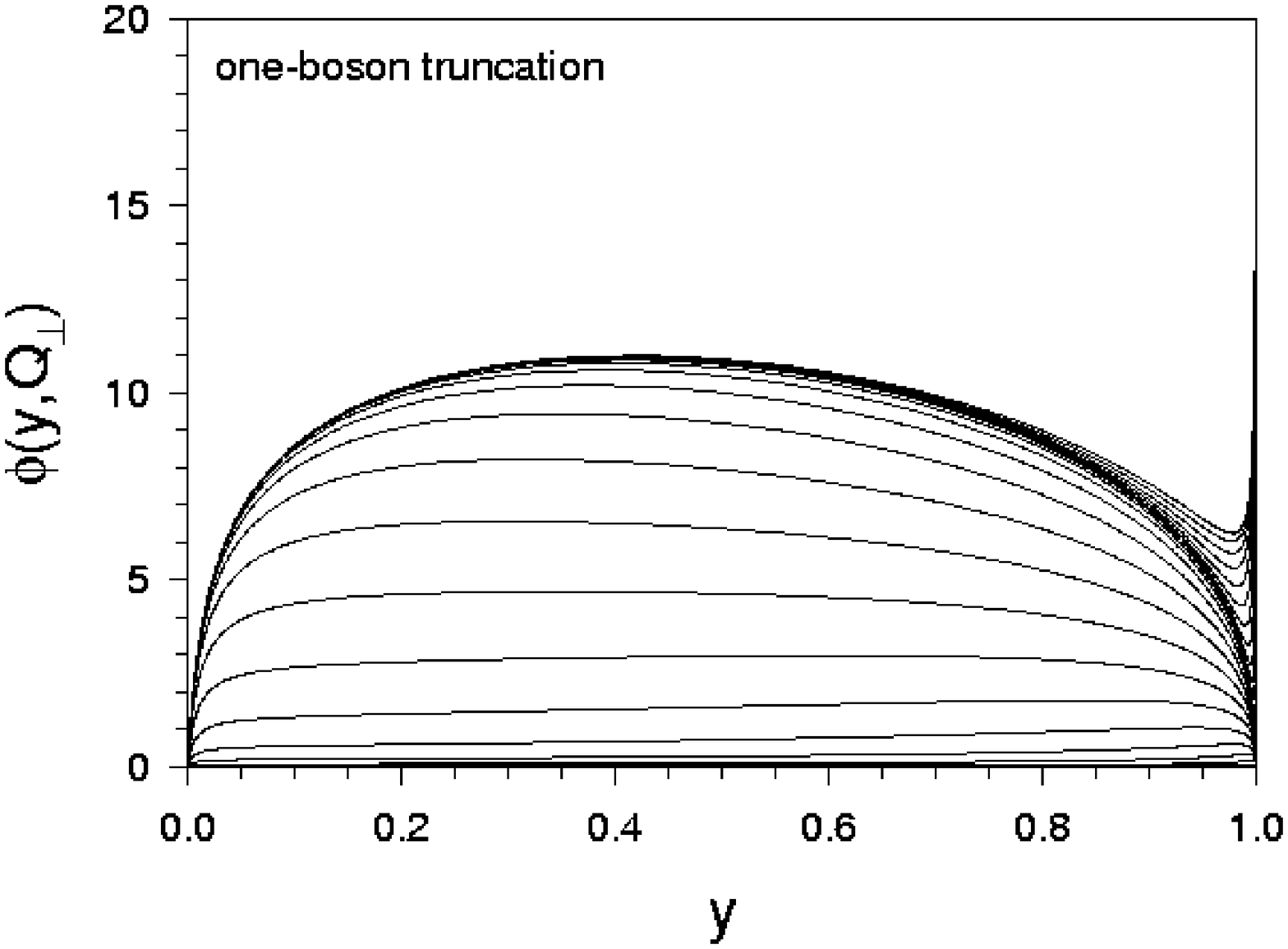} \\
(a) & (b)
\end{tabular}
\caption{Same as Fig.~\ref{fig:1bfb} but for the 
distribution amplitude $\phi(x,Q_\perp)$ defined in
Eq.~(\ref{eq:phi}) of the text.
The individual lines correspond to different values of $Q_\perp$.
For (a) the PV masses are $m_1=\mu_1=2000\mu_0$,
and for (b) they are $m_1=50000\mu_0$ and $\mu_1=500\mu_0$.
}
\label{fig:1bphi}
\end{center}
\end{figure}

\begin{figure}[hpbt]
\begin{center}
\begin{tabular}{cc}
\includegraphics[width=6.5cm]{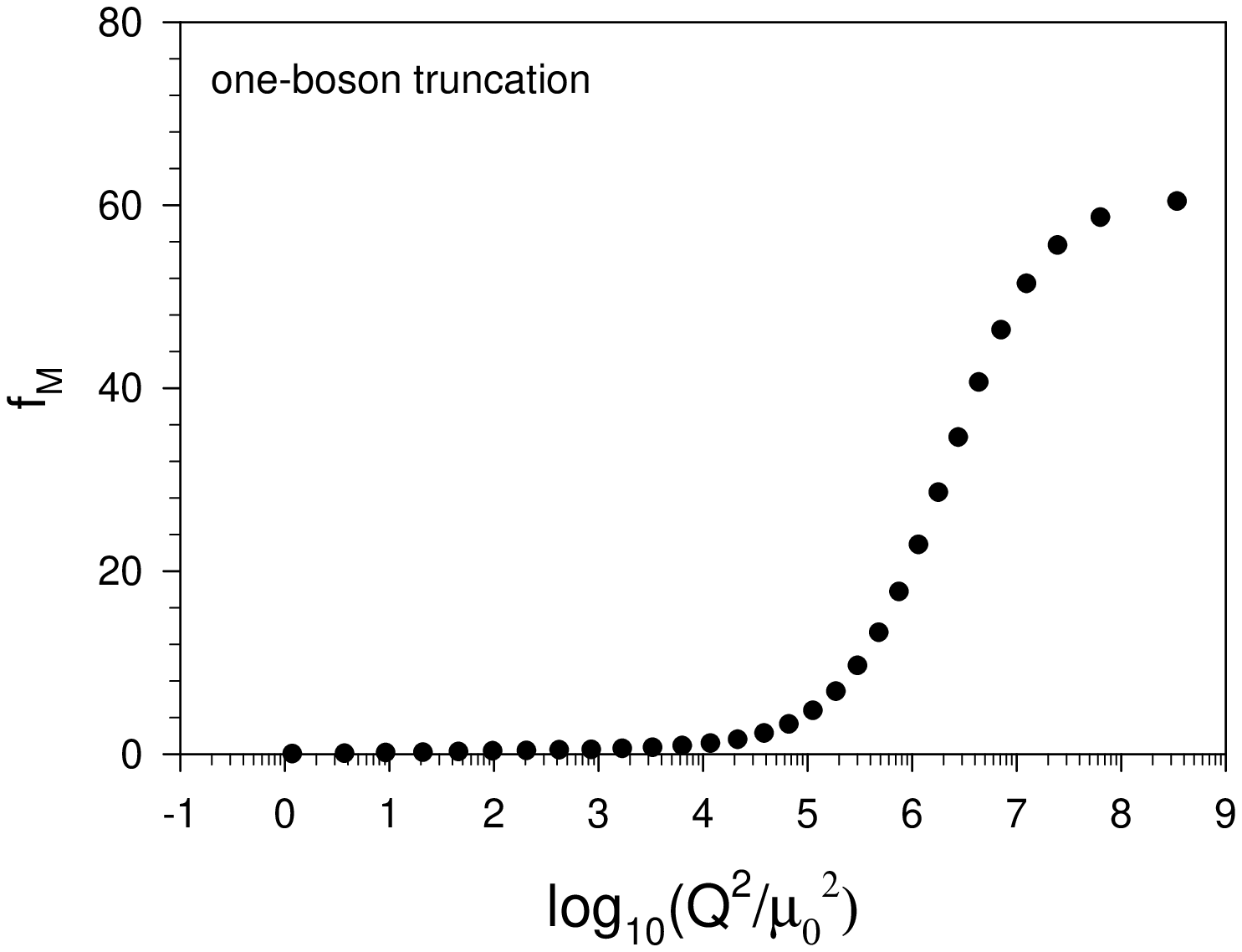} &
\includegraphics[width=6.5cm]{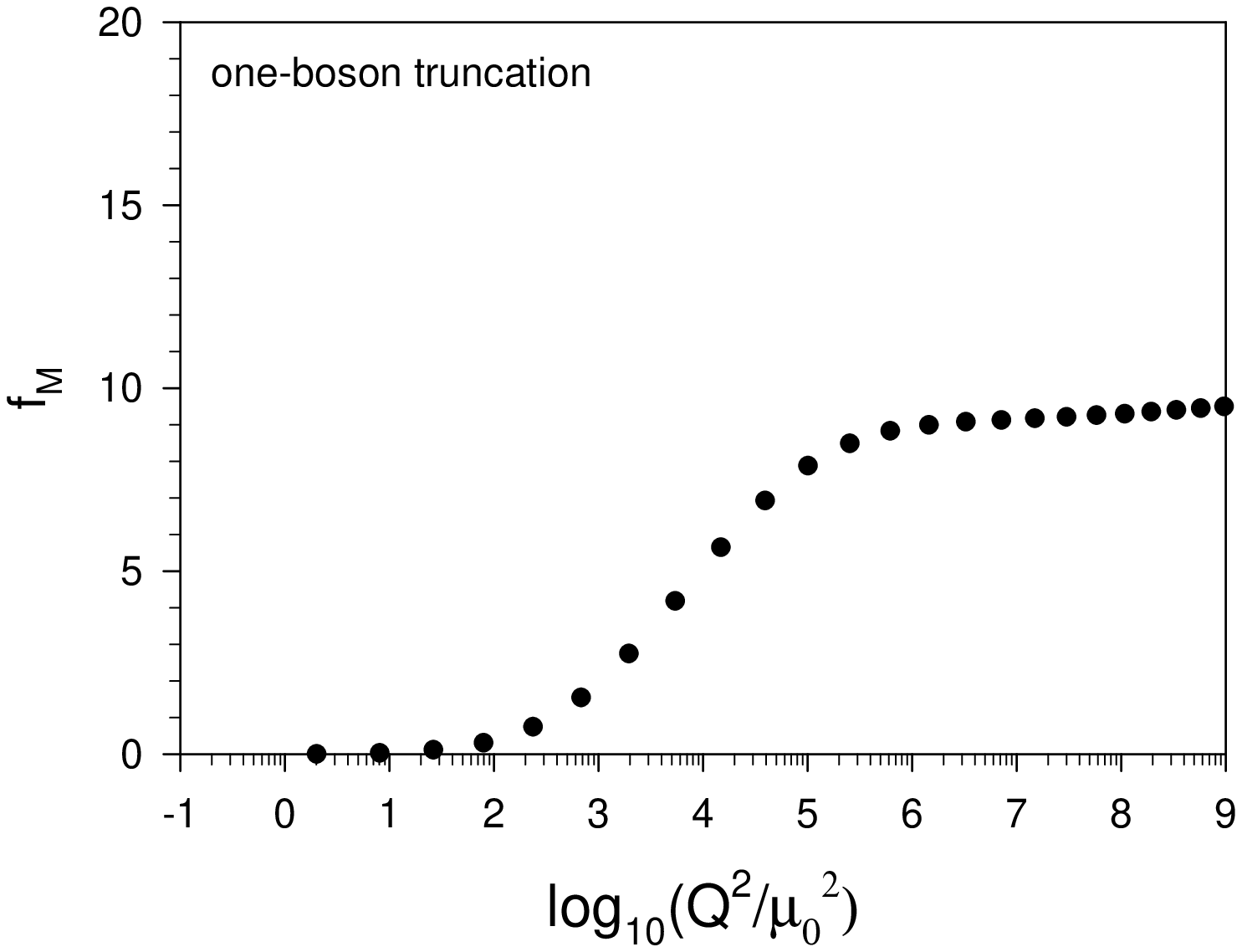} \\
(a) & (b) \\
\\
\includegraphics[width=6.5cm]{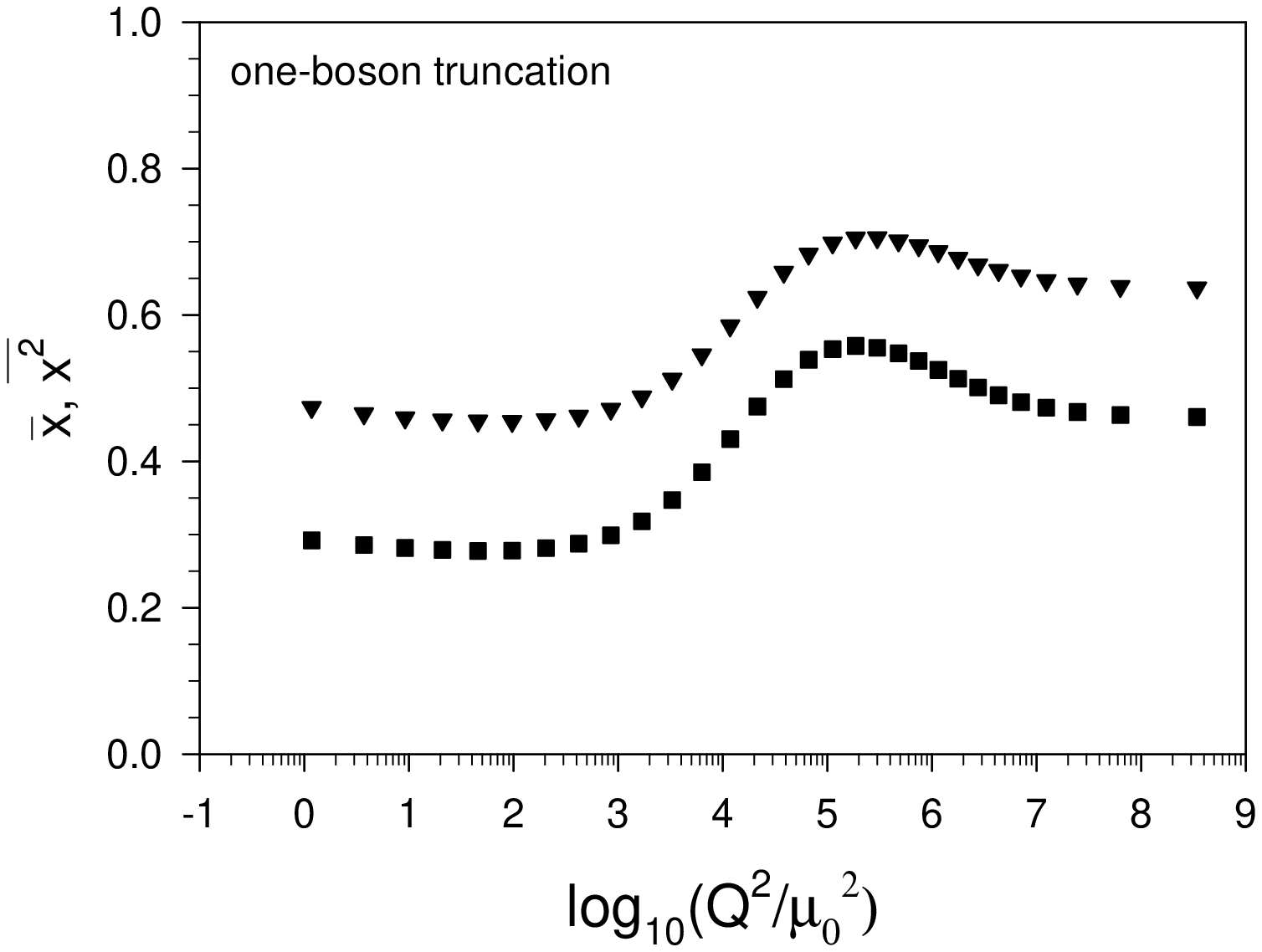} &
\includegraphics[width=6.5cm]{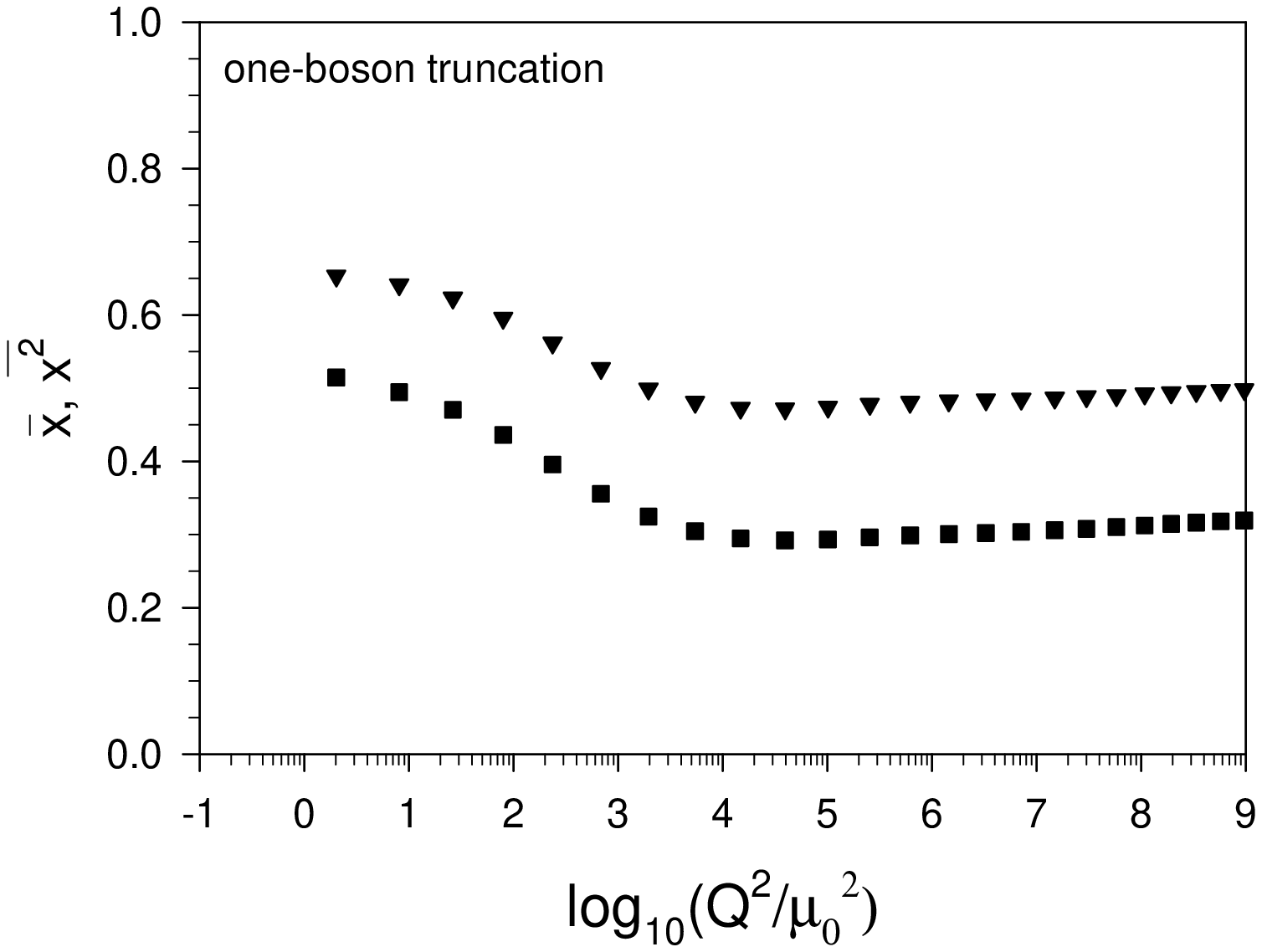} \\
(c) & (d)
\end{tabular}
\caption{Moments of the distribution function,
as defined in Eq.~(\ref{eq:moments}) of the text.  
The filled circles correspond
to $f_M$, the triangles to $\bar x$, and the squares to $\bar{x^2}$.
For (a) and (c), the PV masses are $m_1=\mu_1=2000\mu_0$,
and for (b) and (d), they are $m_1=50000\mu_0$ and $\mu_1=500\mu_0$.
}
\label{fig:1bmoments}
\end{center}
\end{figure}

As an alternative to traditional DLCQ, we can explicitly 
truncate the system with respect to the total number of
bosons in any Fock state.  We have already studied the
case of the one-boson truncation~\cite{TwoParticles}
where the system of two equations can be solved analytically.
The structure of the solution is as follows.
{}From the second equation in the system, Eq.~(\ref{eq:oneboson}),
the one-boson wave functions are immediately found to be
\bea
f_{ij+}(\ub{q})&=&
   \frac{P^+}{M^2-\frac{m_i^2+q_\perp^2}{1-q^+/P^+}
                -\frac{\mu_j^2+q_\perp^2}{q^+/P^+}}
\left[\left\{\sum_k (-1)^kz_k\right\}U_i(\ub{P}-\ub{q},\ub{q})
\right. \nonumber \\
 && \rule{2in}{0in}  \left.
      +\sum_k (-1)^kz_kU_k(\ub{P},\ub{q})\right],
 \nonumber \\
f_{ij-}(\ub{q})&=&
   \frac{P^+}{M^2-\frac{m_i^2+q_\perp^2}{1-q^+/P^+}
                -\frac{\mu_j^2+q_\perp^2}{q^+/P^+}}
\left\{\sum_k (-1)^kz_k\right\}V_+^*(\ub{P}-\ub{q},\ub{q}).
\eea
Substitution into the first equation, Eq.~(\ref{eq:noboson}),
yields algebraic equations for the bare-fermion amplitudes, 
which are
\bea \label{eq:onefermion}
(M^2-m_i^2)z_i &=&
 g^2\mu_0^2 (z_0-z_1)J+g^2 m_i(z_0m_0-z_1m_1) I_0
\nonumber \\
  &&+g^2\mu_0[(z_0-z_1)m_i+z_0m_0-z_1m_1] I_1,
\eea
with
\bea
I_n(M^2)&=&\int\frac{dy dq_\perp^2}{16\pi^2}
   \sum_{jk}\frac{(-1)^{j+k}}{M^2-\frac{m_j^2+q_\perp^2}{1-y}
                                   -\frac{\mu_k^2+q_\perp^2}{y}}
   \frac{(m_j/\mu_0)^n}{y(1-y)^n}, \;\; (n=0,1), \label{eq:In} \\
J(M^2)&=&\int\frac{dy dq_\perp^2}{16\pi^2}
   \sum_{jk}\frac{(-1)^{j+k}}{M^2-\frac{m_j^2+q_\perp^2}{1-y}
                                   -\frac{\mu_k^2+q_\perp^2}{y}}
   \frac{(m_j^2+q_\perp^2)/\mu_0^2}{y(1-y)^2}=\frac{M^2}{\mu_0^2}I_0.
   \label{eq:J}
\eea
The presence of the PV regulators makes these integrals finite
and allows $I_0$ and $J$ to satisfy
the identity $\mu_0^2 J(M^2)=M^2 I_0(M^2)$.
Because $M$ is held fixed, these equations can be viewed
as an eigenvalue problem for $g^2$.
The solution to this eigenvalue problem is
\be \label{eq:gofm}
g^2=-\frac{(M\mp m_0)(M\mp m_1)}{(m_1-m_0)(\mu_0 I_1\pm MI_0)}, \;\;
\frac{z_1}{z_0}=\frac{M \mp m_0}{M \mp m_1}.
\ee
Structure functions and distribution amplitudes can then be computed and 
the PV-mass limits studied. Figures~\ref{fig:1bfb}, \ref{fig:1bphi}, 
and \ref{fig:1bmoments} show typical results.  The two-boson contribution
is computed perturbatively.  The forms show a significant sensitivity
to the Pauli--Villars masses.
Further results can be found in Ref.~\cite{TwoParticles}.

\section{Two-boson truncation}
\label{sec:TwoBoson}

\subsection{Effective equations}

\begin{figure}[hbpt]
\begin{center}
\includegraphics[width=10cm]{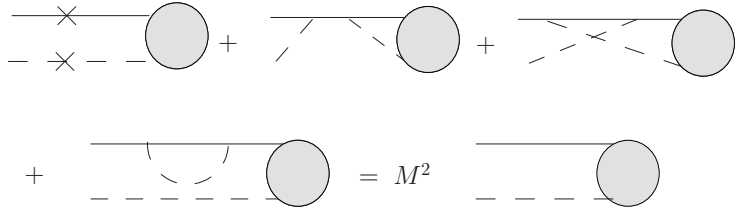} 
\caption{Same as Fig.~\ref{fig:coupledeqns} but for 
the effective equations in (\ref{eq:EffEq}) of the text.
}
\label{fig:effeqn}
\end{center}
\end{figure}

For a two-boson truncation, the solution is no longer 
analytic, but the coupled equations (\ref{eq:noboson})-(\ref{eq:twoboson}) 
can be reduced to eight equations for the two-particle amplitudes only,
which are of the form
\bea  \label{eq:EffEq}
\lefteqn{\left[M^2
  -\frac{m_i^2+q_\perp^2}{1-y}-\frac{\mu_j^2+q_\perp^2}{y}\right]
f_{ijs}(y,q_\perp)=
\frac{g^2}{16\pi^2}\sum_a\frac{I_{ija}(y,q_\perp)}{1-y}f_{ajs}(y,q_\perp)}
 \nonumber \\
 &&  +\frac{g^2}{16\pi^2}\sum_{abs'}\int_0^1dy'dq_\perp^{\prime 2}
   J_{ijs,abs'}^{(0)}(y,q_\perp;y',q'_\perp)f_{abs'}(y',q'_\perp)  \\
   && +\frac{g^2}{16\pi^2}\sum_{abs'}\int_0^{1-y}dy'dq_\perp^{\prime 2}
   J_{ijs,abs'}^{(2)}(y,q_\perp;y',q'_\perp)f_{abs'}(y',q'_\perp),
   \nonumber
\eea
with the angular dependence removed via
$\sqrt{P^+}f_{ij+}(\ub{q})=f_{ij+}(y,q_\perp)$ and
$\sqrt{P^+}f_{ij-}(\ub{q})=f_{ij-}(y,q_\perp)e^{i\phi}$.
Here $I_{ija}$ is a computable self-energy and 
$J_{ijs,abs'}^{(n)}$ is the kernel due to $n$-boson intermediate states.
Specific forms for $I$ and $J^{(n)}$ can be found in Appendix~A\@.
A diagrammatic representation is given in Fig.~\ref{fig:effeqn}.

There is a difficulty with the structure of this eigenvalue problem,
because the original eigenvalue $M^2$ appears in the denominator of
integrands that go into finding the matrix elements.  If attacked
directly, this nonlinearity requires an additional layer of 
numerical effort, to get a self-consistent solution.  An indirect
approach is to again convert the problem to one where $g^2$ is
the eigenvalue.  To maintain symmetry (up to the indefinite norm), 
this conversion is done by defining a new function 
\be
\tilde{f}_{ij\pm}=f_{ij\pm}
     \sqrt{\frac{m_i^2+q_\perp^2}{1-y}+\frac{\mu_j^2+q_\perp^2}{y}-M^2}
\ee
before completing the rearrangement.  The smallest real $g$ that
is obtained for a given $M$ is taken to be the coupling value for
which $M$ is the mass of the lowest state.  Obviously, this works well 
with the renormalization condition where $M$ is fixed as input.

\subsection{Method of solution}

We solve the reduced integral equations (\ref{eq:EffEq}) numerically
by converting them to a discrete matrix equation via Gauss--Legendre
quadrature of the integrals.  The resolution of the quadrature is
characterized by the order of the underlying polynomial, which we
denote by $K$ in the longitudinal direction and by $2N+1$ in the
transverse direction.  For the transverse quadrature, only odd
orders are used, to keep $q_\perp=0$ as a quadrature point, and
only $N+1$ points are used.  (The other $N$ points correspond
to negative $q_\perp$.)  Thus $K$ and $N$ characterize the resolution
of the approximation, and we consider large values in order to be
close to the continuum limit.
 
Before applying the quadrature rules,
the variables $y'$ and $q'_\perp$ are transformed in such a way 
as to emphasize those regions most important to the approximation.
In the case of $q'_\perp$, the transformation is chosen also to
produce a finite
range of integration without introduction of a cutoff.  Since the
PV contributions make the integrals finite, no cutoff is needed
for the continuum problem, and any cutoff would only be an artifact 
of the numerical approximation if introduced.
A more complete discussion of the transformations is given in
Appendix~B.

The matrix eigenvalue problem is solved by applying the
Lanczos diagonalization scheme developed previously~\cite{bhm3}.
This particular scheme was designed to efficiently handle the
present situation where the matrix is self-adjoint with respect
to an indefinite norm.  What is different here, compared to the
case in \cite{bhm3}, is that the matrix is not sparse.
Nevertheless, the Lanczos approach is much faster than standard
diagonalization algorithms for nonsymmetric matrices, because
we are interested in only one eigenstate.  Typical matrix
sizes are on the order of 20,000 by 20,000, but provide as
many nonzero entries as the much larger sparse matrices
considered in \cite{bhm3}.

Use of the Lanczos technique is important for another reason.
Standard diagonalization routines have difficulty when there
are multiple mass scales in the problem; specifically, when
$\mu_1$ is intermediate between $\mu_0$ and $m_1$, such
that $\mu_0\ll \mu_1\ll m_1$, the standard diagonalization
can fail.  For the Lanczos process, we need only an accurate
representation of the product of the matrix and a vector.
The contribution of $J^{(0)}$ to the matrix can then be
written in a factorized form
$\sum_{i'=0}^1 (-1)^{i'}\vec{v}_{i'} \vec{v}_{i'}^T \eta/(M^2-m_{i'}^2)$,
where $\eta$ is a diagonal matrix that represents the signature
of the norm, as determined by the factors $(-1)^{a+b}$ given
as part of the definition of $J^{(0)}$ in Eq.~(\ref{eq:J0}).
This factorized form provides a more faithful numerical
representation of the contribution from $J^{(0)}$ by restricting 
the $J^{(0)}$-vector product to a linear combination of the
vectors $\vec{v}_{i'}$.

\subsection{Results}

\begin{figure}[hpbt]
\begin{center}
\begin{tabular}{cc}
\includegraphics[width=6.5cm]{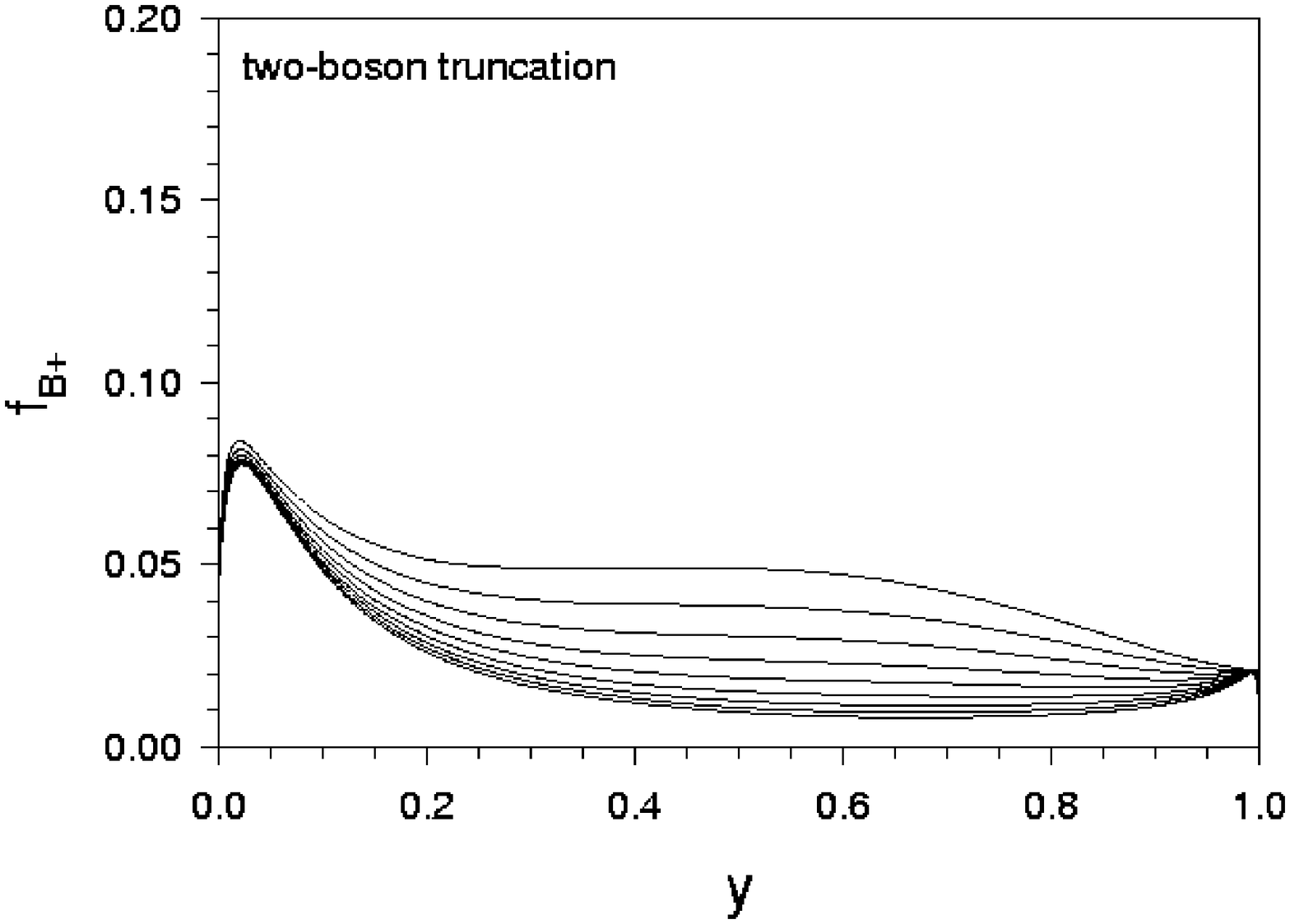} &
\includegraphics[width=6.5cm]{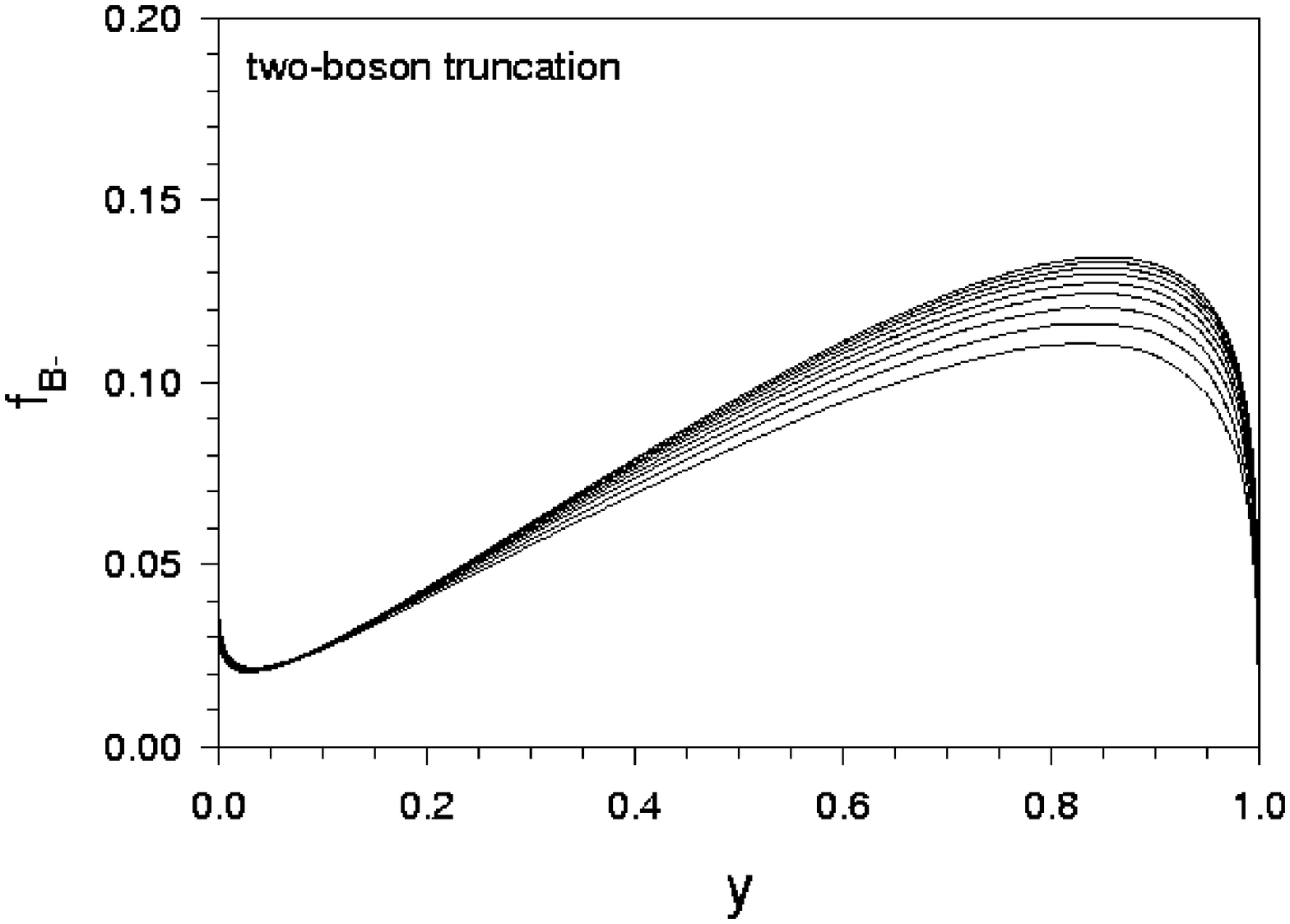} \\
(a) &  (b) \\ \vspace{0.05in} \\
\includegraphics[width=6.5cm]{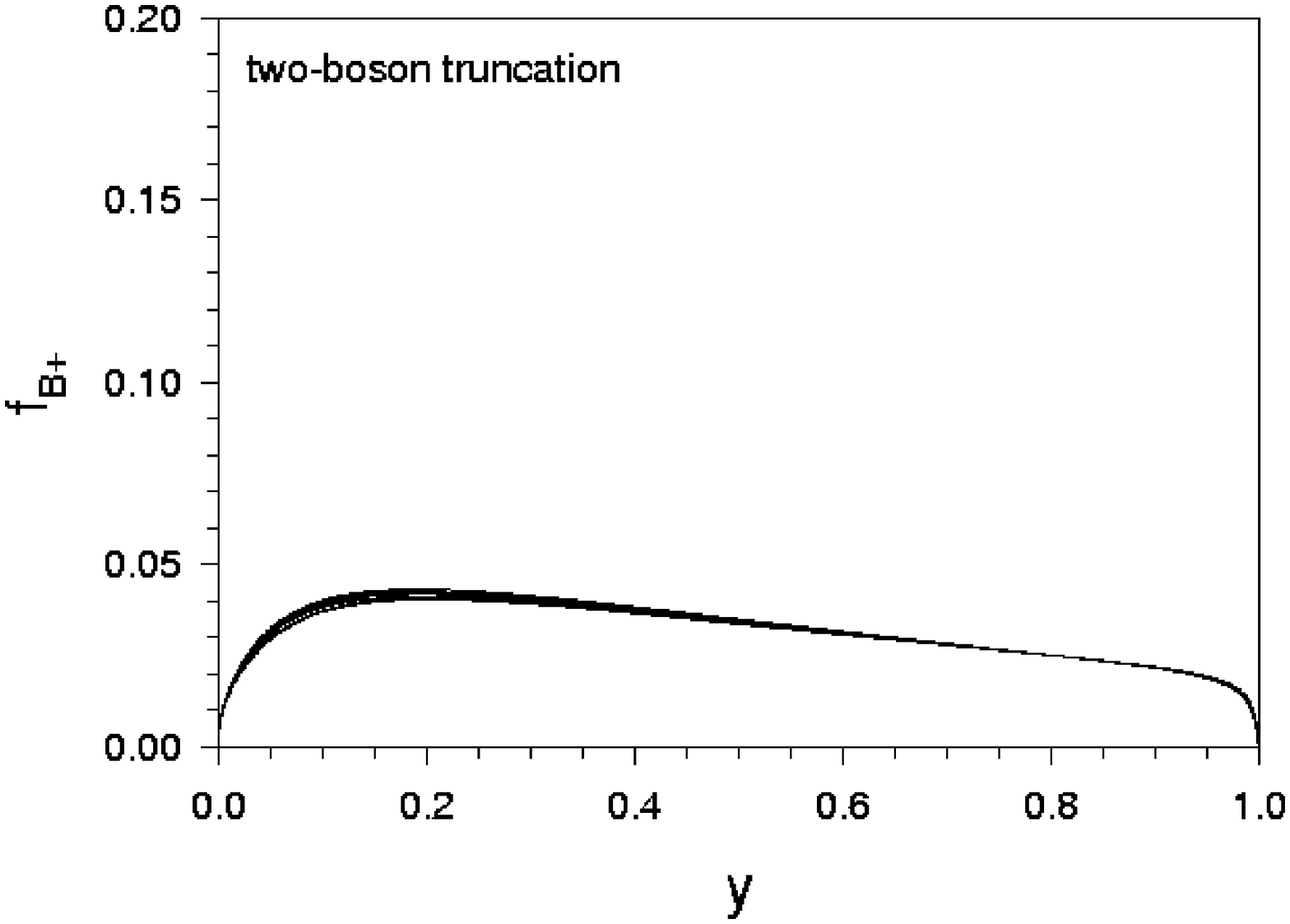} &
\includegraphics[width=6.5cm]{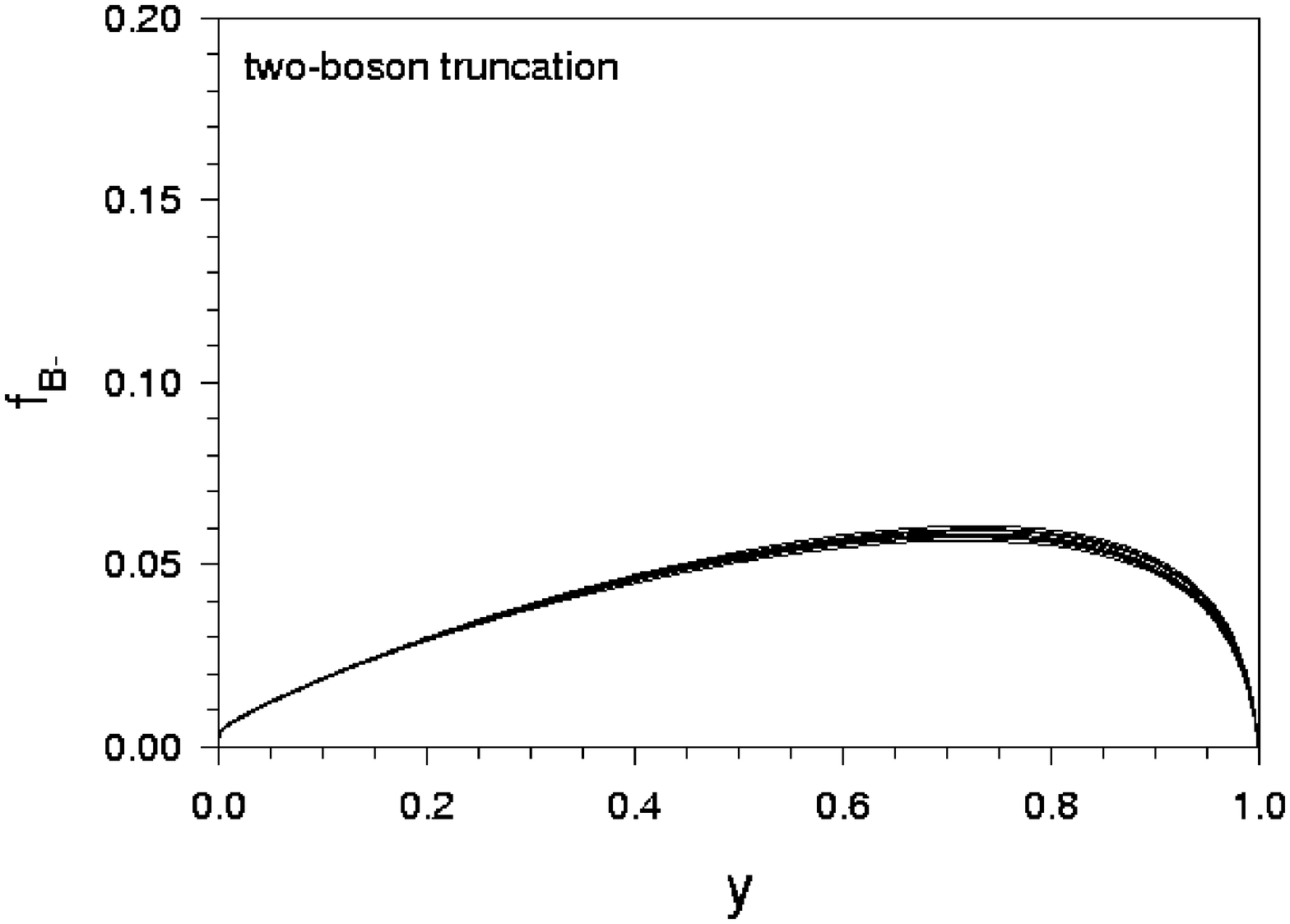} \\
(c) &  (d)
\end{tabular}
\caption{Bosonic structure functions $f_{B\pm}(y)$, as defined in
Eq.~(\ref{eq:fB}) of the text, for the dressed-fermion state
with mass $M=\mu_0$.  The wave functions are computed from
a truncation to two constituent bosons.  The longitudinal
resolution is varied from $K=30$ to $K=70$, with the transverse
resolution fixed at $N=30$.  The bare coupling is $g=2$.
For (a) and (b) the PV masses are $m_1=\mu_1=1000\mu_0$,
and for (c) and (d) they are $m_1=10000\mu_0$ and $\mu_1=100\mu_0$.
In (a) the amplitude of $f_{B+}$ decreases
with increasing $K$, and in (b) the amplitude of $f_{B-}$ increases.
}
\label{fig:fbvsk}
\end{center}
\end{figure}

The convergence with resolution is indicated in 
Fig.~\ref{fig:fbvsk}.
For most quantities, and for the range of PV masses considered,
resolutions of $K=50$ and $N=30$ are sufficient for convergence,
and we use these resolutions for most of our calculations.

\begin{figure}[hpbt]
\begin{center}
\begin{tabular}{c}
\\
\includegraphics[width=13cm]{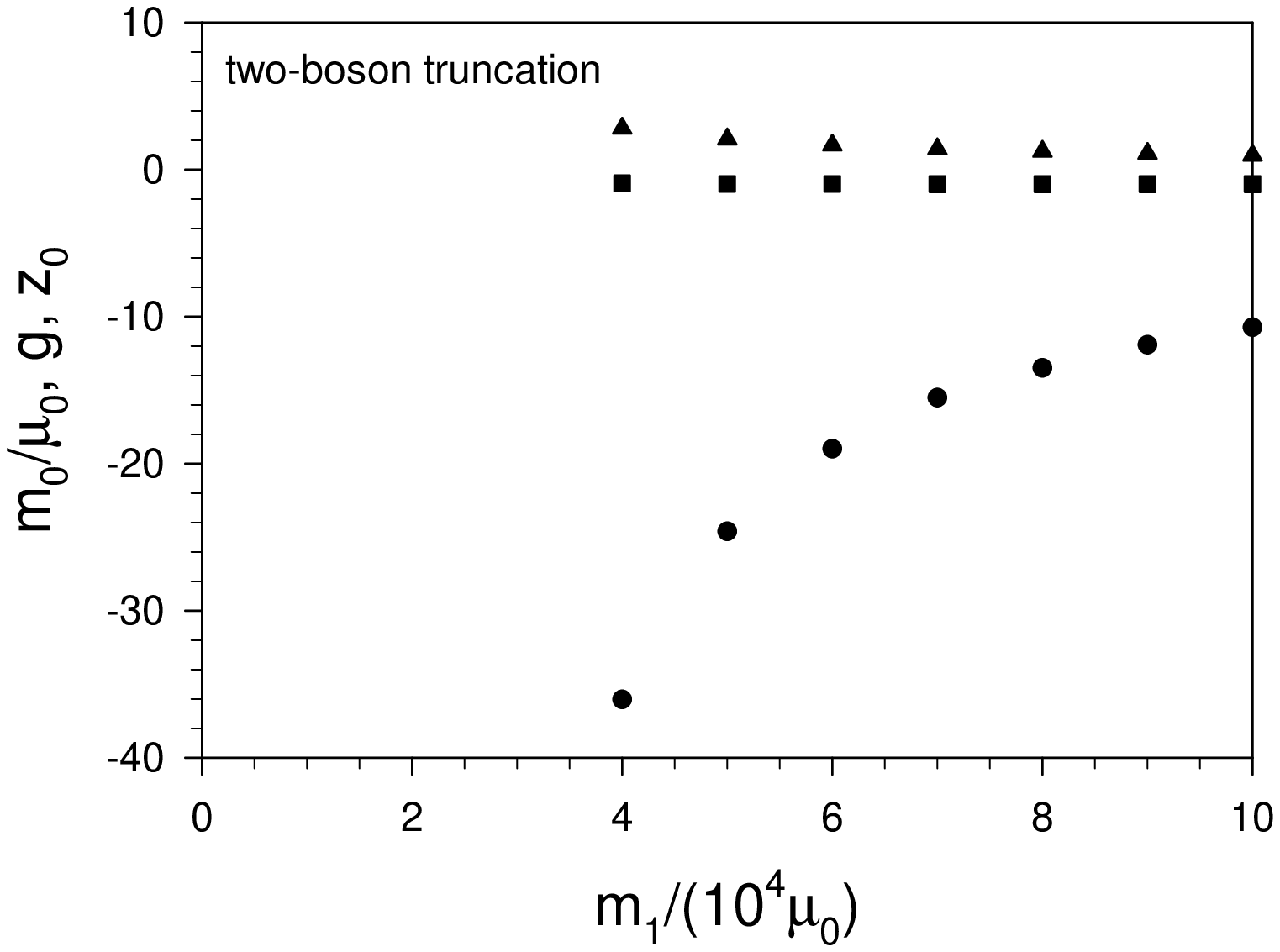}
\end{tabular}
\caption{The bare mass $m_0$ of the constituent fermion
(filled circles), 
the bare coupling $g$ (triangles),
and the bare-fermion amplitude $z_0$ (squares)
as functions of the PV fermion mass $m_1$.  
The ratio $\mu_1/m_1$ of PV boson mass
to fermion mass is fixed at 0.01.
The dressed mass is $M=\mu_0$, the radius is $R=0.01/\mu_0$, and
the numerical resolutions are $K=50$ and $N=30$.}
\label{fig:gm0z0vsm1}
\end{center}
\end{figure}

\begin{figure}[hpbt]
\begin{center}
\begin{tabular}{c}
\includegraphics[width=13cm]{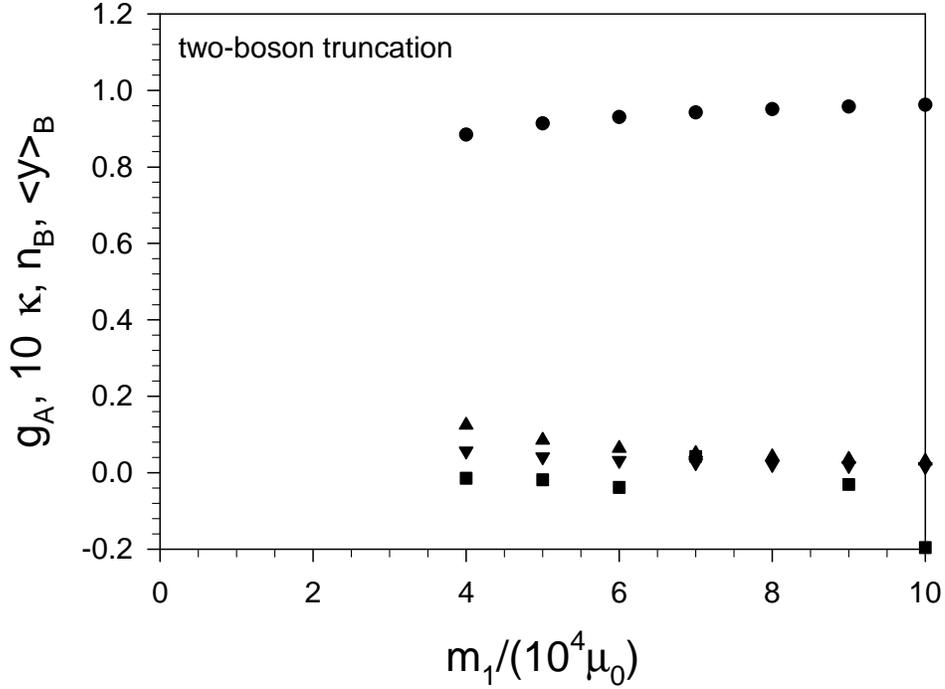}
\end{tabular}
\caption{Same as Fig.~\ref{fig:gm0z0vsm1} but for 
the axial coupling $g_A$ (filled circles), 
the anomalous moment $\kappa$ (squares),
the average number of bosons $n_B$ (upward triangles), 
and the average momentum fraction $\langle y\rangle_B$ (downward triangles).
}
\label{fig:nbkapvsm1}
\end{center}
\end{figure}

\begin{figure}[hpbt]
\vspace{0.1in}
\begin{center}
\begin{tabular}{cc}
\includegraphics[width=6.5cm]{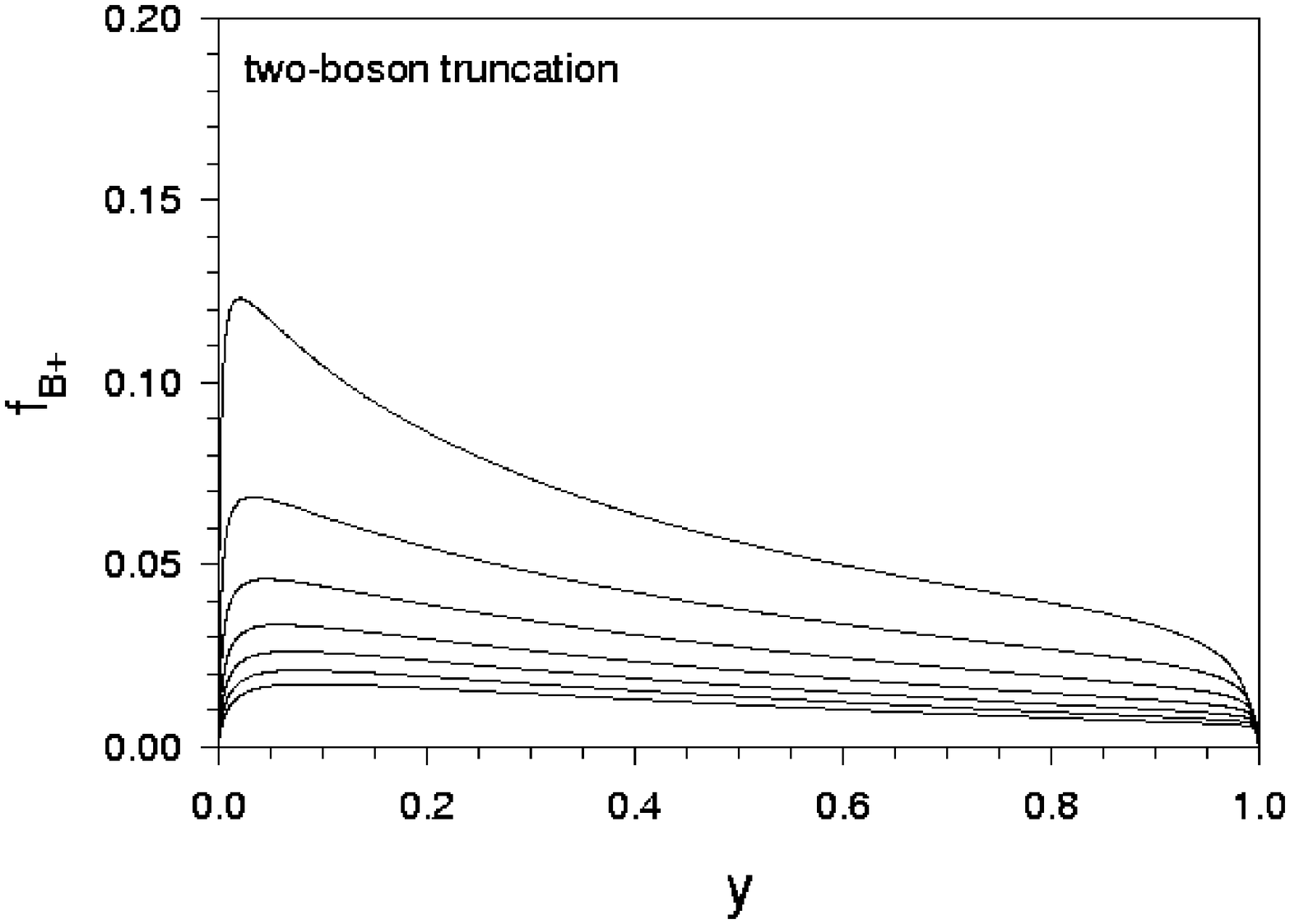} &
\includegraphics[width=6.5cm]{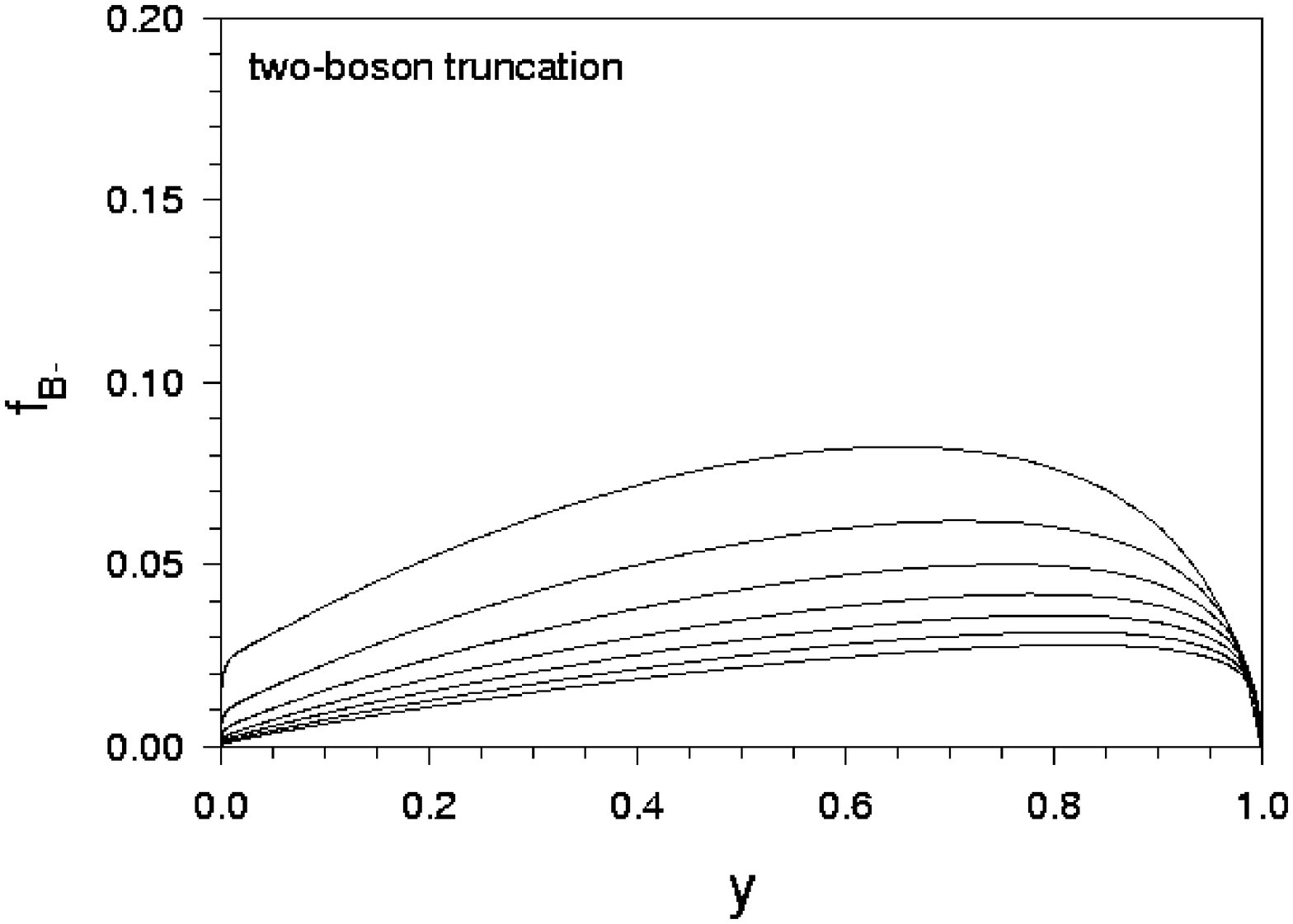} \\
(a) &  (b)
\end{tabular}
\caption{
Bosonic structure functions $f_{B\pm}$ for the dressed-fermion
state with mass $M=\mu_0$ and radius $R=0.01/\mu_0$.
The different curves correspond to increasing values of $m_1$.
The ratio $\mu_1/m_1$ is fixed at 0.01.  The numerical
resolutions are $K=50$ and $N=30$.
}
\label{fig:fbvsm1}
\end{center}
\end{figure}

\begin{figure}[hpbt]
\begin{center}
\begin{tabular}{cc}
\includegraphics[width=6.5cm]{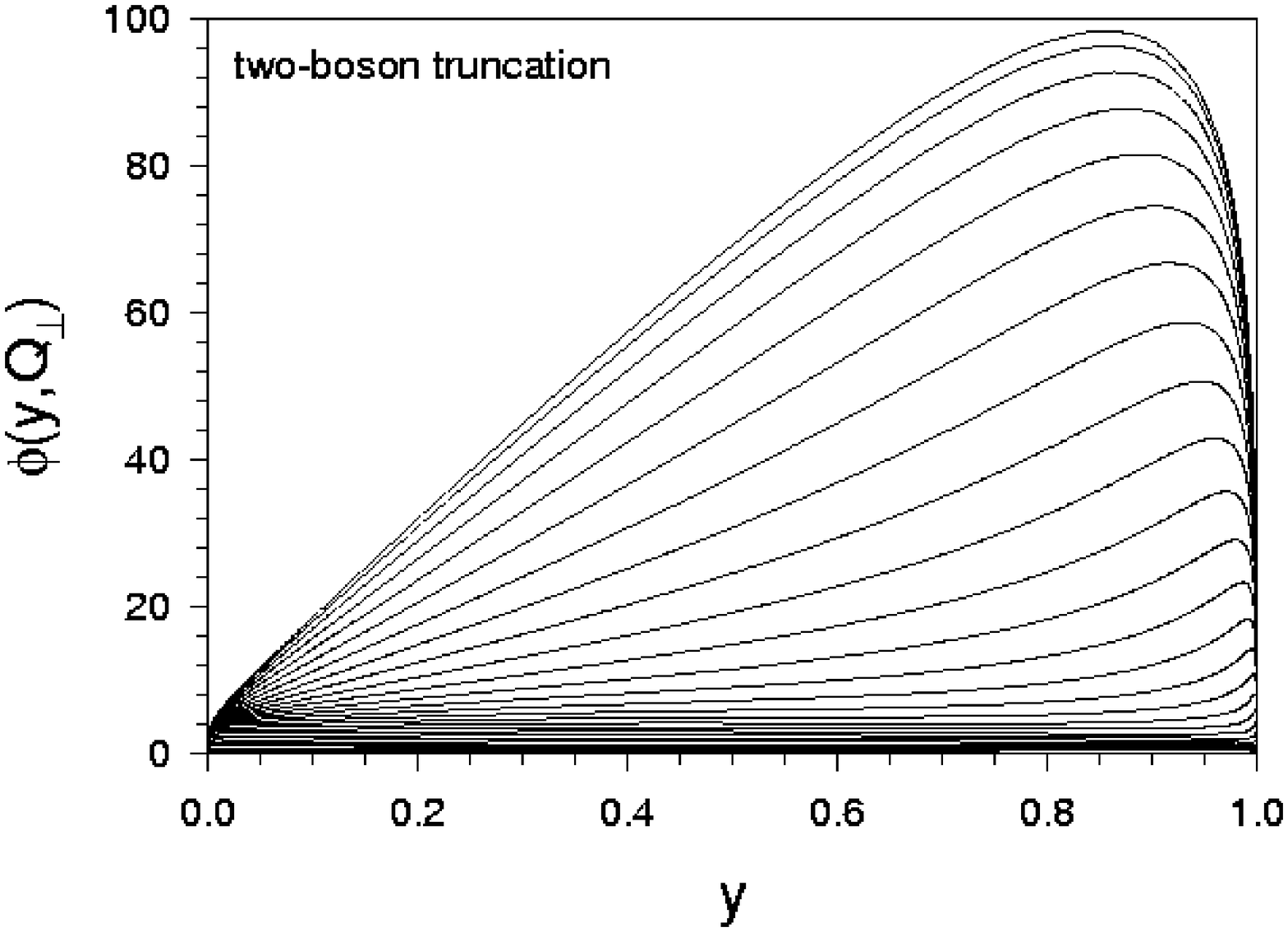} &
\includegraphics[width=6.5cm]{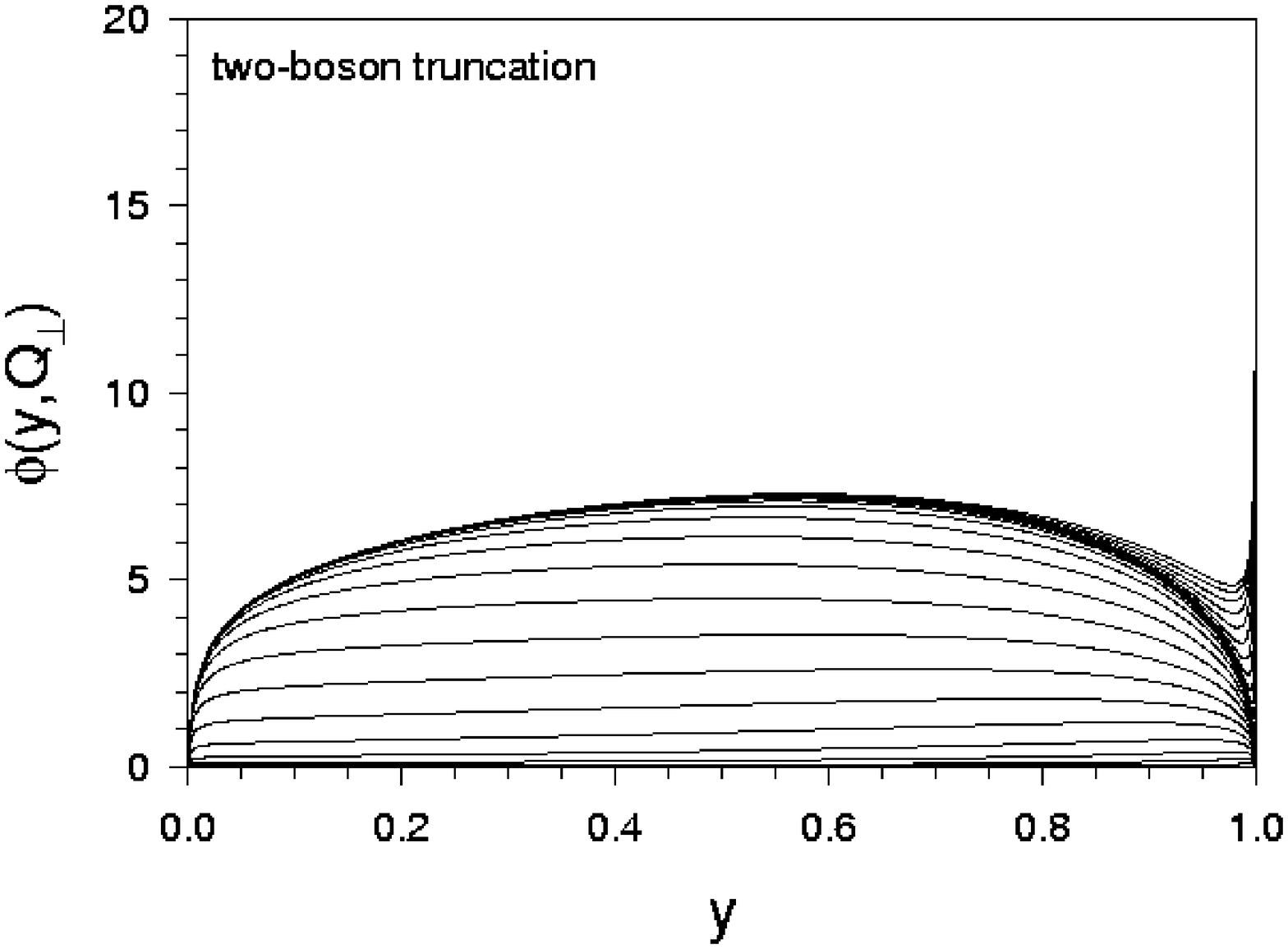} \\
(a) & (b)
\end{tabular}
\caption{Same as Fig.~\ref{fig:1bphi} but for the 
two-boson truncation.}
\label{fig:2bphi}
\end{center}
\end{figure}

\begin{figure}[hpbt]
\begin{center}
\begin{tabular}{cc}
\includegraphics[width=6.5cm]{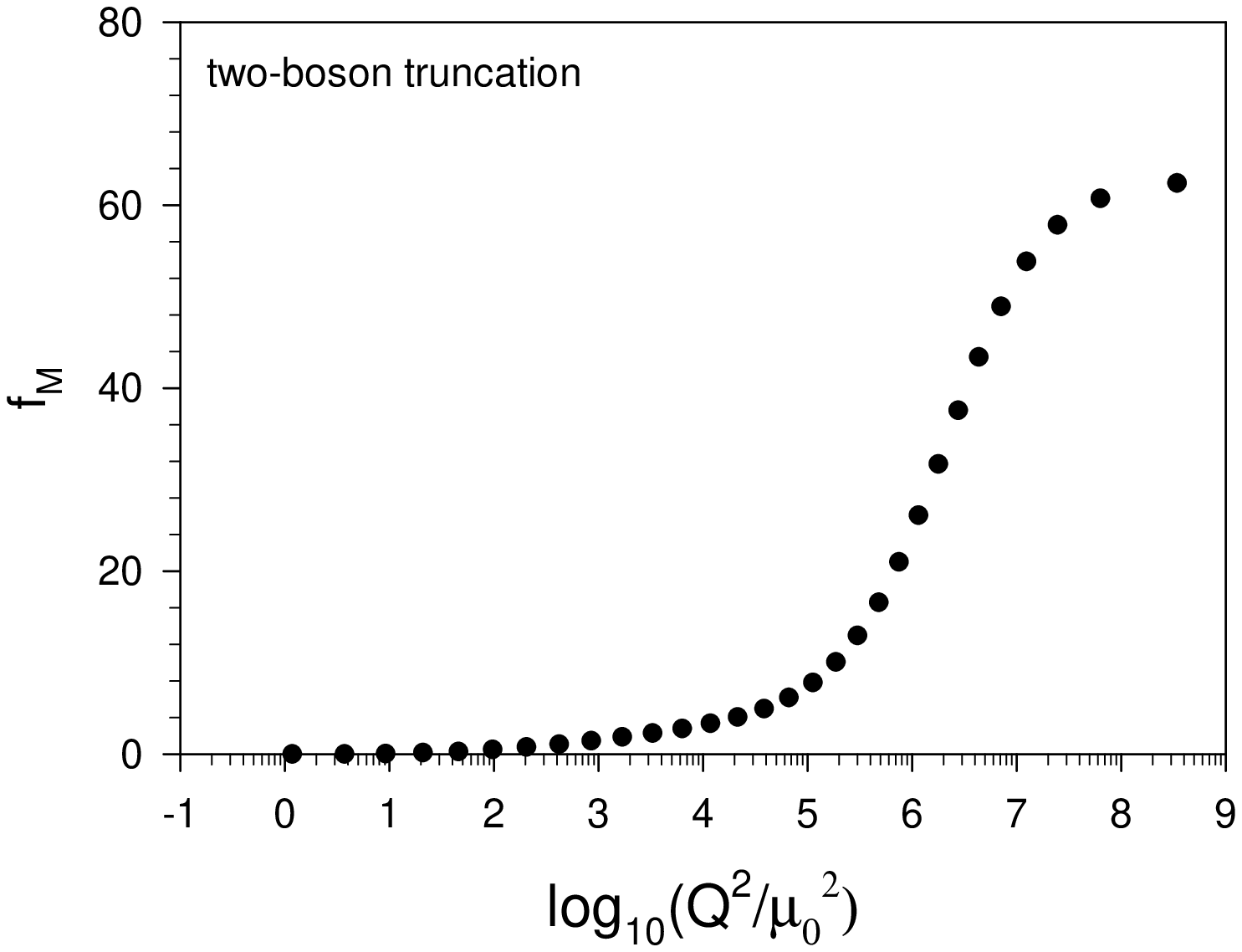} &
\includegraphics[width=6.5cm]{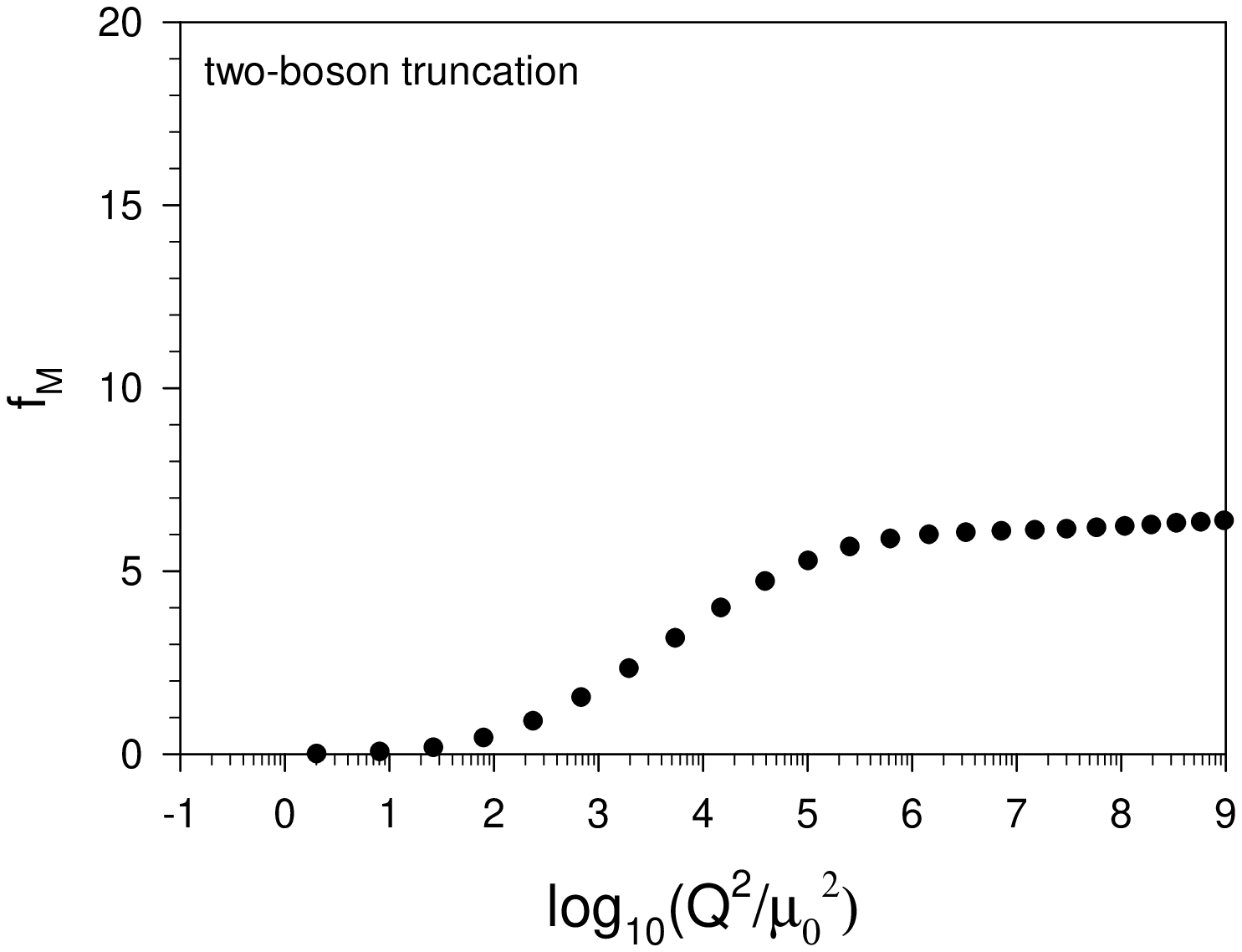} \\
(a) & (b) \\
\\
\includegraphics[width=6.5cm]{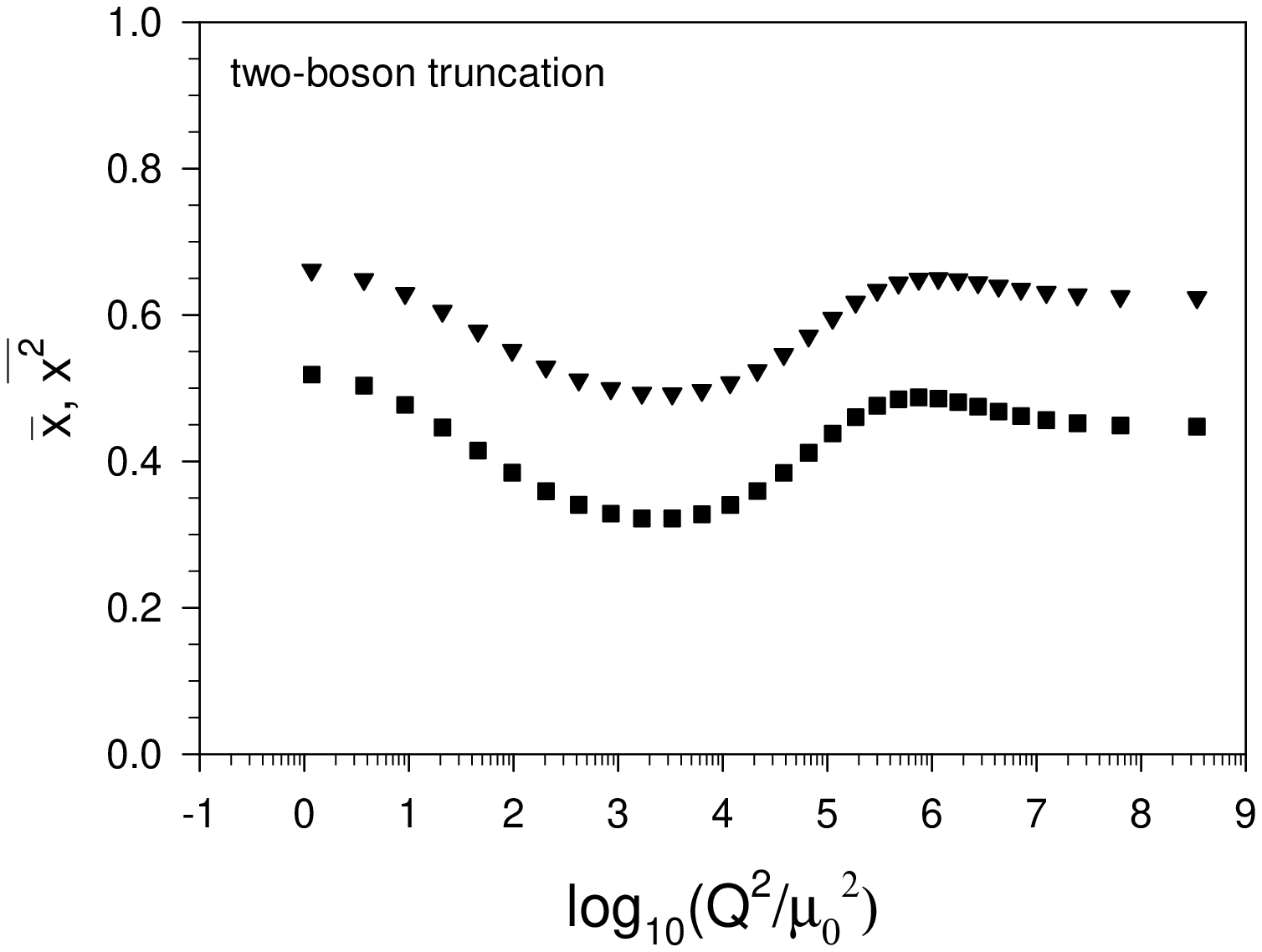} &
\includegraphics[width=6.5cm]{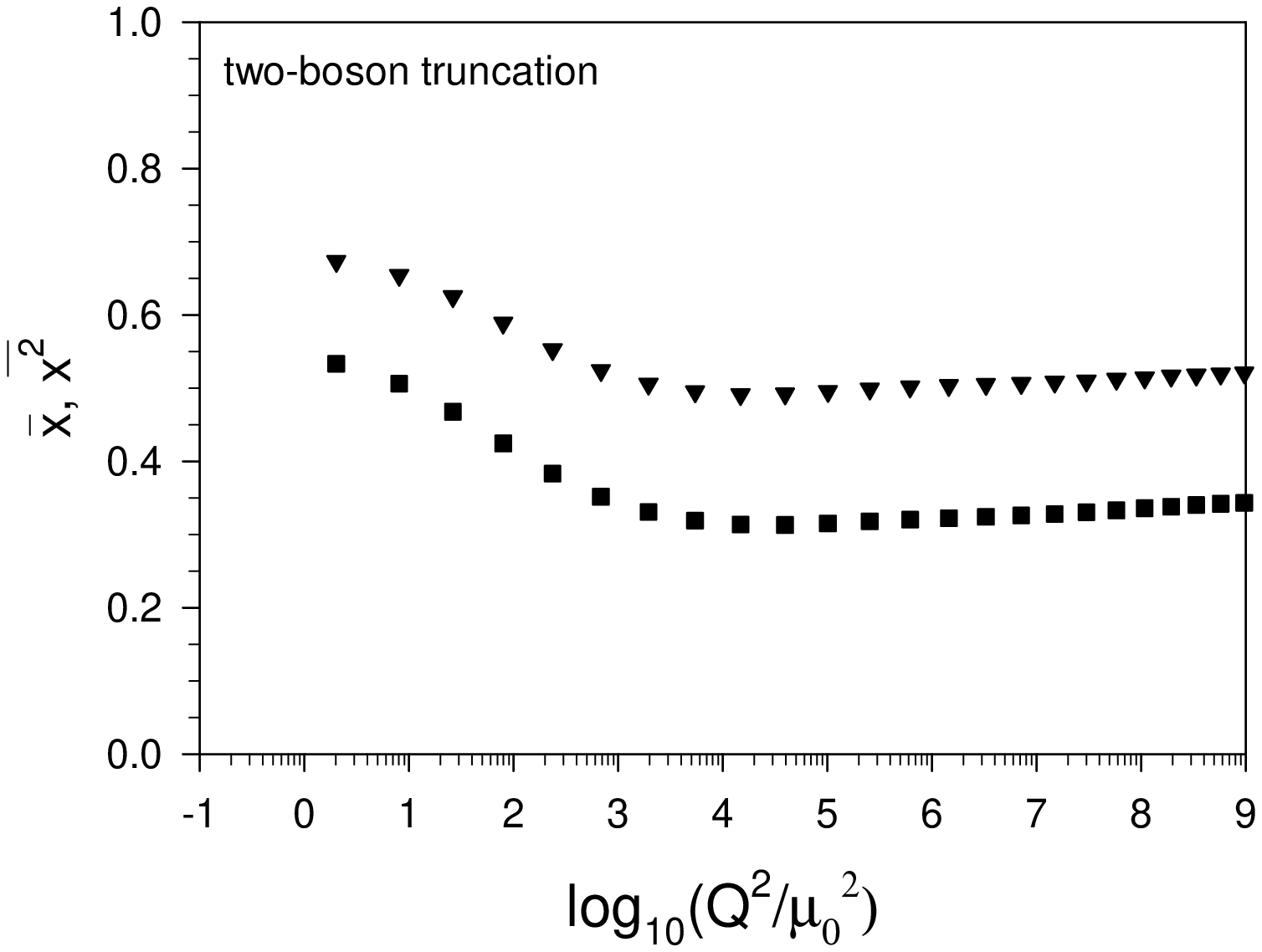} \\
(c) & (d)
\end{tabular}
\caption{Same as Fig.~\ref{fig:1bmoments} but for the 
two-boson truncation.}
\label{fig:2bmoments}
\end{center}
\end{figure}

The variation of the structure functions and some characteristic
quantities with respect to PV mass can be seen in 
Figs.~\ref{fig:gm0z0vsm1}, \ref{fig:nbkapvsm1}, and \ref{fig:fbvsm1}.
When the PV masses are equal, the radius of the state with mass $M=\mu_0$
is driven to zero as the PV mass is made large; thus in this limiting
case a fixed radius cannot be maintained.
The distribution amplitude and its moments are shown in Figs.~\ref{fig:2bphi}
and \ref{fig:2bmoments} for two sets of PV masses.

\begin{figure}[hpbt]
\begin{center}
\begin{tabular}{c}
\includegraphics[width=12cm]{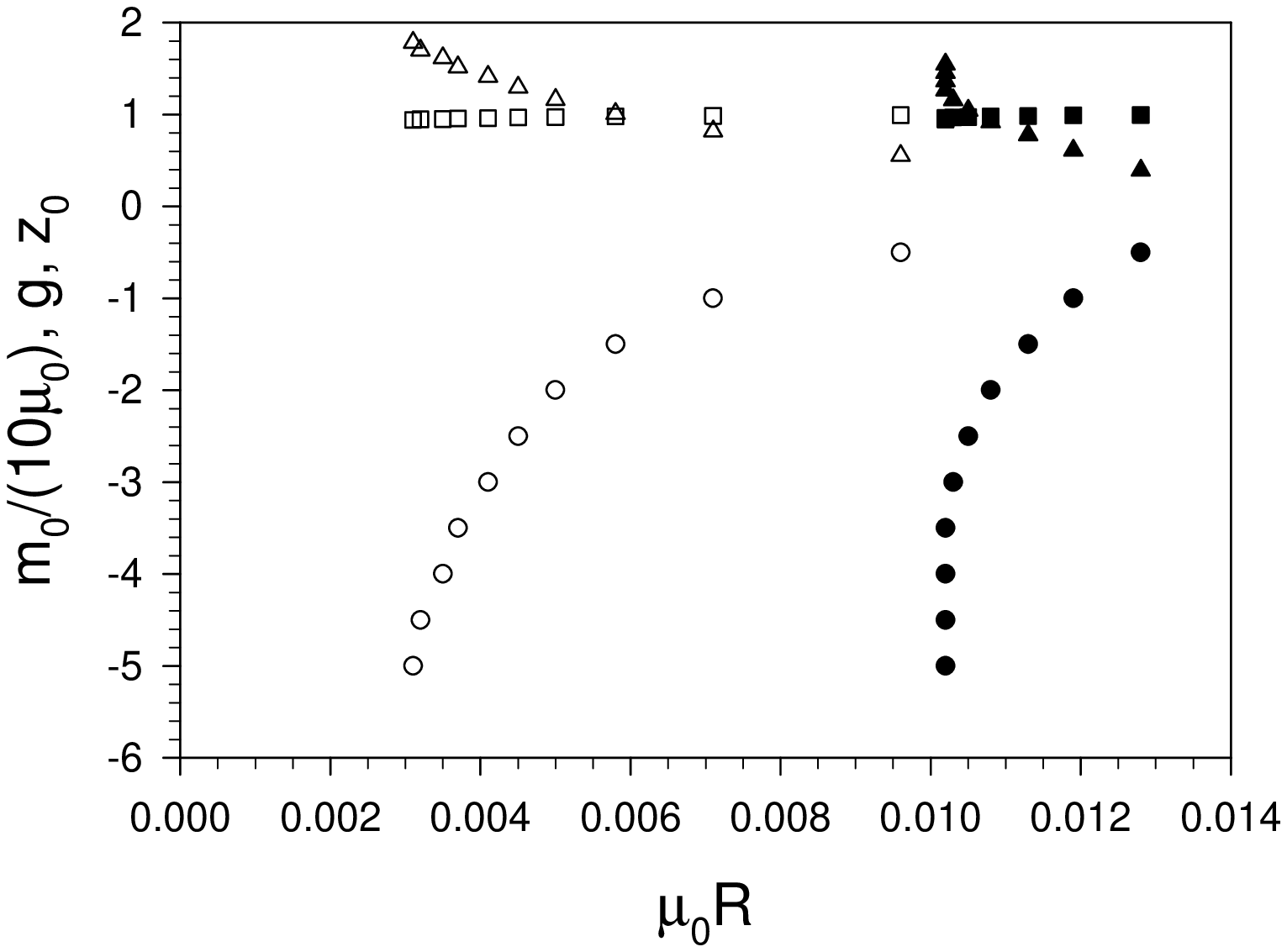} \\
(a) \\
\\
\includegraphics[width=12cm]{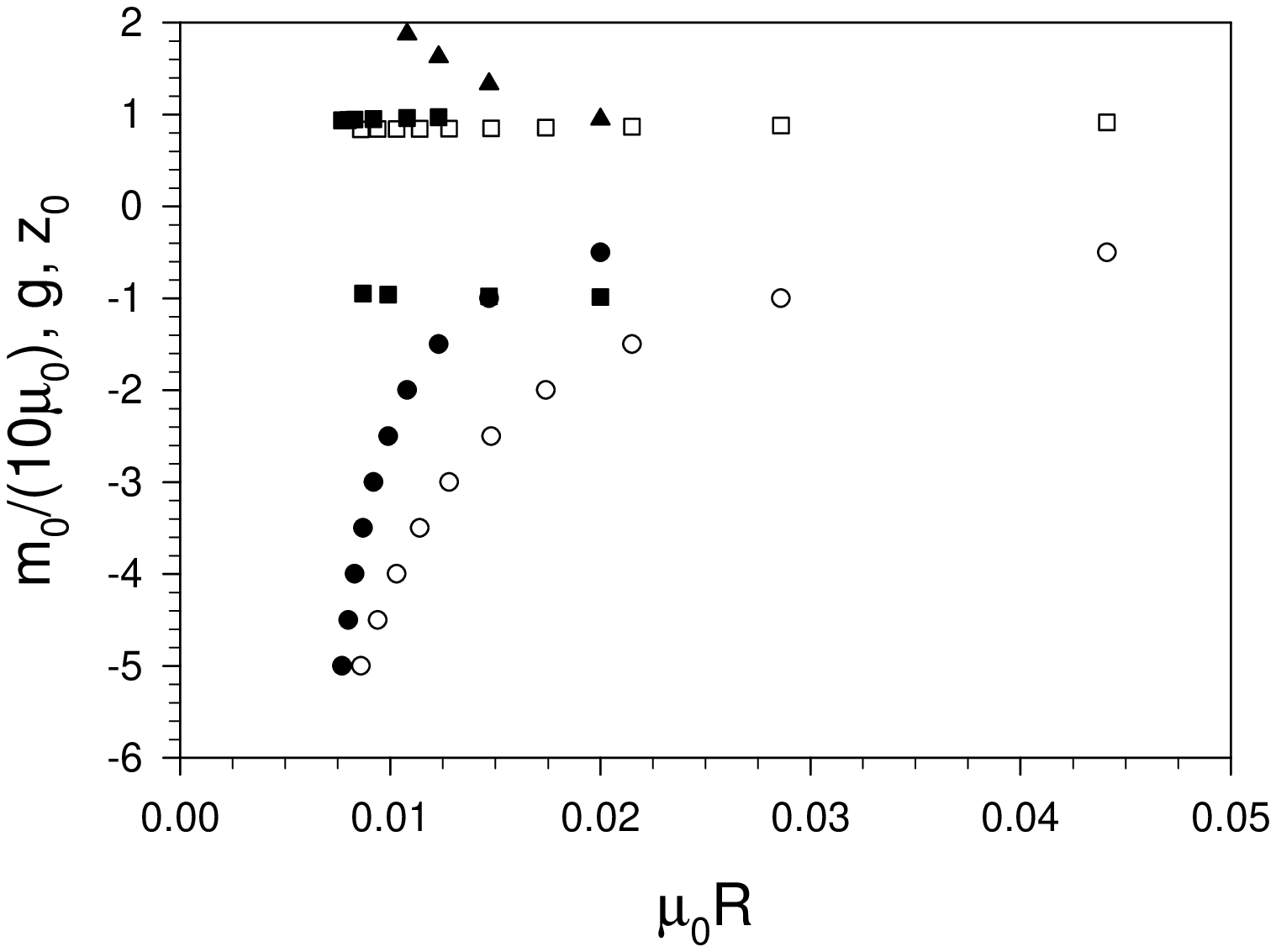} \\
(b)
\end{tabular}
\caption{The bare fermion mass $m_0$ (circles),
the Yukawa coupling $g$ (triangles), and 
the bare fermion amplitude $z_0$ (squares)
as functions of the radius $R$. 
For (a) the PV masses are $m_1=\mu_1=2000\mu_0$, 
and for (b) they are $m_1=50000\mu_0$ and $\mu_1=500\mu_0$.
The filled symbols correspond to the two-boson truncation, and the
open symbols to the one-boson truncation.  The dressed-fermion 
mass is $M=\mu_0$, and the resolutions are $K=50$ and $N=30$.
For the one-boson truncation, the two-boson contribution is
computed perturbatively.
}
\label{fig:gm0vsR}
\end{center}
\end{figure}

\begin{figure}[hpbt]
\begin{center}
\vspace{0.1in}
\begin{tabular}{c}
\includegraphics[width=12cm]{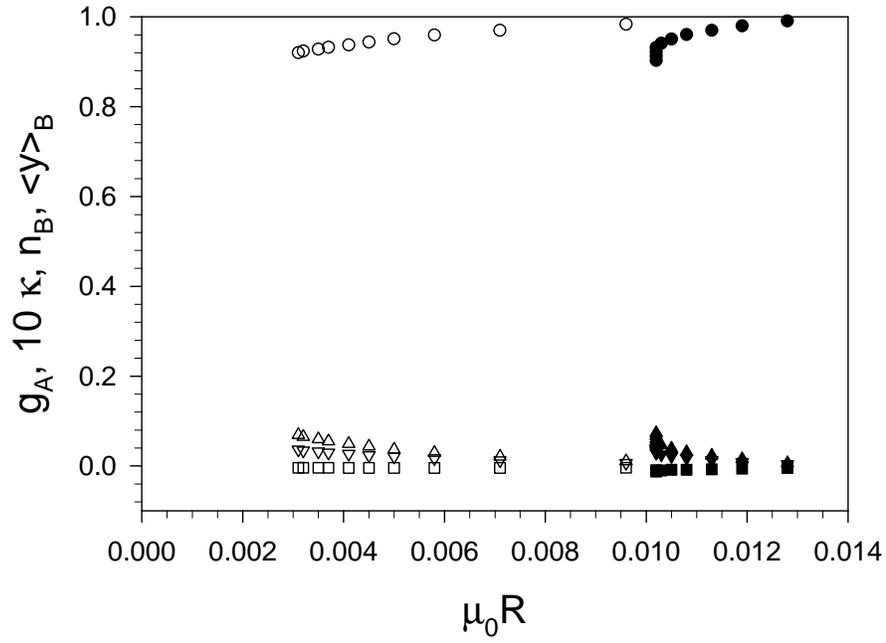} \\
(a) \\
\\
\includegraphics[width=12cm]{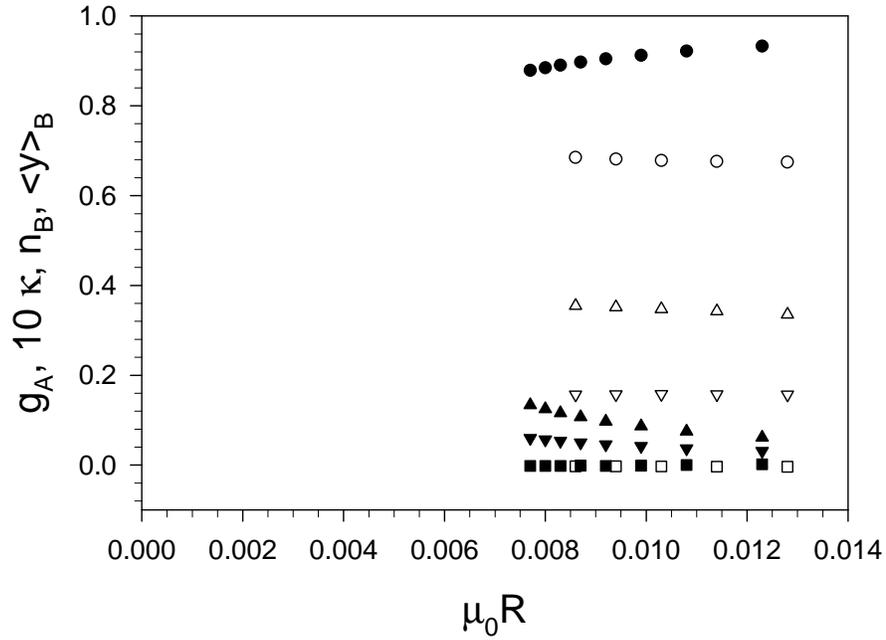} \\
(b)
\end{tabular}
\caption{Same as Fig.~\ref{fig:gm0vsR} but for 
the axial coupling $g_A$ (circles),
the anomalous moment $\kappa$ (squares),
the average number of bosons $n_B$ (upward triangles), 
and the average momentum fraction $\langle y\rangle_B$ (downward triangles).
The filled symbols correspond to the two-boson truncation, and the
open symbols to the one-boson truncation.
}
\label{fig:nbkapvsR}
\end{center}
\end{figure}

\begin{figure}[hpbt]
\begin{center}
\begin{tabular}{cc}
\includegraphics[width=6.5cm]{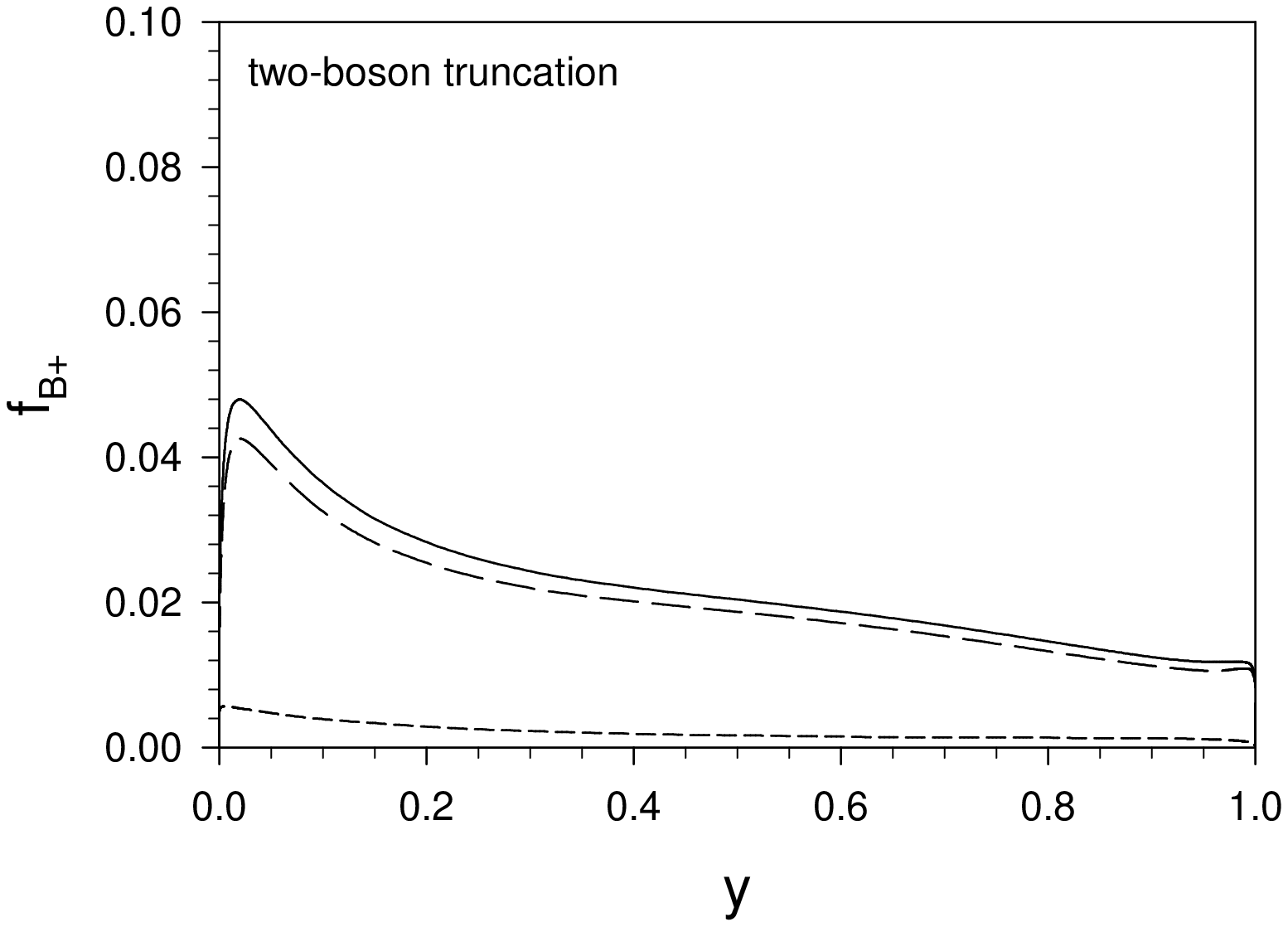} &
\includegraphics[width=6.5cm]{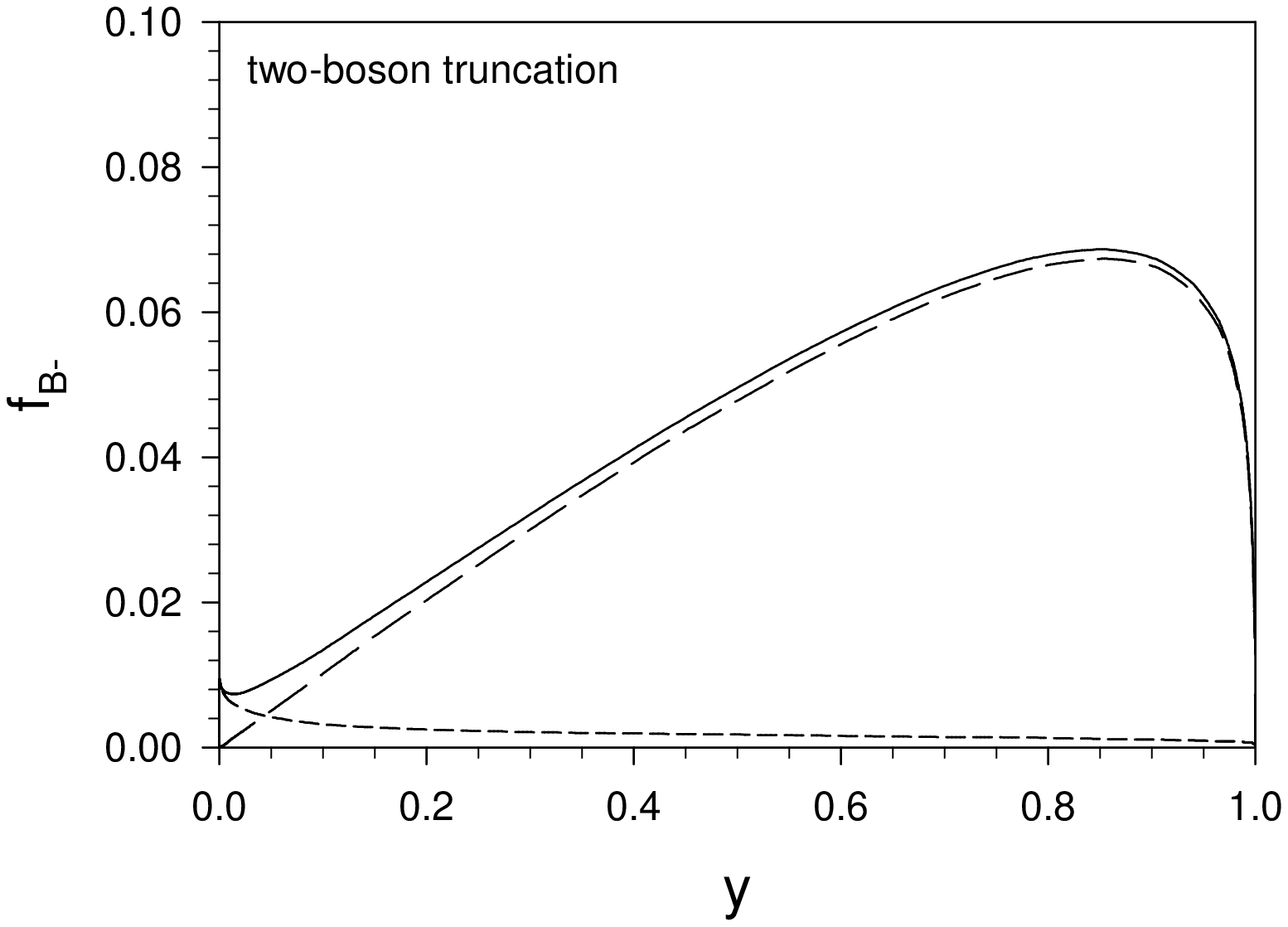} \\
(a) &  (b) \\ \vspace{0.05in} \\
\includegraphics[width=6.5cm]{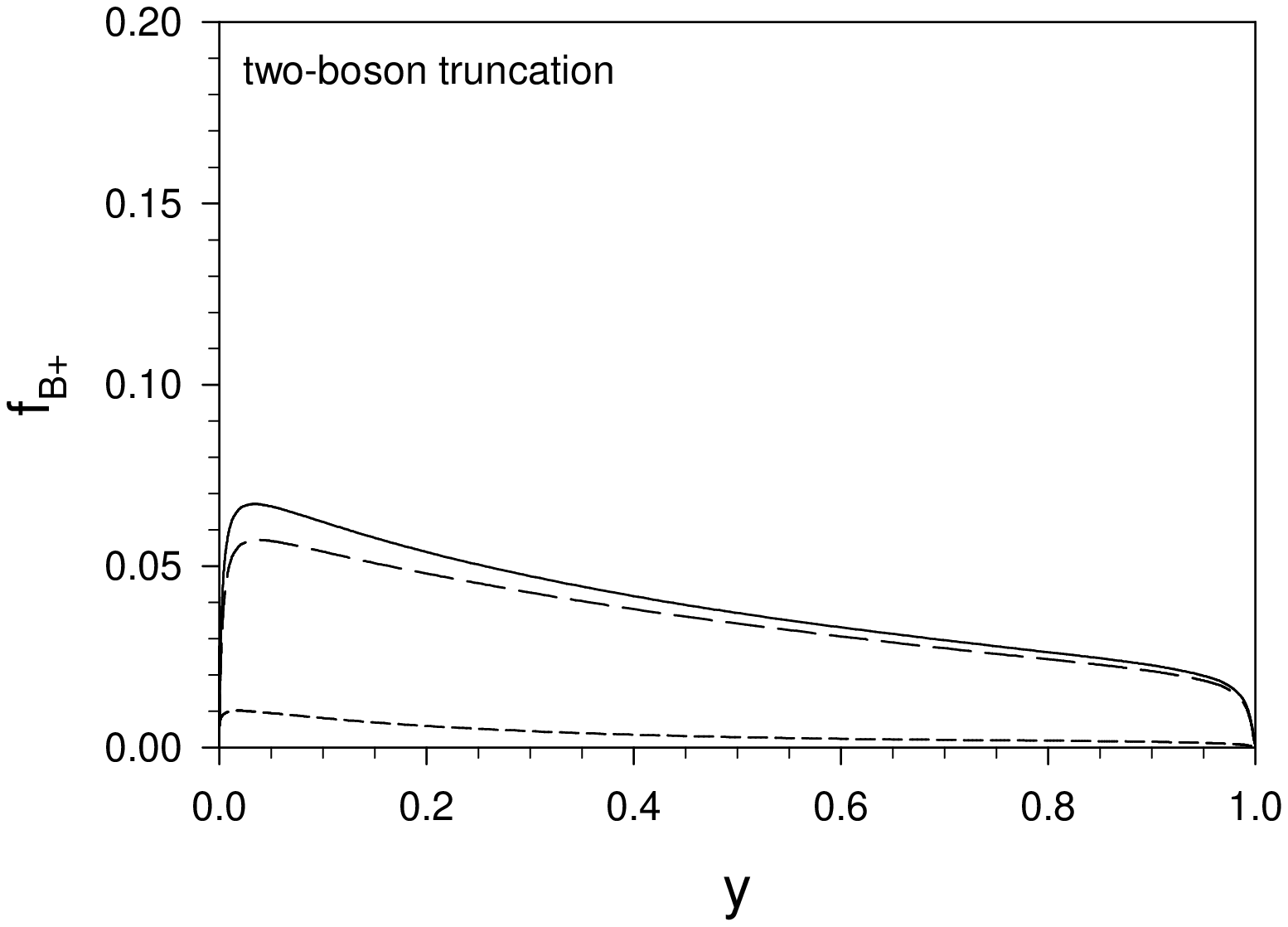} &
\includegraphics[width=6.5cm]{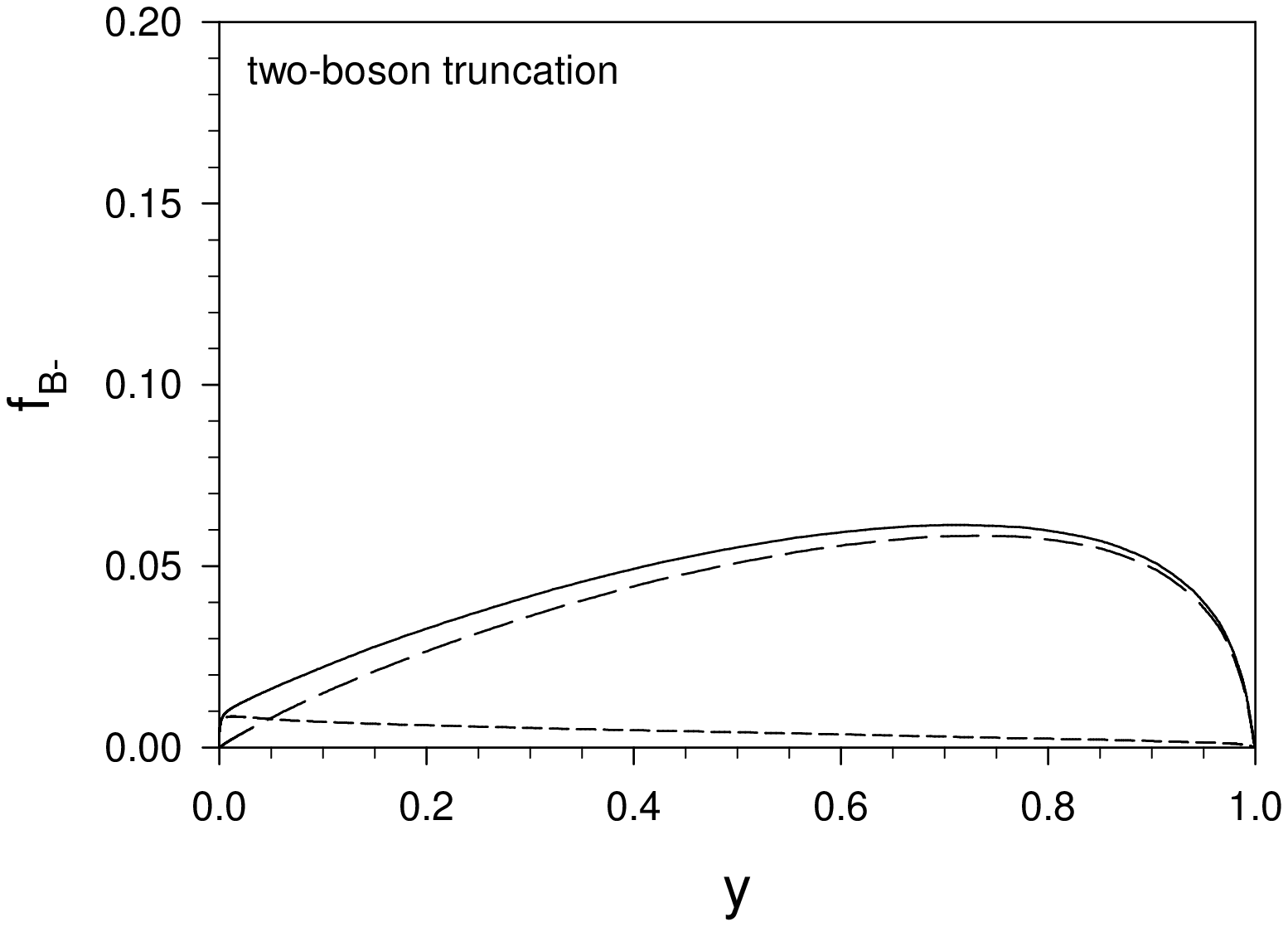} \\
(c) &  (d)
\end{tabular}
\caption{Same as Fig.~\ref{fig:1bfb} but for the wave functions
computed from a truncation to two constituent bosons instead of one,
with the bare coupling $g$ and bare-fermion mass $m_0$ adjusted to 
maintain $M=\mu_0$ and $R=0.01/\mu_0$.  
The resolutions are $K=50$ and $N=30$.}
\label{fig:2bfb}
\end{center}
\end{figure}

\begin{figure}[hpbt]
\begin{center}
\begin{tabular}{cc}
\includegraphics[width=6.5cm]{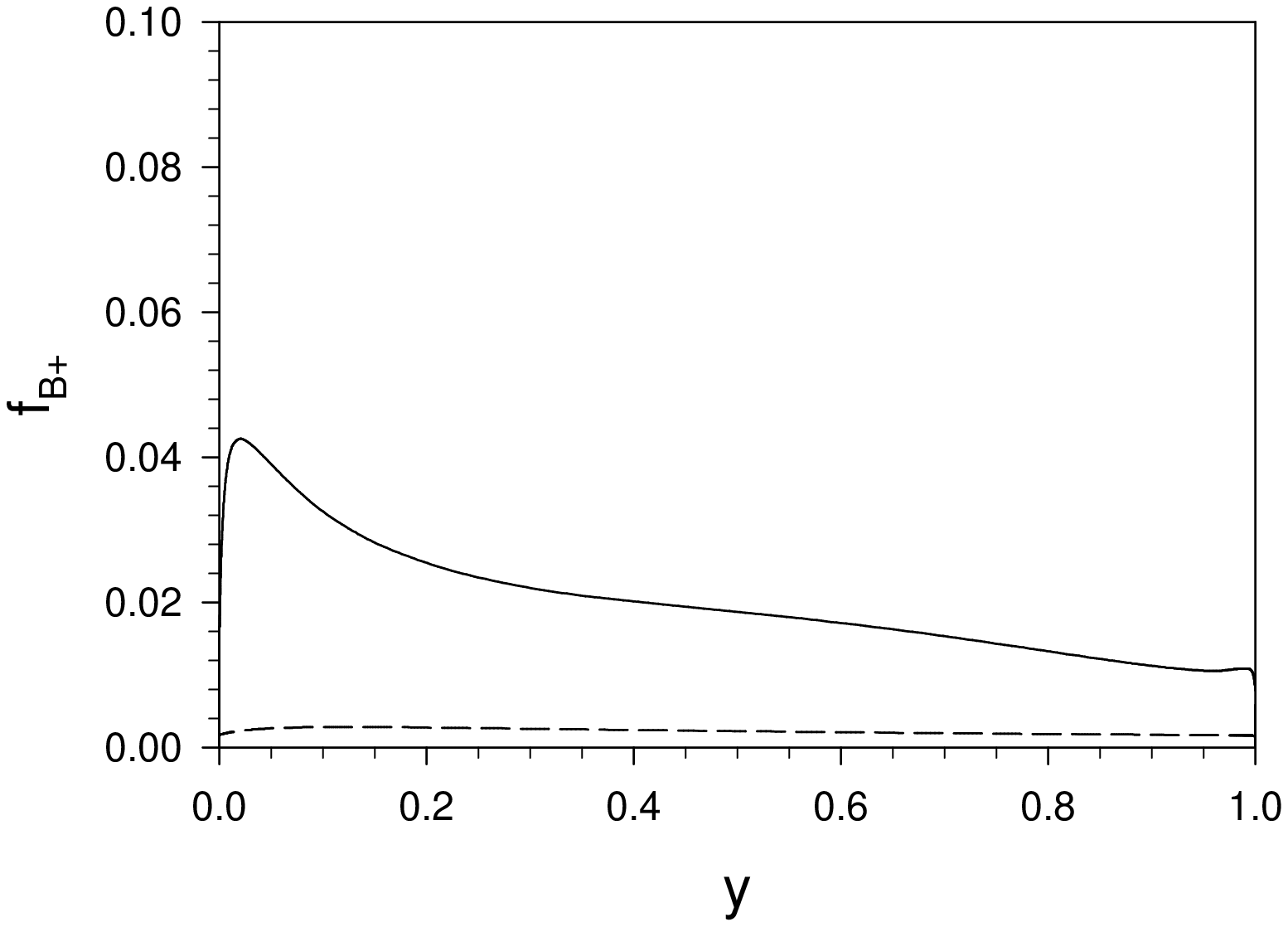} &
\includegraphics[width=6.5cm]{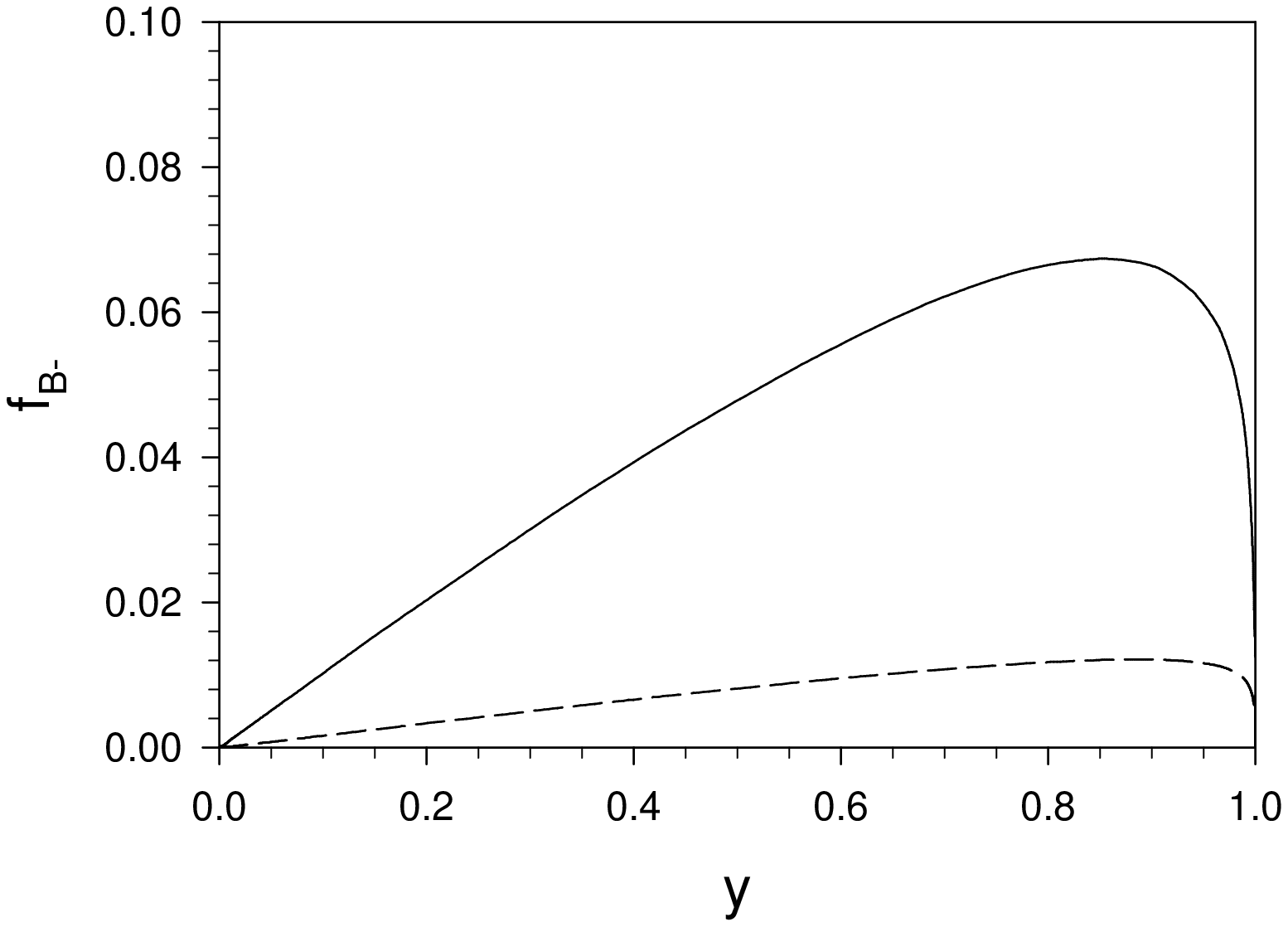} \\
(a) &  (b) \\ \vspace{0.05in} \\
\includegraphics[width=6.5cm]{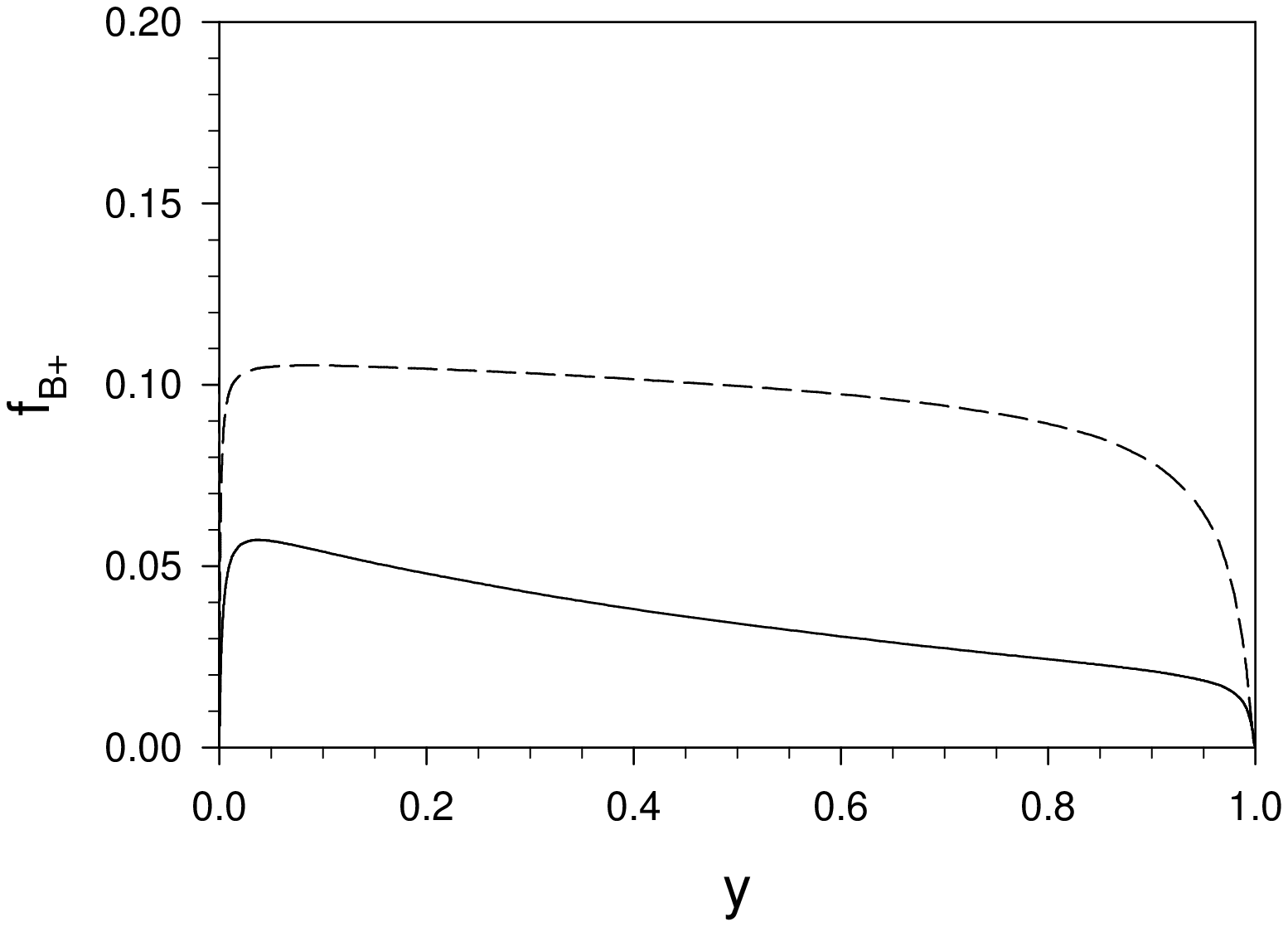} &
\includegraphics[width=6.5cm]{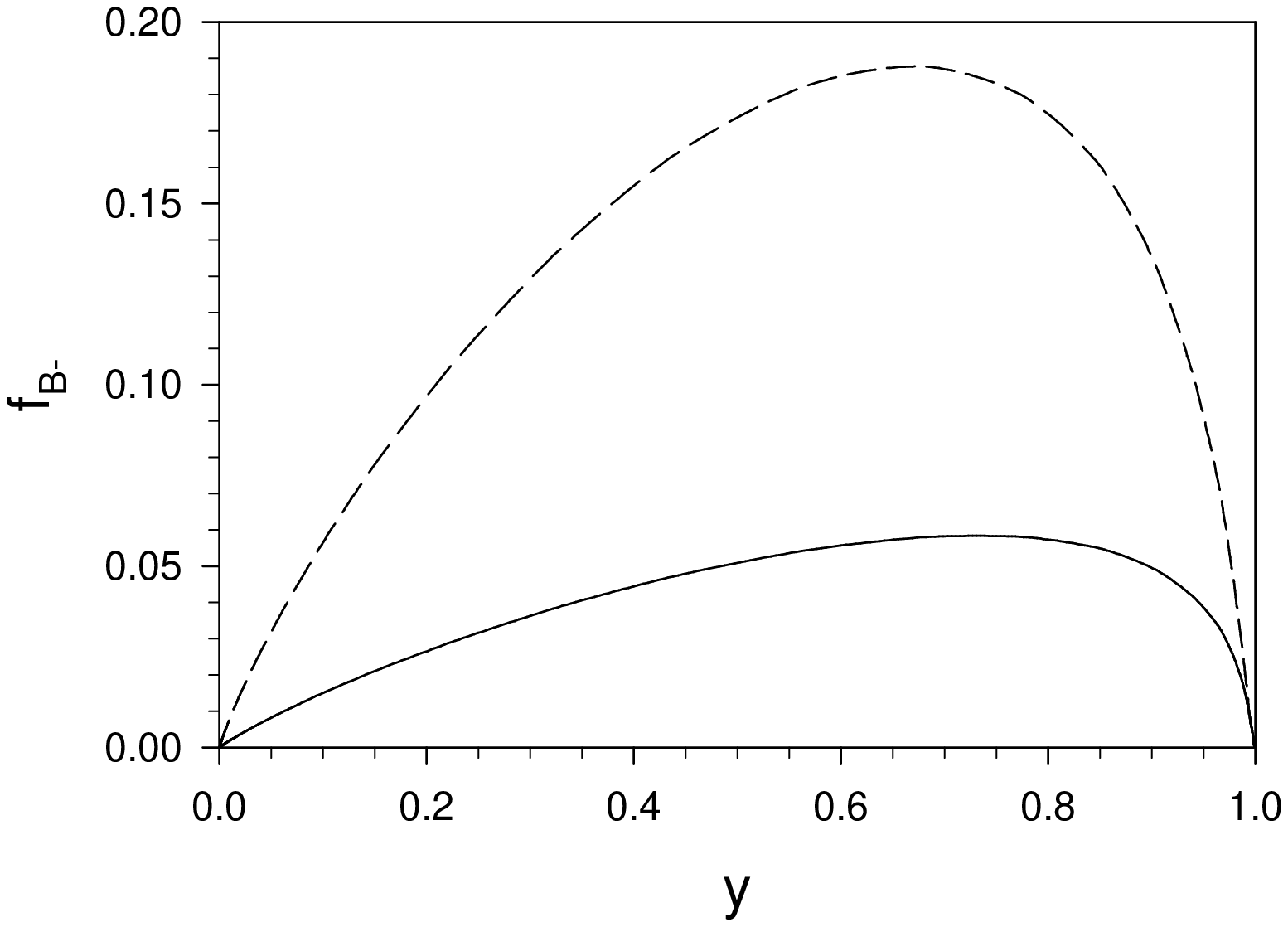} \\
(c) &  (d)
\end{tabular}
\caption{One-boson contributions to the structure functions $f_{B\pm}(y)$
as computed in the one-boson (dashed) and two-boson (solid)
truncations.  The dressed-fermion mass is $M=\mu_0$, and the radius is
$R=0.01/\mu_0$.  
For (a) and (b), the PV masses are $m_1=\mu_1=2000\mu_0$,
and for (c) and (d), they are $m_1=50000\mu_0$ and $\mu_1=500\mu_0$.
The two-boson truncation is calculated with resolutions $K=50$ 
and $N=30$.}
\label{fig:fbone}
\end{center}
\end{figure}

\begin{figure}[hpbt]
\begin{center}
\begin{tabular}{cc}
\includegraphics[width=6.5cm]{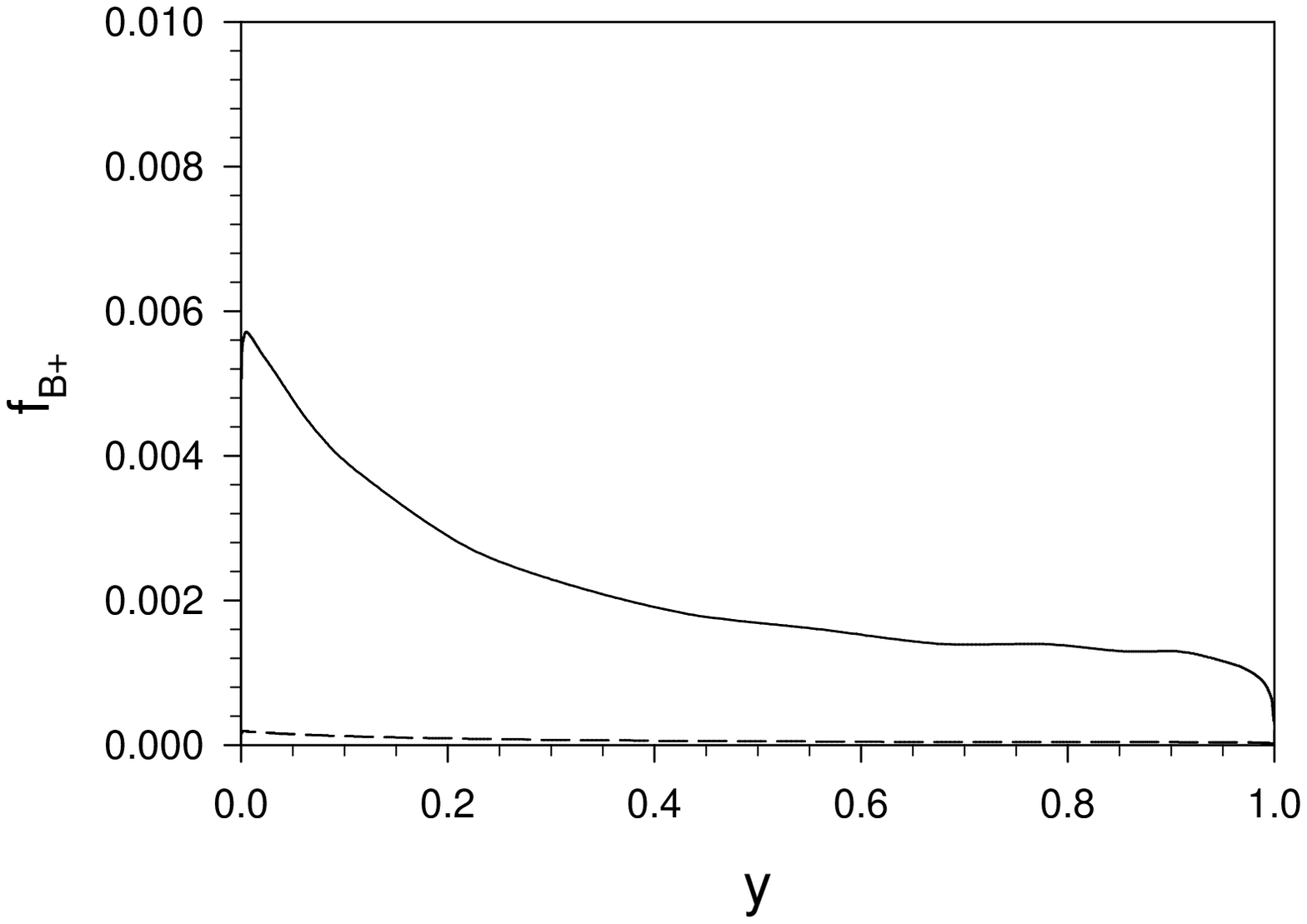} &
\includegraphics[width=6.5cm]{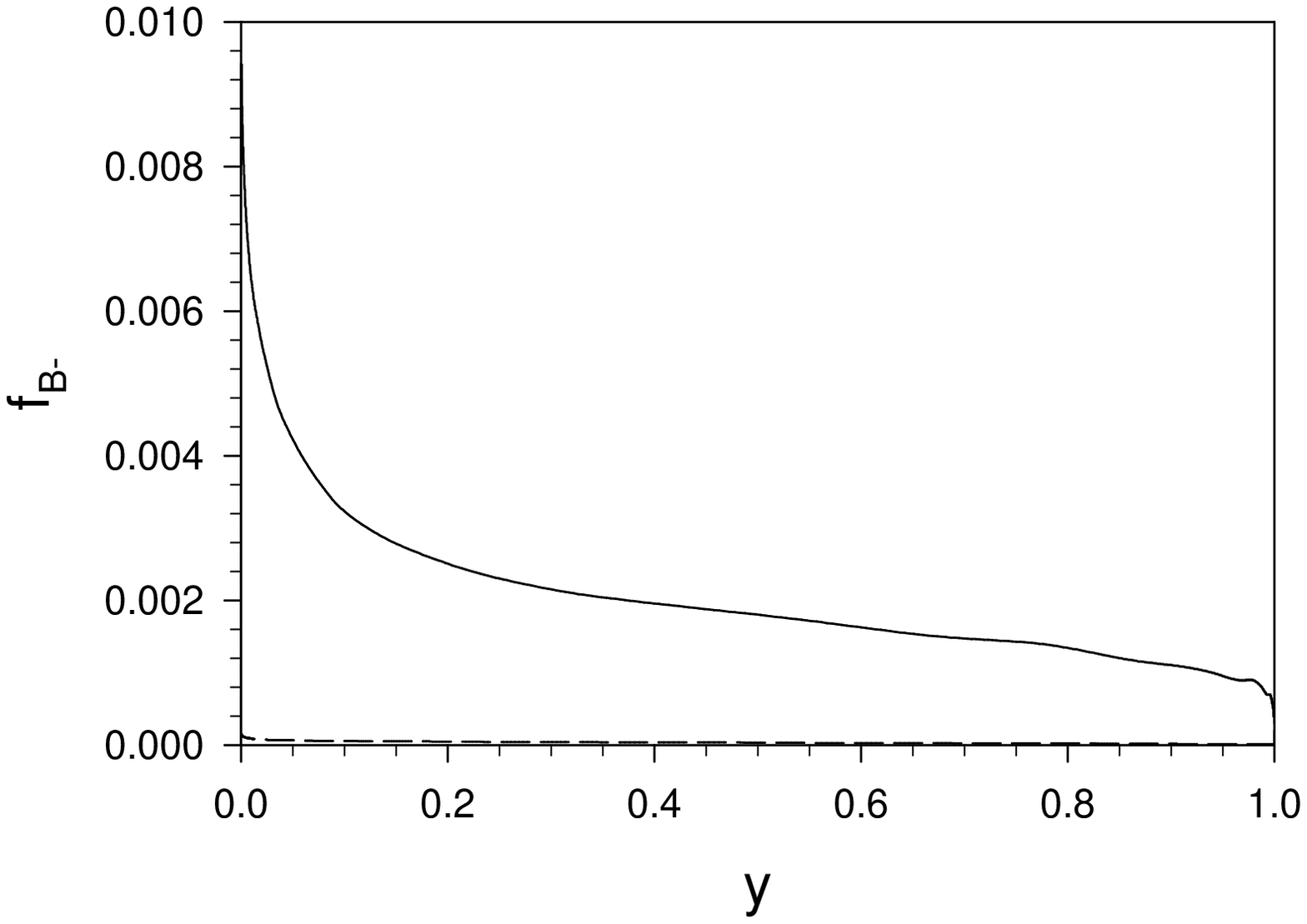} \\
(a) &  (b) \\ \vspace{0.05in} \\
\includegraphics[width=6.5cm]{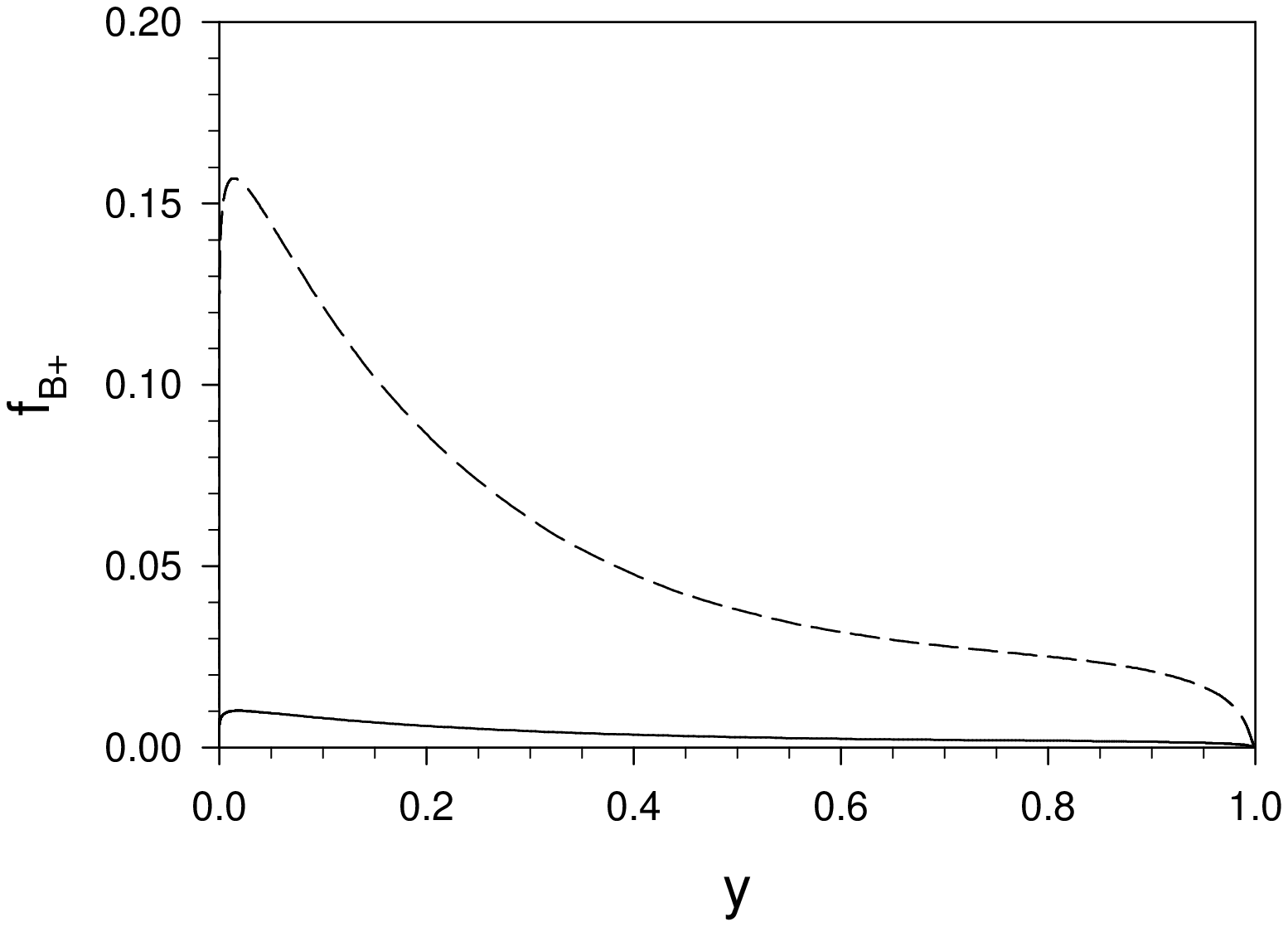} &
\includegraphics[width=6.5cm]{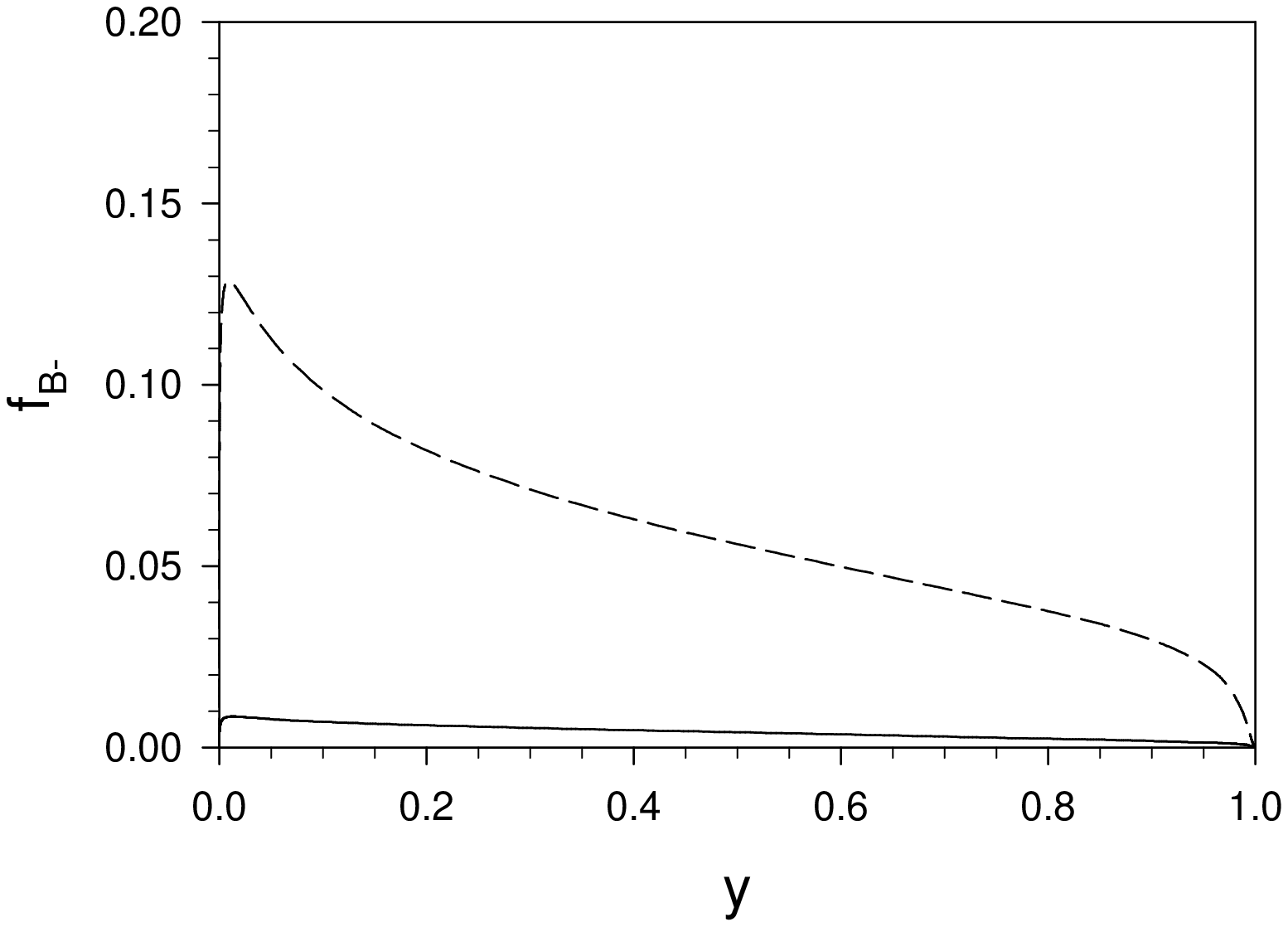} \\
(c) &  (d)
\end{tabular}
\caption{Same as Fig.~\ref{fig:fbone}, but for the two-boson contribution.
In the case of the one-boson truncation, this is computed perturbatively.}
\label{fig:fbtwo}
\end{center}
\end{figure}

A comparison of results for the one and two-boson truncations
at fixed resolution and fixed PV masses is 
given in Figs.~\ref{fig:gm0vsR} and \ref{fig:nbkapvsR}.
The dressed-fermion radius $R$ is held fixed at the same value
of $0.01/\mu_0$ for both truncations.
The one and two-boson contributions to the structure functions
are shown in Fig.~\ref{fig:2bfb}; these should be compared
with the results in Fig.~\ref{fig:1bfb} for the one-boson
truncation.  To see the comparison more easily, the one
and two-boson contributions in each case are plotted separately
in Figs.~\ref{fig:fbone} and \ref{fig:fbtwo}.
The differences in the results, between the one-boson and
two-boson truncations, reflects not only the effects of including
the two-boson contributions to the kernel in the one-boson
equation, but also the differences in the bare fermion mass
and the bare coupling.
In the case of unequal PV masses, as in Figs.~\ref{fig:fbone}(c) and
(d) and \ref{fig:fbtwo}(c) and (d), the perturbative calculation
significantly overestimates the two-boson contribution.

\begin{figure}[hpbt]
\begin{center}
\begin{tabular}{c}
\includegraphics[width=13cm]{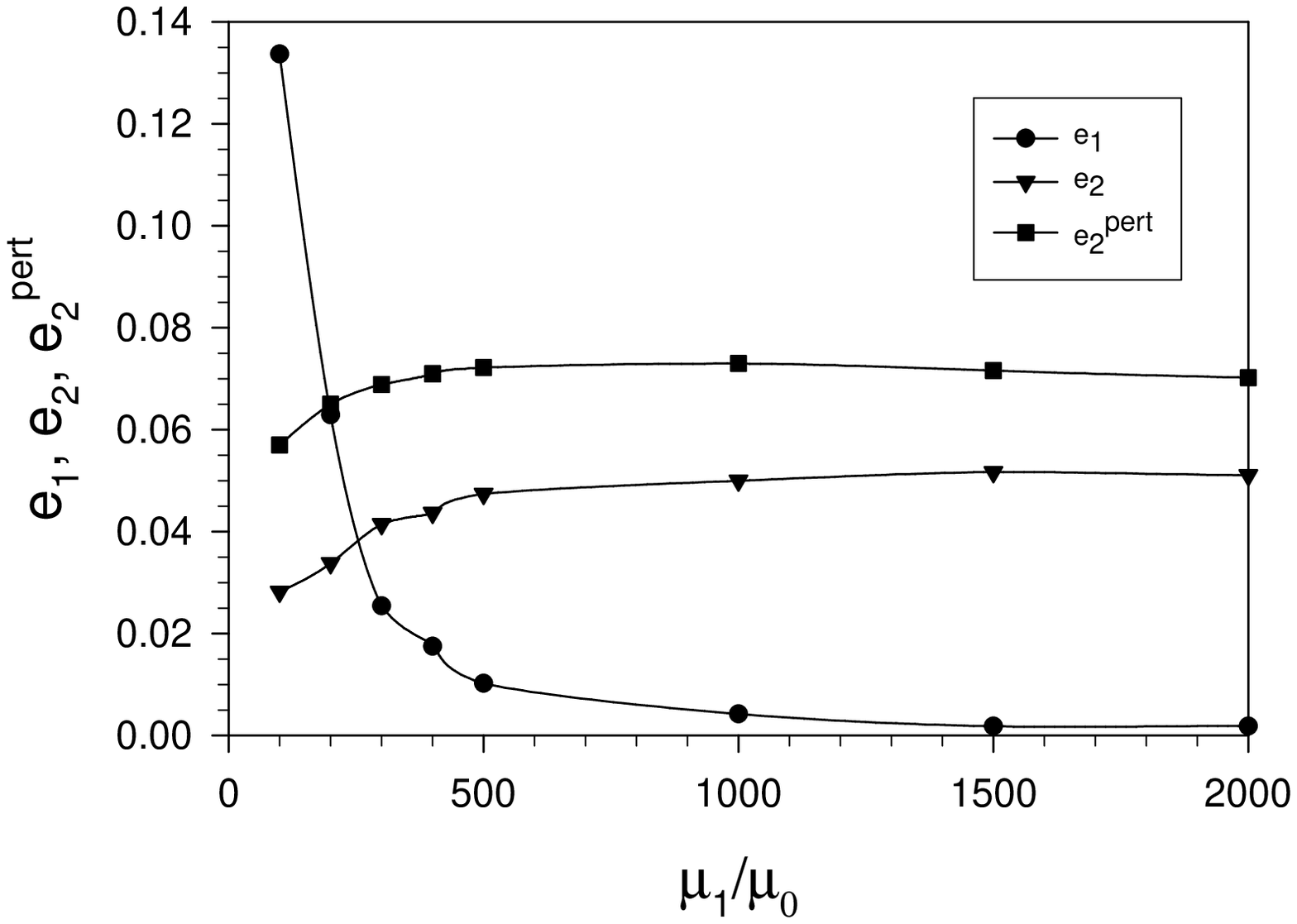}
\end{tabular}
\caption{Measures of error in the truncations used as functions
of the PV boson mass $\mu_1$.  The quantity $e_1$ is a measure of
the PV contribution to the probability of the one-boson sector,
and $e_2$ is a measure of the truncation error, defined as the 
ratio of the total probability of the two-boson sector to that
of the one-boson sector.  The latter is 
computed both in the two-boson truncation and perturbatively in the one-boson
truncation.
The PV fermion mass is fixed at a large value, $m_1=50000\mu_0$,
and the coupling is fixed at $g=2$.  The bare fermion mass $m_0$
is allowed to vary, to maintain the constraint of $M^2=\mu_0^2$
on the dressed-fermion mass.  The dressed-fermion radius is 
allowed to vary.  
The resolutions are $K=50$ and $N=30$.
}
\label{fig:errors}
\end{center}
\end{figure}

In principle,
these solutions can be used to estimate the optimal PV mass for the
one-boson truncation.  As the PV masses are increased, we expect
the PV contributions to the one-boson Fock sector to decrease, but 
the magnitude of the total contribution from the two-boson sector to
increase.  The latter measures the truncation error
for a one-boson calculation
that does not include two-boson states even perturbatively.
(To estimate the optimal PV mass for a two-boson truncation,
we would need to estimate the three-boson contribution, which
would require a much larger calculation.)
These two measures, the PV 
contribution to the probability of the one-boson sector, $e_1$, and 
the ratio of the total probability of the two-boson sector to that
of the one-boson sector, $e_2$, are plotted in 
Fig.~\ref{fig:errors}, along with $e_2^{\rm pert}$, a perturbative
estimate of $e_2$.  We compute $e_1$ as the ratio of the 
PV contribution to the physical contribution, each of which is
individually divergent but regulated by 
a transverse cutoff at $q_\perp^2=(3\mu_1)^2$.  The fermion mass
is held fixed but very large. 
The optimal PV boson mass would be chosen to make the two errors equal;
this apparently corresponds to a mass on the order of $250\mu_0$.
%

\section{Discussion}
\label{sec:Discussion}

We have solved for the dressed fermion state in Yukawa theory using
Pauli--Villars regularization and a truncation to no more than one
fermion and two bosons in the Fock expansion.  The solution yields
the bare coupling $g$, the bare mass $m_0$ of the constituent fermion,
and the Fock-state wave functions as functions of the PV masses 
$\mu_1$ and $m_1$ and the dressed mass $M$ and radius $R$.  A 
limited region of the parameter space
has been explored; however, a wide range of PV mass values was
studied in order to observe the limiting behavior as these masses
approached infinity.  We also studied the effects of truncation
by comparing the solutions obtained previously with no more than
one boson in the basis.

{}From the wave functions, we calculated various properties of the
dressed fermion.  These are shown in Figs.~\ref{fig:gm0vsR} and
\ref{fig:nbkapvsR}.  The one and two-boson truncations agree in
the weak-coupling, large-radius limit, as expected, 
but show significant differences
for stronger coupling, even though a two-boson contribution is 
included perturbatively in the case of the one-boson truncation.
The presence of the two-boson intermediate states in the kernel
of the effective integral equation does have important effects.

The use of light-cone quantization,
Pauli--Villars regularization, and carefully crafted
quadrature schemes can produce accurate solutions for bound
states in quantum field theory.  These techniques should be
applicable to more interesting situations, such as the
dressed electron in quantum electrodynamics~\cite{qed}
and two-fermion bound states in Yukawa theory and QED.
As applications to a gauge theory become better understood,
we can hope to develop methods sufficiently robust to
solve for the bound states of quantum chromodynamics.

\section*{Acknowledgments}
This work was supported by the Department of Energy
through contracts DE-AC02-76SF00515 (S.J.B.),
DE-FG02-98ER41087 (J.R.H.), and DE-FG03-95ER40908 (G.M.)
and by the Minnesota Supercomputing Institute through
grants of computing time.

\section*{Appendix A: Interaction kernels} \label{sec:A}

The self-energy in the reduced, coupled system Eq.~(\ref{eq:EffEq}) is
\bea
I_{ija}&=&\sum_{i',b}(-1)^{i'+a+b}(1-y)\int_0^{1-y}
    \frac{dy'}{y'}\frac{d\phi'}{2\pi}dq_\perp^{\prime 2}
      \frac{-1}{D_{i'jb}+F\cos\phi'} \\
  &&\times\left[\left(\frac{m_{i'}}{1-y-y'}+\frac{m_i}{1-y}\right)
      \left(\frac{m_{i'}}{1-y-y'}+\frac{m_a}{1-y}\right) \right. \nonumber \\
  && \left. +\frac{1}{(1-y-y')^2}\left(
      \frac{y^{\prime 2} q_\perp^2}{(1-y)^2}
      +q_\perp^{\prime 2}
      +\frac{2y'q_\perp q'_\perp \cos\phi'}{(1-y)}\right)\right]. \nonumber 
\eea
The bare-fermion kernel is
\bea \label{eq:J0}
J_{ij+,ab+}^{(0)}&=&\sum_{i'}\frac{(-1)^{i'+a+b}}{\sqrt{yy'}}
  \frac{1}{M^2-m_{i'}^2}\left(\frac{m_i}{1-y}+m_{i'}\right)
      \left(\frac{m_a}{1-y'}+m_{i'}\right), \\
J_{ij+,ab-}^{(0)}&=&\sum_{i'}\frac{(-1)^{i'+a+b}}{\sqrt{yy'}}
  \frac{1}{M^2-m_{i'}^2}\left(\frac{m_i}{1-y}+m_{i'}\right)
      \frac{q'_\perp}{1-y'}, \nonumber \\
J_{ij-,ab+}^{(0)}&=&\sum_{i'}\frac{(-1)^{i'+a+b}}{\sqrt{yy'}}
  \frac{1}{M^2-m_{i'}^2}\frac{q_\perp}{1-y}
                   \left(\frac{m_a}{1-y'}+m_{i'}\right),\nonumber \\
J_{ij-,ab-}^{(0)}&=&\sum_{i'}\frac{(-1)^{i'+a+b}}{\sqrt{yy'}}
  \frac{1}{M^2-m_{i'}^2}\frac{q_\perp}{1-y}\frac{q'_\perp}{1-y'}, \nonumber 
\eea
and the two-boson kernel is
\bea
J_{ij+,ab+}^{(2)}&=&\sum_{i'}\frac{(-1)^{i'+a+b}}{\sqrt{yy'}}
\int\frac{d\phi'}{2\pi}\frac{-1}{D_{i'jb}+F\cos\phi'} \\
&&\times\left(\left[\frac{m_{i'}}{1-y-y'}+\frac{m_a}{1-y'}\right]
      \left[\frac{m_{i'}}{1-y-y'}+\frac{m_i}{1-y}\right] \right. \nonumber \\
  &&  +\frac{1}{(1-y-y')^2}\left[
      \frac{y'q_\perp^2}{1-y}+\frac{yq_\perp^{\prime 2}}{1-y'} \right.
      \nonumber \\
  &&  \left. \left. \rule{1.25in}{0in}
    +\frac{(1-y-y'+2yy')q_\perp q'_\perp }{(1-y)(1-y')}
        \cos\phi'\right]\right),  \nonumber \\
J_{ij+,ab-}^{(2)}&=&\sum_{i'}\frac{(-1)^{i'+a+b}}{\sqrt{yy'}}
\int\frac{d\phi'}{2\pi}\frac{-1}{D_{i'jb}+F\cos\phi'} \\
&&\times\left(-\left[\frac{m_{i'}}{1-y-y'}+\frac{m_i}{1-y}\right]
    \frac{yq'_\perp}{(1-y-y')(1-y')}\right.\nonumber \\
  && +\left[\frac{m_{i'}}{1-y-y'}+\frac{m_a}{1-y'}\right]
     \frac{q'_\perp}{1-y-y'}  \nonumber \\
  && -\left[\frac{m_{i'}}{1-y-y'}+\frac{m_i}{1-y}\right]
      \frac{q_\perp\cos\phi'}{1-y-y'} \nonumber \\
  &&  \left.  +\left[\frac{m_{i'}}{1-y-y'}+\frac{m_a}{1-y'}\right]
      \frac{y'q_\perp\cos\phi'}{(1-y-y')(1-y)}
      \right),  \nonumber \\
J_{ij-,ab+}^{(2)}&=&\sum_{i'}\frac{(-1)^{i'+a+b}}{\sqrt{yy'}}
\int\frac{d\phi'}{2\pi}\frac{-1}{D_{i'jb}+F\cos\phi'} \\
&&\times\left(-\left[\frac{m_{i'}}{1-y-y'}+\frac{m_a}{1-y'}\right]
    \frac{y'q_\perp}{(1-y-y')(1-y)} \right. \nonumber \\
  &&  +\left[\frac{m_{i'}}{1-y-y'}+\frac{m_i}{1-y}\right]
     \frac{q_\perp}{1-y-y'}  \nonumber \\
  && -\left[\frac{m_{i'}}{1-y-y'}+\frac{m_a}{1-y'}\right]
      \frac{q'_\perp\cos\phi'}{1-y-y'} \nonumber \\
  &&  \left.  +\left[\frac{m_{i'}}{1-y-y'}+\frac{m_i}{1-y}\right]
      \frac{yq'_\perp\cos\phi'}{(1-y-y')(1-y')}
      \right),  \nonumber \\
J_{ij-,ab-}^{(2)}&=&\sum_{i'}\frac{(-1)^{i'+a+b}}{\sqrt{yy'}}
\int\frac{d\phi'}{2\pi}\frac{-1}{D_{i'jb}+F\cos\phi'} \\
&&\times\left(\left[\frac{m_{i'}}{1-y-y'}+\frac{m_a}{1-y'}\right]
      \left[\frac{m_{i'}}{1-y-y'}+\frac{m_i}{1-y}\right]\cos\phi'
       \right. \nonumber \\
  &&  +\frac{1}{(1-y-y')^2}\left[
      \frac{y'q_\perp^2}{1-y}+\frac{yq_\perp^{\prime 2}}{1-y'}\right]
      \cos\phi'\nonumber \\
  && \left.-\frac{q_\perp q'_\perp }{(1-y-y')(1-y)(1-y')}
      +\frac{2q_\perp q'_\perp\cos^2\phi'}{(1-y-y')^2}\right).  \nonumber
\eea
Here we have used
\bea
D_{ijk}&=&\frac{m_i^2+q_\perp^2+q_\perp^{\prime 2}}{1-y-y'}
    +\frac{\mu_j^2+q_\perp^2}{y}+\frac{\mu_k^2+q_\perp^{\prime 2}}{y'}-M^2, 
\\
F&=&\frac{2q_\perp q'_\perp}{1-y-y'}, \nonumber
\eea
as well as a shift in the azimuthal angle $\phi'-\phi\rightarrow\phi'$.  
The angular integrals can be done analytically; they are
\bea
\int_0^{2\pi}\frac{d\phi'}{2\pi}\frac{1}{D+F\cos\phi'}
  &=&\frac{1}{\sqrt{D^2-F^2}}, \\
\int_0^{2\pi}\frac{d\phi'}{2\pi}\frac{\cos\phi'}{D+F\cos\phi'}
  &=&\frac{1}{F}\left(1-\frac{D}{\sqrt{D^2-F^2}}\right), \\
\int_0^{2\pi}\frac{d\phi'}{2\pi}\frac{\cos^2\phi'}{D+F\cos\phi'}
  &=&-\frac{D}{F^2}\left(1-\frac{D}{\sqrt{D^2-F^2}}\right).
\eea

The integrals in the self-energy $I_{ija}$ also can be done analytically.
After a change of variables to $\xi=q^{\prime +}/q^+=y'/(1-y)$ and
$\vec{k}_\perp=\vec{q}_\perp^{\,\prime}+\xi\vec{q}_\perp$, we obtain
\bea
I_{ija}
&=&-\sum_{i',b}(-1)^{i'+a+b}\int_0^1\frac{d\xi}{\xi}\frac{d^2k_\perp}{\pi}
\frac{m_i m_a + \frac{m_i+m_a}{1-\xi}m_{i'}+\frac{m_i'^2+k_\perp^2}{(1-\xi)^2}}
{M_j^2+\frac{m_{i'}^2+k_\perp^2}{1-\xi}+\frac{\mu_b^2+k_\perp^2}{\xi}} \\
&=&16\pi^2(-1)^a\left[m_i m_a I_0(-M_j^2)+\mu_0(m_i+m_a)I_1(-M_j^2)
            +\mu_0^2 J(-M_j^2)\right],  \nonumber
\eea
with
\be
M_j^2\equiv \frac{\mu_j^2+q_\perp^2}{y}-\mu_j^2-(1-y)M^2 > 0,
\ee
and $I_0$, $I_1$, and $J$ defined in Eqs.~(\ref{eq:In}) and (\ref{eq:J}).
For $J$ we have $\mu_0^2 J(-M_j^2)=-M_j^2 I_0(-M_j^2)$.  For $I_0$ 
and $I_1$, integration over $\vec{k}_\perp$ yields
\bea \label{eq:I0sum}
16\pi^2 I_0&=&\sum_{i'b}(-1)^{i'+b}(L_{0i'b}-L_{1i'b}), \\
\label{eq:I1sum}
16\pi^2 I_1&=&\sum_{i'b}(-1)^{i'+b}\frac{m_{i'}}{\mu_0} L_{0i'b},
\eea
where
\be \label{eq:Ldef}
L_{ni'b}\equiv\int_0^1 dz z^n \ln[z m_{i'}^2/M_j^2 + (1-z)\mu_b^2/M_j^2+z(1-z)].
\ee
Integration over $z$ obtains
\bea \label{eq:L0}
L_{0i'b}&=& (1-y_{ji'b})\ln\left(\frac{m_{i'}^2}{M_j^2}\right)
   +y_{ji'b}\ln\left(\frac{\mu_b^2}{M_j^2}\right)-2 \\
 && -\sqrt{x_{ji'b}}
     \ln\left[\frac{\sqrt{x_{ji'b}}+y_{ji'b}-1}
                  {\sqrt{x_{ji'b}}-y_{ji'b}+1}
              \frac{\sqrt{x_{ji'b}}-y_{ji'b}}
                  {\sqrt{x_{ji'b}}+y_{ji'b}}\right],\nonumber \\
\label{eq:L1}
L_{1i'b}&=& y_{ji'b} L_{0i'b}
  -\frac{1}{2}\left[\frac{m_{i'}^2}{M_j^2}\ln\left(\frac{m_{i'}^2}{M_j^2}\right)
                             -\frac{m_{i'}^2}{M_j^2}
                    -\frac{\mu_b^2}{M_j^2}\ln\left(\frac{\mu_b^2}{M_j^2}\right)
                             +\frac{\mu_b^2}{M_j^2}\right],
\eea
with $y_{ji'b}\equiv(m_{i'}^2-\mu_b^2+M_j^2)/(2M_j^2)$
and $x_{ji'b}\equiv y_{ji'b}^2+\mu_b^2/M_j^2$.

In practice one needs to use an alternate form when $M_j^2$ is
small, as happens when $q_\perp^2$ is zero and $y$ is near 1.  This
alternate form is obtained as an expansion in powers of $M_j^2$
of the integrand in (\ref{eq:Ldef}), written as
\be \label{eq:L-smallMj}
L_{ni'b}=\int_0^1 dz z^n \ln[z m_{i'}^2 + (1-z)\mu_b^2+M_j^2z(1-z)]
               -\frac{\ln\left(M_j^2\right)}{n+1}.
\ee
The dependence on $\ln\left(M_j^2\right)$ cancels in the sums in
(\ref{eq:I0sum}) and (\ref{eq:I1sum}).  When 
$M_j^2z(1-z)/\left(z m_{i'}^2 + (1-z)\mu_b^2\right)$ is of order
0.01 or smaller at $z=1/2$, two-term expansions of the first term
in (\ref{eq:L-smallMj}) in powers of
$M_j^2$ are sufficient as replacements for (\ref{eq:L0}) and
(\ref{eq:L1}).

There is also a special form needed when $M_j^2$ is large,
such as when $z$ is near zero.  When $M_j^2$ is greater than
$10m_1^2$ or $10\mu_1^2$, expansions
of the analytic expressions (\ref{eq:L0}) and (\ref{eq:L1})
to order $\mu_0^2/M_j^2$ and $\ln(M_j/\mu_0)\mu_0^2/M_j^2$ are used.

\section*{Appendix B: Quadrature schemes} \label{sec:B}

To solve the integral equations (\ref{eq:EffEq}), we convert them
to a matrix equation via quadrature in $y'$ and $q_\perp^{\prime 2}$
and then diagonalize the matrix.  For integrals with an upper
limit of $y'=1-y$, the form of the integrand at $1-y$ is obtained
by explicitly taking the limit; the PV counterterms insure that
this limit is finite.  The quadrature weight is reduced by 1/2
at that point, to take into account the edge effect.
 
A particularly useful set
of quadrature schemes is based on Gauss--Legendre quadrature
combined with variable transformations to allocate quadrature
points to important regions and to reduce the $q_\perp^{\prime 2}$ 
integral to a finite range.  The transformations for the $y'$
integral are
\bea
y'(t)&=&t^3(1+dt)/[1+d-(3+4d)t+(3+6d)t^2-4dt^3+2dt^4] ,  \\
t(u)&=&2u-1.
\eea
The new variable $u$ ranges between -1 and 1, which is the
nominal range for Gauss--Legendre quadrature.  The transformation
from $y'$ to $t$ is motivated by the need for an accurate approximation
to the integral $J$, defined in (\ref{eq:J}).  This integral is
largely determined by contributions near the endpoints whenever the
PV masses are large.  The transformation $y'(t)$ places many of
the quadrature points near 0 and 1.  It was found empirically, 
beginning with a transformation constructed to compute the
integral $\int[\ln(y+\epsilon_0)-\ln(y+\epsilon_1)] dy$ exactly
from the quadrature formula, with $\epsilon_0$ and $\epsilon_1$
small.  The final form is restricted to be symmetric under the
transformation $t\rightarrow(1-t)$.  The parameter $d$ is chosen
such that $y'\simeq 0.01 t^3$ for small $t$.

For the transverse integral, the transformation is
\be
q_\perp^{\prime 2}(v)=a^2\frac{1-\left(b^2/a^2\right)^v}
                         {\left(b^2/a^2\right)^{v-1}-1},
\ee
with $v$ in the range 0 to 1.  Only the
positive Gauss--Legendre quadrature points of an odd order
are used for $v$ between -1 and 1, so that $v=0$, and 
therefore $q'_\perp=0$, is always a quadrature point.  
The points in the negative half of the range are discarded.
To maintain the accuracy of the underlying sum, the weight of
the point at zero is reduced by a factor of 1/2.
The transformation from $q_\perp^{\prime 2}$ to $v$
is motivated by its ability to obtain an exact result
for the integral $\int [1/(a^2+q^2)-1/(b^2+q^2)]dq^2$.
For use in the integral equation (\ref{eq:EffEq}),
the parameters $a$ and $b$ are chosen to be the smallest 
and largest mass scales, i.e. $a=\mu_0$ and $b=m_1$.


\end{document}